\documentclass[9pt]{article}
\usepackage{arxiv}

\onecolumn 

\usepackage{graphicx}
\usepackage{amsmath}
\usepackage{natbib}
\usepackage{amssymb}
\usepackage{amsbsy}
\usepackage{lmodern,bm} 
\usepackage{enumitem}
\usepackage{relsize}
\usepackage{exscale}
\newlist{steps}{enumerate}{1}
\setlist[steps, 1]{label = Step \arabic*:}
\usepackage[]{subfig}
\usepackage{url} 
\usepackage{rotating}

\usepackage{mathtools,tikz,caption}
\DeclareRobustCommand\sampleline[1]{%
  \tikz\draw[#1] (0,0) (0,\the\dimexpr\fontdimen22\textfont2\relax)
  -- (2em,\the\dimexpr\fontdimen22\textfont2\relax);%
}

\usepackage{float}
\usepackage{bigints}
\usepackage{algorithm}
\usepackage{bigints}

\usepackage{caption}
\DeclareCaptionType[within=section]{mat}[Matrix]

\newcommand{\norm}[1]{\left\lVert#1\right\rVert}

\newcommand{\inv}{^{\raisebox{.2ex}{$\scriptscriptstyle-1$}}}

\usepackage{kbordermatrix}
\usepackage{float}

\title{Linking climate and dengue in the Philippines using a two-stage Bayesian spatio-temporal model}

\author{
 Stephen Jun Villejo \\
  School of Mathematics and Statistics/School of Statistics\\
  University of Glasgow/University of the Philippines\\
  \texttt{stephen.villejo@glasgow.ac.uk} \\
  orcid.org/0000-0002-0510-3143\\
   \And
 Sara Martino \\
  Department of Mathematical Sciences\\
  Norwegian University of Science and Technology\\
\texttt{sara.martino@ntnu.no} \\
orcid.org/0000-0003-4326-9029\\
  \And
 Janine Illian \\
  School of Mathematics and Statistics\\
  University of Glasgow\\
  \texttt{janine.illian@glasgow.ac.uk} \\
  orcid.org/0000-0002-6130-2796\\
}

\begin{document}

\fontsize{10pt}{10pt}\selectfont

\maketitle

\begin{abstract}
    Dengue is an infectious disease which poses significant socioeconomic and disease burden in many tropical and subtropical regions of the world. This work aims to provide additional insight into the association between dengue and climate in the Philippines. We employ a two-stage modelling framework: the first stage fits climate models, while the second stage fits a health model that uses the climate predictions from the first stage as inputs. We postulate a Bayesian spatio-temporal model and use the integrated nested Laplace approximation (INLA) approach for inference. To account for the uncertainty in the climate models, we perform posterior sampling and then perform Bayesian model averaging to compute the final posterior estimates of second-stage model parameters. The results indicate that temperature is positively associated with dengue, although extremely hot conditions tend to have a negative effect. Moreover, the relationship between rainfall and dengue varies in space. In areas with uniform amounts of rainfall all year round, rainfall is negatively associated with dengue. In contrast, in regions with pronounced dry and wet season, rainfall shows a positive association with dengue. Finally, there remains unexplained structured variation in space and time after accounting for the impact of climate variables and other covariates.
\end{abstract}


\section{Introduction}\label{sec:intro}

Dengue fever, an infectious disease caused by the dengue arbovirus and commonly transmitted by two mosquito species (\textit{Aedes aegypti} and \textit{Aedes albopictus}), poses a strong public health threat as well as a substantial socioeconomic and disease burden in many tropical and subtropical areas around the world \citep{murray2013epidemiology}. Over half of the world’s population live in areas at risk of the disease \citep{USCDCP}. From 1990 to 2019, the estimated increase in dengue incidence is 85.5\%; and from 2021, the number of global cases has been reported to double each year \citep{DengueEClinicalMed}. It is estimated that by 2085, 50 -- 60\% of the world population will be at risk of dengue due to climate change scenarios, holding all risk factors constant \citep{hales2002potential}. According to the European Centre for Disease Prevention and Control, dengue is the most important mosquito-borne viral disease affecting humans worldwide. Each year, tens of millions of cases are reported, resulting in approximately 20 000 -- 25 000 deaths, mainly among children \citep{ECDPC_factsheet}. 
This work focuses on dengue in the  Philippines, a tropical country in Southeast Asia  that has consistently been among the nations with the highest dengue incidence in the region \citep{undurraga2013use, undurraga2017disease}. In the period 2008 -- 2012, the country's Health Department reported an annual average of 117,065 dengue cases, and a fatality rate of 0.55\%  \citep{edillo2015economic}. The last dengue epidemic in the country occurred in 2019, with 437,563 recorded cases,  the highest number ever recorded worldwide \citep{ong2022perspectives}.

  Dengue is classified as a neglected tropical disease (NTD), a group of diseases caused by pathogens and primarily affecting  impoverished areas in tropical countries \citep{WHO_NTD}.   \textit{Aedes} mosquitoes breed in small water bodies -- even as small as containers, car tyres, etc., in and around houses. This allows the virus to be easily transmitted and spread among and within communities in urban locations despite the relatively limited flight range of the vector.
 Its complex epidemiology presents significant challenges for public health control. It is important to understand and identify risk factors of dengue, in order to inform public health policy aimed at controlling disease transmission and predicting future outbreaks.  
 
Since the pathogen causing dengue spends part of its life cycle in the external environment \citep{mcmichael2003climate}, its abundance is particularly influenced by climatic factors. Hence, the association between dengue and climate variables, particularly temperature, rainfall, and relative humidity (RH), has been extensively studied in the literature \citep{naish2014climate,abdullah2022association,murray2013epidemiology, couper2021will, xu2017climate, colon2021projecting}; details are presented in Section \ref{sec:literature}.  

Studies relating dengue to climate factors in the Philippines have mainly employed relatively simple statistical methodology and limited the analysis to a subarea rather than the entire country, relying on rather sparse covariate data measured at a small number of weather stations.  These analyses include correlation analysis or classical linear models such as MANOVA \citep{edillo2024detecting,murphy2022climate,su2008correlation, dulay2013climate, duque2020correlation, edillo2022temperature, marigmen2022forecasting,marigmen2022climatic}, generalized additive modelling  \citep{cruz2024current,cawiding2025disentangling, carvajal2018machine}, deterministic climate-dengue risk functions \citep{xu2020high},  generalized linear models like a Poisson regression \citep{iguchi2018meteorological,francisco2021dengue},  classical time series approaches such as ARIMA models \citep{pineda2019modeling}, or spectral analysis methods, which also focus on the temporal and/or seasonal variations of the relationship between climate and dengue \citep{sumi2017effect, subido2022correlation, francisco2021dengue}, or machine learning algorithms \citep{carvajal2018machine, buczak2014prediction}. \cite{seposo2023socio} employed a mixed modelling framework, but they only considered unstructured random intercepts in space and time. 

This work aims to contribute to the literature on the association between climate variables and dengue using data from the entire area of the Philippines. In particular, the novelty of this work lies in the use of a complex two-stage Bayesian spatio-temporal model that incorporates both structured and unstructured random effects across space and time, including their interactions. The models account for the complex spatial structure of the Philippines, an archipelago consisting of more than 7000 islands, while employing a data-fusion approach that uses climate data from weather stations as well as a weather prediction model.

More specifically, the first stage fits climate models, and produces predicted surfaces of the climate variables, while  the second stage fits a health model, where dengue incidence is the outcome, and climate predictions from the first-stage model are the primary covariates of interest. The climate model uses data from both weather stations and outcomes of a numerical prediction model called the \textit{Global Spectral Model}, in a process referred to as \textit{data fusion}. In order to propagate the uncertainty from the first-stage model to the second-stage model, we use a resampling approach \citep{blangiardo2016two,cameletti2019bayesian, liu2017incorporating, zhu2003hierarchical, villejo2023data}. This approach generates several samples from the first-stage model posteriors, where each sample is used as an input to the second-stage model. The final second-stage model posteriors are then obtained via model averaging. 

A two-stage modelling framework is typically used in spatial analysis, particularly in cases where the response variable and predictors are spatially misaligned, and is common in spatial epidemiology \citep{szpiro2011efficient, gryparis2009measurement, cameletti2019bayesian, blangiardo2016two, lee2017rigorous, liu2017incorporating}. In this work, dengue incidence (the response) is areal, while the climate variables  are point-referenced, resulting in  spatial misalignment. A joint modelling approach -- also known as a fully Bayesian approach -- which simultaneously fits the climate and health model, is  computationally expensive \citep{gryparis2009measurement}. Moreover, it is not practical when there are several epidemiological models of interest, as it would require refitting the  complex climate model multiple times \citep{liu2017incorporating, blangiardo2016two}. Another potential issue of a joint model is occurrence of \textit{feedback effects}, where the health data (dengue incidence) distort the prediction for the climate variables, particularly when the first-stage data are sparse \citep{wakefield2006health, shaddick2002modelling, gryparis2009measurement}. 

In most studies examining the relationship between dengue and climate, including those focusing on the Philippines, data from meteorological stations are the primary source of climate data. The measurements taken at these stations are considered the gold standard due to their high accuracy. However, weather station networks are typically sparse,  as is the case in the Philippines (see Figure \ref{fig:climate_data}a). Models trained on sparse data often produce predictions with high uncertainty and potential biases \citep{lawson2016handbook}. To mitigate these challenges, additional data sources, such as satellite imagery and outputs from numerical models, may be  used together with the stations data \citep{lawson2016handbook, fuentes2005model, fuentes2001model, cameletti2019bayesian, villejo2023data, villejo2025data}. While these sources provide broader spatial coverage, they may introduce higher biases \citep{lawson2016handbook}.
In climate prediction, commonly used supplementary data sources include global circulation models (GCMs) and numerical weather prediction models \citep{naish2014climate}. The process of integrating multiple data sources to improve predictions is known as data fusion or data assimilation \citep{gettelman2022future, bauer2015quiet, lawson2016handbook}. Previous research has demonstrated that data fusion improves model accuracy \citep{villejo2025data, forlani2020joint}. In this work, we mainly use a data fusion model as input to the health model.

For inference, we use integrated nested Laplace approximation (INLA), as it provides fast and accurate posterior estimates \citep{rue2009approximate}. In addition, we use the stochastic partial differential equation (SPDE) approach \citep{lindgren2011explicit} to represent spatial Gaussian Mat\'ern fields  in the climate models. The combination of INLA and the SPDE approach has proven to be a powerful tool for spatial or spatio-temporal analysis \citep{bakka2018spatial,cameletti2013spatio, blangiardo2015spatial, schrodle2011spatio, lindgren2015bayesian}.  

This paper is structured as follows: Section \ref{sec:Data} presents the data sources and initial data exploration. Section \ref{sec:literature} presents a literature review on the link between climate and dengue. The proposed models are discussed in Section \ref{sec:Model}. The estimation approach, particularly INLA, is discussed in Section \ref{sec:modelestimation}. Section \ref{subsec:uncertainty} presents the approach for uncertainty propagation.  Results and discussion are presented in Section \ref{sec:results}, followed by conclusions and future work in Section \ref{sec:conclusions}.

\section{Data}\label{sec:Data}

\begin{figure}[]
    \centering
    \includegraphics[trim={0 1.2cm 0 0},clip,scale=0.43]{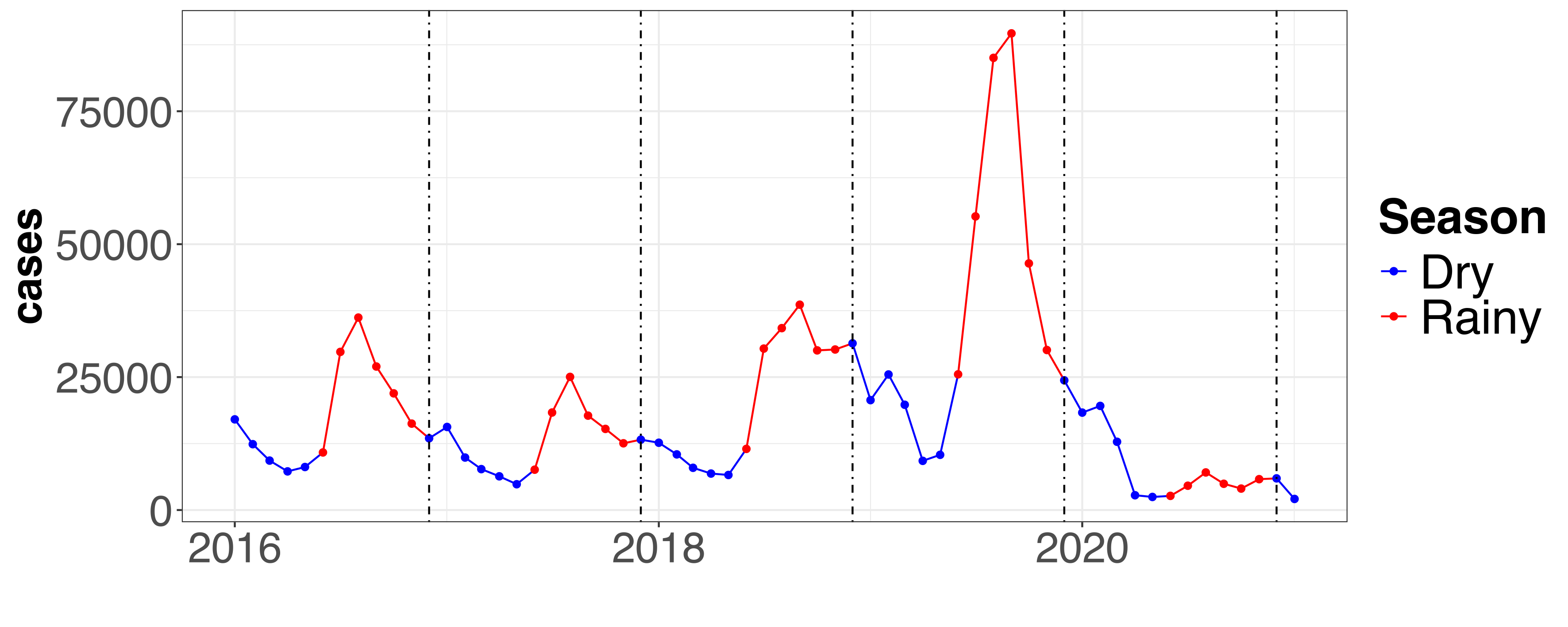}
    \caption{Time series plot of the number of dengue cases in the Philippines from January 2016 to January 2021}
    \label{fig:TP_Dengue}
\end{figure}

\begin{figure}[b]
    \centering
    \includegraphics[scale=0.35]{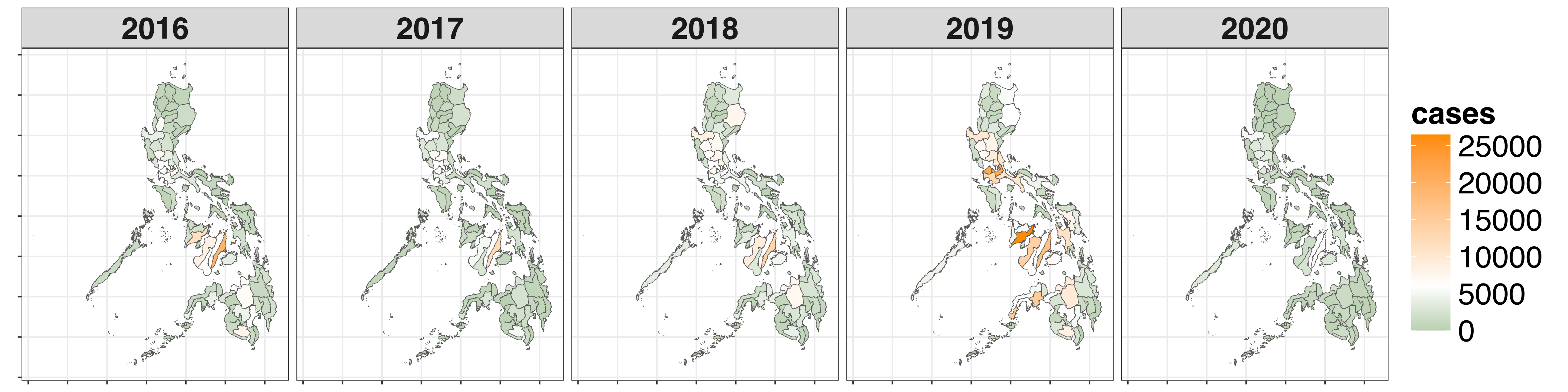}
    \caption{Plot of total dengue cases yearly (2016--2020) in the Philippines}
    \label{fig:DengueYearly}
\end{figure}

Data on dengue cases were retrieved from the United Nations Office for the Coordination of Humanitarian Affairs, whose primary mandate is to support humanitarian organizations worldwide through data collection, information dissemination, funding and resource mobilization, development of policies to meet the needs of crisis-affected people, and implementation of campaigns that advocate for humanitarian action. The data consist of weekly dengue case counts at  provincial level from 2016 to 2020. Data prior to 2016 were excluded due to differences in surveillance and reporting practices. Specifically, by 2015, the country's Health Department mandated nationwide reporting to include probable cases, whereas previously only  clinically-confirmed cases  were recorded in selected sentinel sites \citep{seposo2021dengue}. For the analysis, we aggregate the counts to monthly level. Figure \ref{fig:TP_Dengue} shows a plot of the monthly dengue cases in the Philippines from January 2016 to January 2021. Seasonality is evident, with cases generally higher during  rainy season (June to November) \citep{PHClimate_PAGASA}. Notably, there is an unusually high number of cases from August to October 2019. In August 2019, the Philippines declared a national dengue alert and epidemic due to a surge in cases and deaths \citep{BBCepidemic}. During 2020,  coinciding with the onset  of the COVID-19 pandemic, the number of reported cases is very low. This global phenomenon \citep{WHODengueSituation} was attributed to both reduced mobility -- several studies have shown that limited household movement is linked to lower transmission \citep{stoddard2013house} -- and reporting hesitancy, as individuals were afraid of contracting COVID-19 when visiting health facilities \citep{seposo2021dengue}.
Moving forward, in 2023, there was an upsurge in the dengue cases globally, with a simultaneous occurrence of multiple outbreaks even in regions previously unaffected by dengue \citep{WHODengueSituation}.

Figure \ref{fig:DengueYearly} shows maps of the annual (2016 to 2020) total number of dengue cases in the country. It is apparent that 2019 had the highest number of cases, and it also shows specific areas with the highest number of recorded cases. In particular, the region most badly hit by the 2019 dengue epidemic is the island in the western central part of the country, which is referred to as the Western Visayas region. Additional areas badly hit by dengue that can be identified in the plot are several contiguous provinces in the north, referred to as the Calabarzon region, and several areas located in the south. These were specific areas identified by the Health Department of the country as requiring immediate emergency attention \citep{BBCepidemic}. 

\begin{figure}[t]
    \centering
    \includegraphics[scale=0.35]{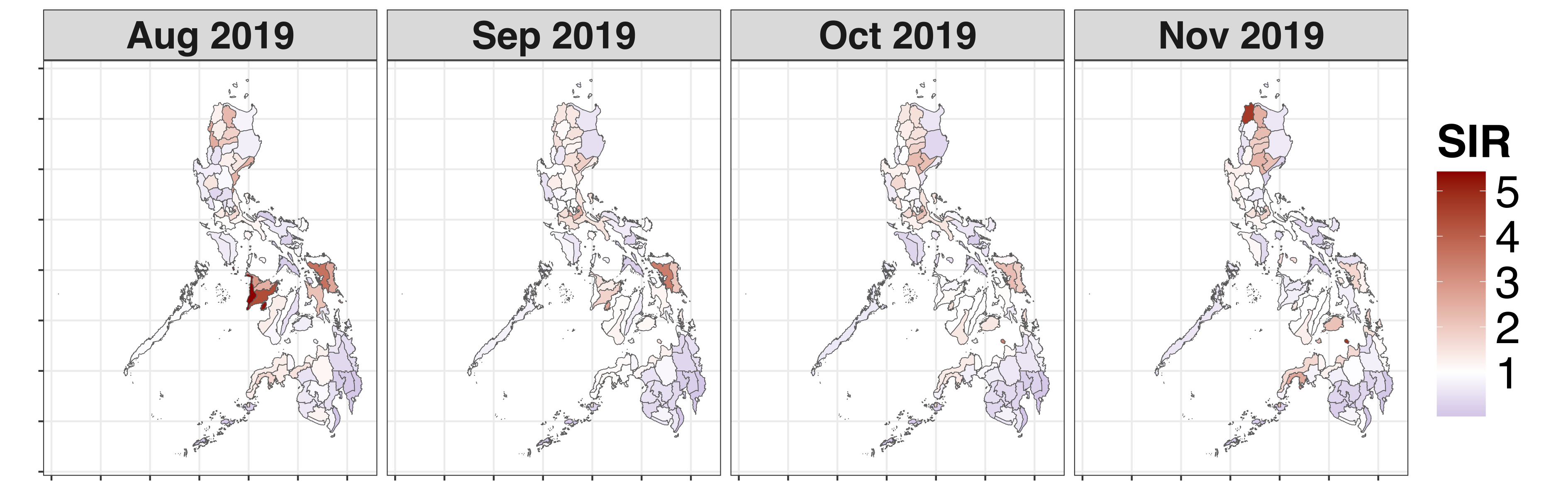}
    \caption{Plot of standardized incidence ratios (SIR) of dengue in the Philippines from August 2019 to November 2019}
    \label{fig:DengueSIRsAugtoNov2019}
\end{figure}

In many epidemiological applications, a measure of risk which accounts for the differences in the sizes and demographic structure of the provinces or areas, is typically mapped \citep{waller2010disease, waller2004applied}. A common measure of risk is the standardized incidence ratio (SIR), and is computed as the ratio of observed and expected cases. The expected cases are typically computed using indirect standardization, which involves applying age-specific rates (incidence proportions) of a standard population on the study population \citep{waller2004applied}. The age-specific national rates can be used, which are then applied on each age stratum for each province. However, a limitation of our data is that we do not have information on age-specific rates. Hence, we used internal standardization to compute the expected cases. This approach uses an estimate for the baseline individual risk based on the national observed disease rate \citep{waller2010disease}. This is further detailed in Section \ref{subsec:poissonmodel}. Figure \ref{fig:DengueSIRsAugtoNov2019} shows a plot of the dengue SIRs for August 2019 to November 2019. The reason for choosing these specific months is that these correspond to the period of high incidence of dengue, based on Figure \ref{fig:TP_Dengue}. The figure shows, specifically for August 2019, that the area with the highest number of reported cases is also the area with the highest SIR. 

\begin{figure}[h!]
     \centering
     \subfloat[][Weather stations]{\includegraphics[trim={0cm 0cm 0 2cm},clip,scale=0.32]{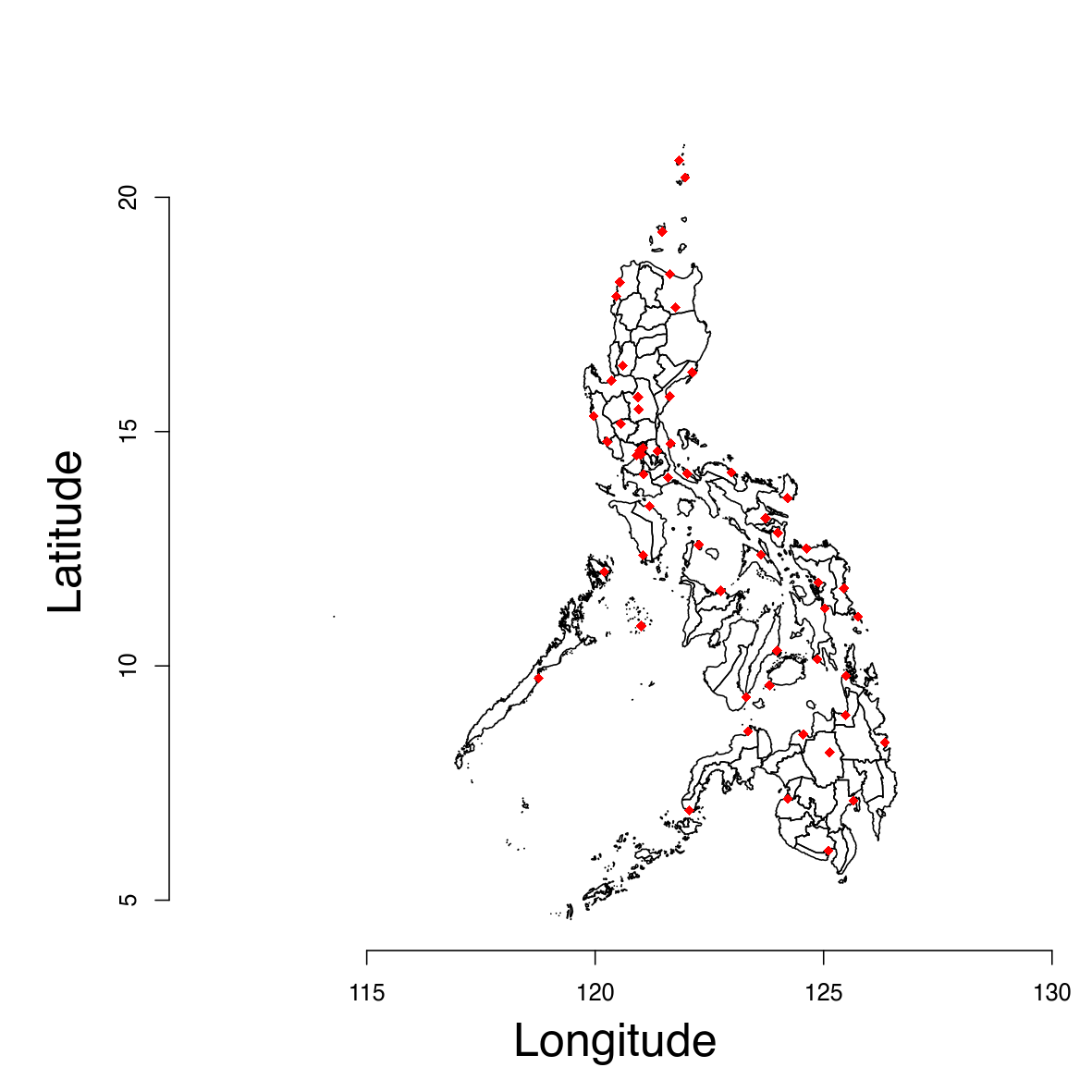}\label{fig:PH_monitors}}\hspace{5mm}
     \subfloat[][Sample GSM outcomes]{\includegraphics[trim={0cm 0cm 0 2cm},clip,scale=0.32]{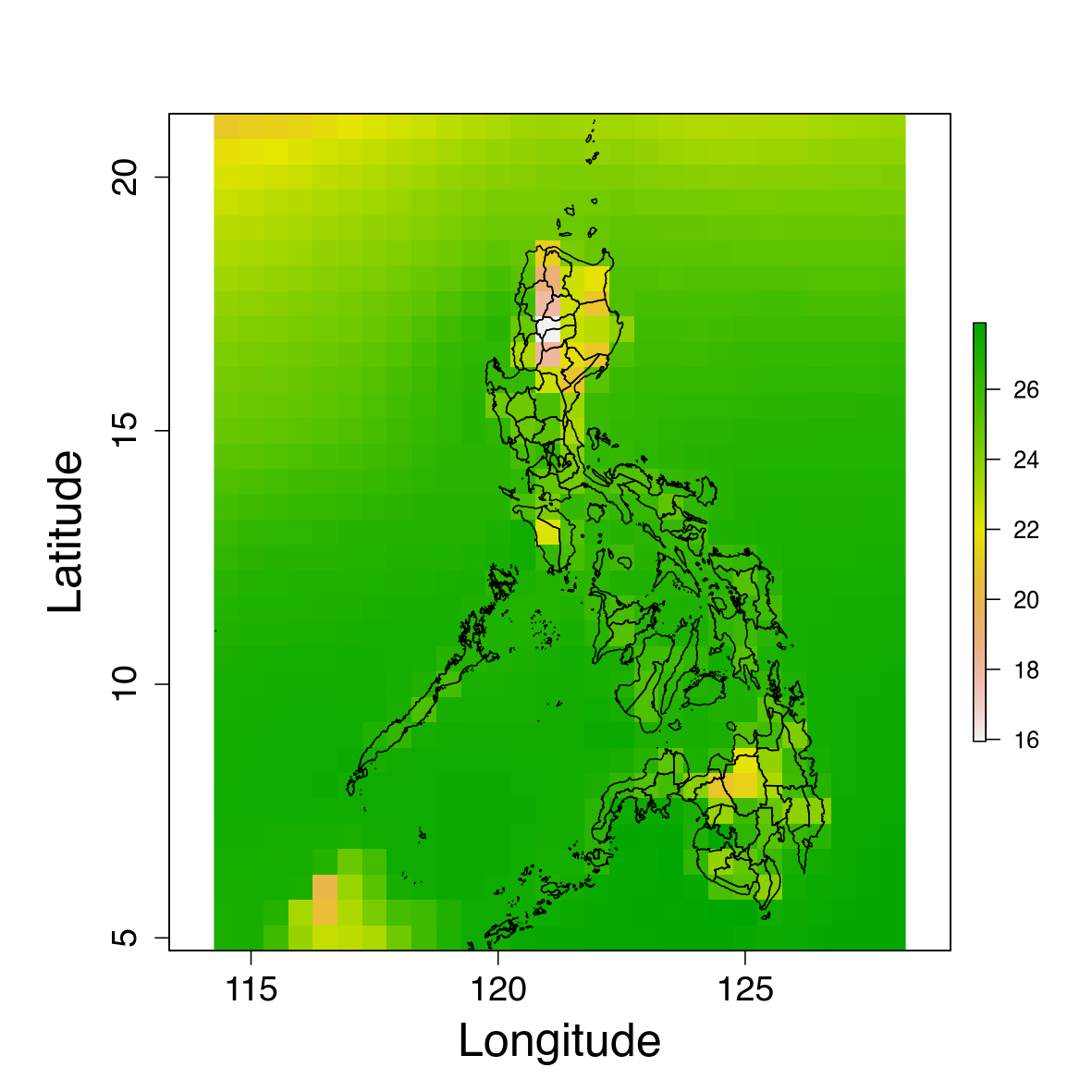}\label{fig:PH_GSM}}     \caption{Climate data sources for Philippines: (a) 57 weather synoptic stations (b) Global Spectral Model (GSM), a numerical weather prediction model maintained by the Japan Meteorological Agency}
     \label{fig:climate_data}
\end{figure}
For the main results in this work, we use existing climate predictions from the data fusion models in \cite{villejo2025data}, since predicted fields based on sparse stations data either have high uncertainty or more biased \citep{lawson2016handbook}. Figure \ref{fig:climate_data}a shows the locations of the sparse network of weather station, while Figure \ref{fig:climate_data}b shows a simulated output of temperature from the Global Spectral Model (GSM), a numerical weather forecast model maintained by the Japan Meteorological Agency, aggregated monthly for August 2019. However, the data fusion models cover only years 2019 to 2020, due to data availability constraints. Thus, we also explore the use of longer monthly time series, albeit for stations only, from 2016 to 2020, for the same climate variables: temperature (in $^{\circ}$C), relative humidity (in \%), and total rainfall (in mm). The stations data were obtained from the Philippine Atmospheric, Geophysical and Astronomical Services Administration (PAGASA), the country's meteorological office. The results from these models are in the Supplementary Material.

\section{Climate and dengue}\label{sec:literature}

Dengue is classified as an \textit{indirectly transmitted disease}, because its pathogen -- the dengue  virus -- is transmitted between humans by mosquito vectors, making it a vector-borne disease \citep{mcmichael2003climate}. Since pathogens of indirectly transmitted diseases spend part of their life cycle in the external environment, they are particularly influenced by climatic factors. 

Temperature affects the extrinsic incubation period of the pathogen, as well as the reproductive rate and biting rate of the mosquito \citep{promprou2005climatic, ewing2016modelling}. Higher temperatures accelerate mosquito breeding and shorten the virus incubation period, allowing mosquitoes to become infectious more quickly \citep{macdonald1957epidemiology}. Furthermore, increasing temperatures enhance the mosquito's biting behavior; raising the risk of virus  transmission. However, there is also a temperature threshold beyond which mosquito survival decreases. The temperature range shown to be optimal for dengue transmission is 21.3--34.0$^\circ$C for \textit{Ae.\ aegypti} and 19.9--29.4$^\circ$C for \textit{Ae.\ albopictus} \citep{ryan2019global}. Rising global temperatures are expected to increase the risk of mosquito-borne diseases, particularly dengue and malaria. Under worst climate-change scenarios, a 1$^\circ$C increase in global mean temperature could put 2.4 billion people at risk of both diseases by 2100 \citep{colon2021projecting}, with the higher risk  concentrated in densely populated areas of Africa, Southeast Asia and the Americas. In the Philippines, temperature has been shown to have a positive effect on dengue incidence \citep{subido2022correlation, francisco2021dengue, cawiding2025disentangling}. In \cite{seposo2024projecting}, they showed that between 2010 and 2019, 72.1\% of reported dengue cases in the Philippines were attributable to temperature, which implies that it is a significant driver of dengue transmission. \cite{edillo2022temperature} looked at the effect of temperature at the vector level, particularly looking at three development-related phenotypes: percent pharate larvae, hatch rates, and reproductive outputs. Their results show that temperature, together with season and latitudinal differences of the islands, significantly influences the phenotypes of \textit{Aedes aegypti}. The latitude of a spatial location brings variation in the amount of sunlight received, which affects the suitability of the breeding sites for mosquitoes.  Lastly, \cite{xu2020high} showed a non-linear association between temperature and dengue.

Increasing rainfall creates more breeding sites for mosquitoes, leading to an increase in the mosquito population and a higher risk of virus transmission \citep{promprou2005climatic, ewing2016modelling}. However, excessive rainfall can wash away breeding sites,  decreasing the risk of dengue. In consistently wet regions, a decrease in rainfall, such as during droughts, can cause water stagnation in rivers and lead to increased water storage, both of which create ideal breeding conditions for \textit{Aedes} mosquitoes \citep{mcmichael2003climate}. In the Philippines, rainfall has also been shown to be positively related to dengue, with some lagged effect \citep{francisco2021dengue}. \cite{cawiding2025disentangling} showed that the effect of rainfall could vary depending on the location. In western areas of the country, which experience pronounced dry and wet season, sporadic rainfall could create new breeding sites, thus increasing dengue incidence. However, for areas in the eastern part of the country, with a uniform amount (low variation) of rainfall, rainfall tends to flush out stagnant water, reducing mosquito breeding sites. 

Since relative humidity is positively related to rainfall \citep{UKMetRH}, it is also linked to dengue transmission. Virus transmission tends to be higher during months of high humidity  \citep{mcmichael2003climate}, as increased humidity favours mosquito survival \citep{gubler2001climate}. In contrast, mosquitoes dessicate easily under dry conditions \citep{focks1995simulation, hales2002potential}. \cite{xu2020high} showed that relative humidity is the only factor to be associated with a future seasonal peak of dengue in the Philippines. 

Other important factors are house structure, human behaviour and general socioeconomic conditions \citep{patz2000potential}. Dengue transmission is further exacerbated by ineffective vector and disease surveillance, inadequate public health infrastructure, population growth, unplanned and uncontrolled urbanization, and increased travel \citep{gubler2014dengue, rigau1998dengue, mcmichael2003climate, ECDC_infog, murphy1994emergence}.

\section{Proposed model}
\label{sec:Model}

As discussed in Section \ref{sec:intro}, we use a two-stage modelling approach to link climate variables and dengue cases. This section mainly presents the proposed second-stage (dengue) models. The first-stage (climate) data fusion models are based on existing results in \cite{villejo2025data}. Section  \ref{subsec:climatemodel} in the Appendix provides a brief description of the data fusion models. Section \ref{subsec:poissonmodel} presents the health model. This section also discusses the formula for computing block-averages of climate field (Section \ref{subsec:blockaverage}), specification of spatial and temporal effects (Section \ref{subsec:spatiotemporaleffs}), intrinsic conditional autoregressive specification for disconnected graphs (Section \ref{subsec:iCARdisconnected}), interaction effects (Section \ref{subsec:interactionmodel}), and prior specification for second-stage model parameters (Section \ref{subsec:priors}). The results from a stations-only model input, which also covers a longer time series (2016 to 2020), are provided in the Supplementary Material.

\subsection{Poisson model for dengue}\label{subsec:poissonmodel}

Let $\text{y}(B_i,t)$ be the number of observed dengue cases in area $B_i$, $i=1,\ldots,N$, and time $t$, $t=1,\ldots,T$. We assume that $\text{y}(B_i,t)$ is Poisson distributed with mean $\mu(B_i,t)$. The model is given by: 
\begin{equation}
    \begin{aligned}\label{eq:poisson}
        \text{y}(B_i,t) &\sim \text{Poisson}\Big(\mu(B_i,t)\Big), \;\;\; \mathbb{E}\Big[\text{y}(B_i,t)\Big] = \mu(B_i,t) = \lambda(B_i,t)\times \text{E}(B_i,t) \\  &\log\Big(\lambda(B_i,t)\Big) = \gamma_0 + \gamma_1\hat{x}(B_i,t) + \bm{\gamma}_2^\intercal\bm{z}_2 + \varphi(B_i,t),
    \end{aligned}
\end{equation}

where $\{\gamma_0$, $\gamma_1$, $\bm{\gamma}_2\}$ are fixed effects, $\hat{x}(B_i,t)$ is the block-level value for a climate variable at time $t$, $\bm{z}_2$ is a set of covariates, and $\varphi(B_i,t)$ is a spatio-temporal random effect. In Equation \eqref{eq:poisson}, the Poisson mean is expressed as a product of $\lambda(B_i,t)$ and $\text{E}(B_i,t)$, where $\lambda(B,t)$ is the disease risk while $\text{E}(B_i,t)$ is the expected cases. $\text{E}(B_i,t)$  are known quantities, which we compute using internal standardization \citep{waller2010disease}. In particular, suppose that $\text{n}(B_i,t)$ is the population at risk at area $B_i$ and time $t$. Moreover, suppose that $r_t$ is the constant baseline risk per person at time $t$, which is estimated using aggregate population data, i.e., $\hat{r}_t = \dfrac{\sum_{\forall B_i} \text{y}(B_i,t)}{\sum_{\forall B_i} \text{n}(B_i,t)}$, and is interpreted as the global observed disease rate for time $t$. The expected number of cases in $B_i$ at time $t$ is then computed as $\text{E}(B_i,t) = \text{n}(B_i,t)\times \hat{r}_t$. From Equation \eqref{eq:poisson}, since we have $\log\Big(\lambda(B_i,t)\Big) = \log\Bigg(\dfrac{\mu(B_i,t)}{\text{E}(B_i,t)}\Bigg)$,  this implies that $\log\Big(\mu(B_i,t)\Big) = \log\Big(\lambda(B_i,t)\Big) + \log\Big(\text{E}(B_i,t)\Big)$
i.e., $\log\Big(\text{E}(B_i,t)\Big)$ operates as an offset parameter in the Poisson model.

\subsection{Block averages of climate variables}\label{subsec:blockaverage}

Since the climate model is point-referenced and  the health model is area referenced, the two models are spatially misaligned. To address this, we use the definition of the average level of $x(\mathbf{s},t)$ for an area $B_i$ at time $t$ in \cite{gelfand2010handbook}, denoted by $x(B_i,t)$, as follows:
\begin{equation}\label{eq:average_x_formula}
    x(B_i,t)=\int_{B_i} x(\mathbf{s},t)p(\mathbf{s})d\mathbf{s}, 
\end{equation}
for a weighting function $p(\mathbf{s})$ such that $\bigintssss_{B_i}p(\mathbf{s})d\mathbf{s}=1$. \cite{cameletti2019bayesian} mentioned two ways to approximate Equation \eqref{eq:average_x_formula}. The first one is a linear combination based on neighbourhood intersections, while the second one is an unweighted (simple) mean of predicted values of $x(\mathbf{s},t)$ that lie inside $B_i$. In this work, we use the latter approach, and is computed as follows:
\begin{equation}\label{eq:blockaverage}
    \hat{x}(B_i,t) = \sum_{\forall \mathbf{s}_i^*\in B_i} \hat{x}(\mathbf{s}_i^*,t)p(\mathbf{s}_i^*) = \dfrac{1}{\#B_i} \sum_{\forall \mathbf{s}_i^* \in B_i} \hat{x}(\mathbf{s}_i^*,t),
\end{equation}
where $\#B_i$ denotes the number of prediction points in block $B_i$ and $\hat{x}(\mathbf{s}_i^*,t)$ is the predicted value of the climate variable at spatial location $\mathbf{s}_i^*$ and time $t$ (see Section \ref{subsec:climatemodel} in the Appendix for details). Shown in Figure \ref{fig:PHgraphs}a is the prediction grid for the Philippines. 

\subsection{Spatio-temporal effects $\varphi(B_i,t)$}\label{subsec:spatiotemporaleffs}

We assume the following form for the spatio-temporal effects:
\begin{equation}
    \label{eq:STeffects}
    \begin{aligned}
        \varphi(B_i,t) &= \psi(B_i) + \zeta(t) + \nu(t) + \upsilon({B_i,t}) \\
    & \psi(B_i) = \bigg[\dfrac{1}{\sqrt{\tau_{\psi}}}\sqrt{1-\phi}\text{v}(B_i) + \sqrt{\phi}\text{u}(B_i) \bigg] \\
    &\text{v}(B_i) \sim \mathcal{N}(0,1)\\
    &\text{u}(B_i) \sim \text{scaled iCAR on a disconnected graph} \\
    &\zeta(t) \overset{\text{iid}}{\sim} \mathcal{N}(0,\sigma^2_{\zeta}) \\
    &\nu(t)\;\text{is a random walk in time of order 2}\\
    &\upsilon(B_i,t) \; \text{is in interaction term between space and time}
    \end{aligned}
\end{equation}

The model specification for $\psi(B_i)$ follows that of \cite{riebler2016intuitive}. It provides a compromise between pure overdispersion, denoted by $\text{v}(B_i)$, and spatially-structured correlation, denoted by $\text{u}(B_i)$. The total (marginal) variance of the spatial main effect $\psi(B_i)$ is $1/\tau_{\psi}$, and the proportion of the variance explained by the structured effect is $\phi$. Thus, $\phi$ is a mixing parameter between the unstructured and structured spatial effect. The structured effect $\text{u}(B_i)$ is defined as an intrinsic conditionally autoregressive (iCAR) process \citep{besag1991bayesian}. In particular, since the Philippines is an archipelago, which means that there are provinces without neighbours (islands),  we define the iCAR process on a disconnected graph, the details of which are provided in Section \ref{subsec:iCARdisconnected}. Similarly, Equation \eqref{eq:STeffects} also specifies an unstructured and structured effect in time, denoted by $\zeta(t)$ and $\nu(t)$, respectively. We assume a random walk of order 2 for the structured time effect. Finally, $\upsilon(B_i,t)$ specifies the interaction term between space and time (see Section \ref{subsec:interactionmodel} for details).

\subsection{Specifying the iCAR process $\text{u}(B_i)$ on a disconnected graph}\label{subsec:iCARdisconnected}

\begin{figure}
    \centering
    \subfloat[][]
    {\includegraphics[width=0.27\linewidth]{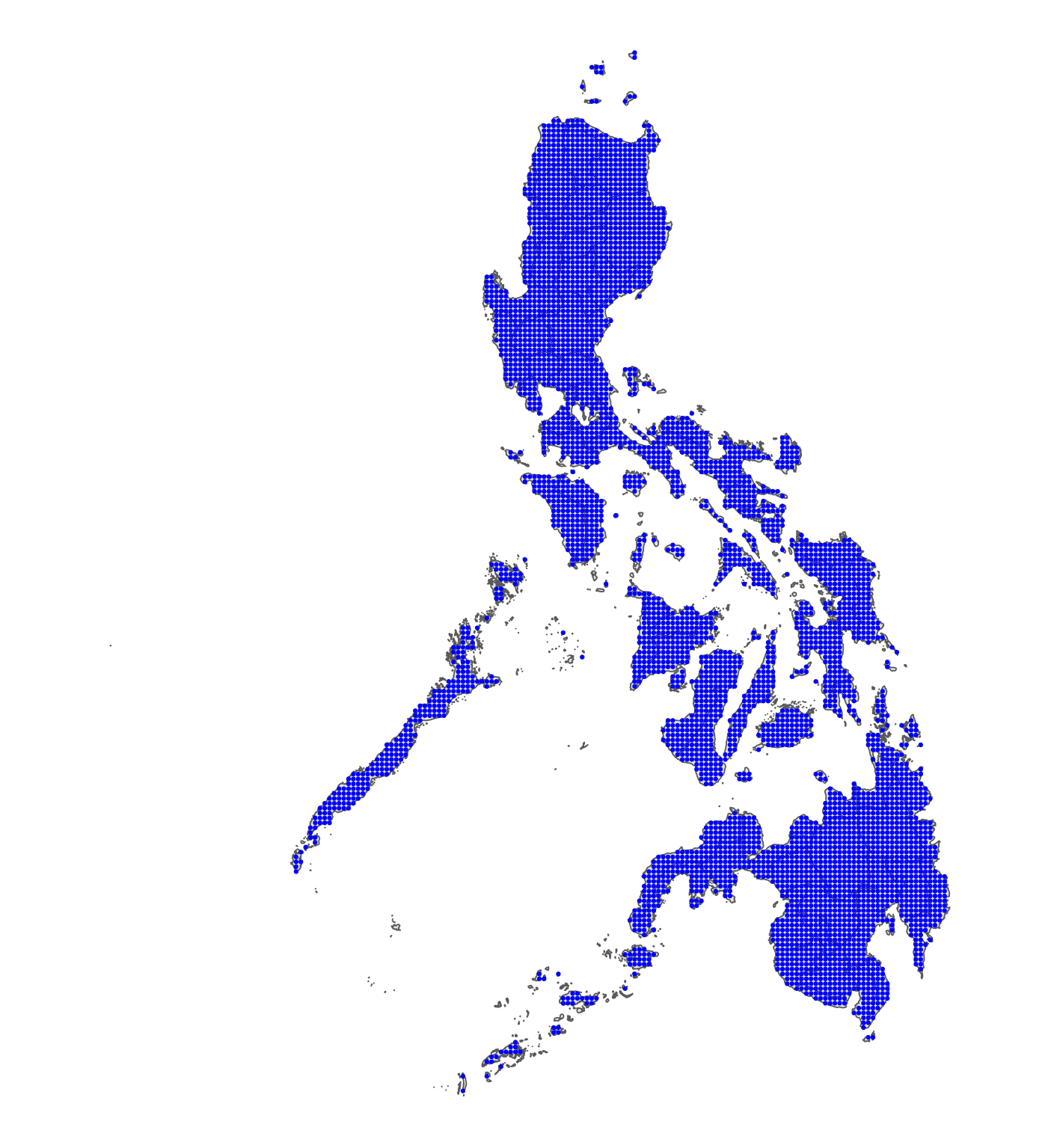}}\hspace{3mm}
    \subfloat[][]
    {\includegraphics[width=0.27\linewidth]{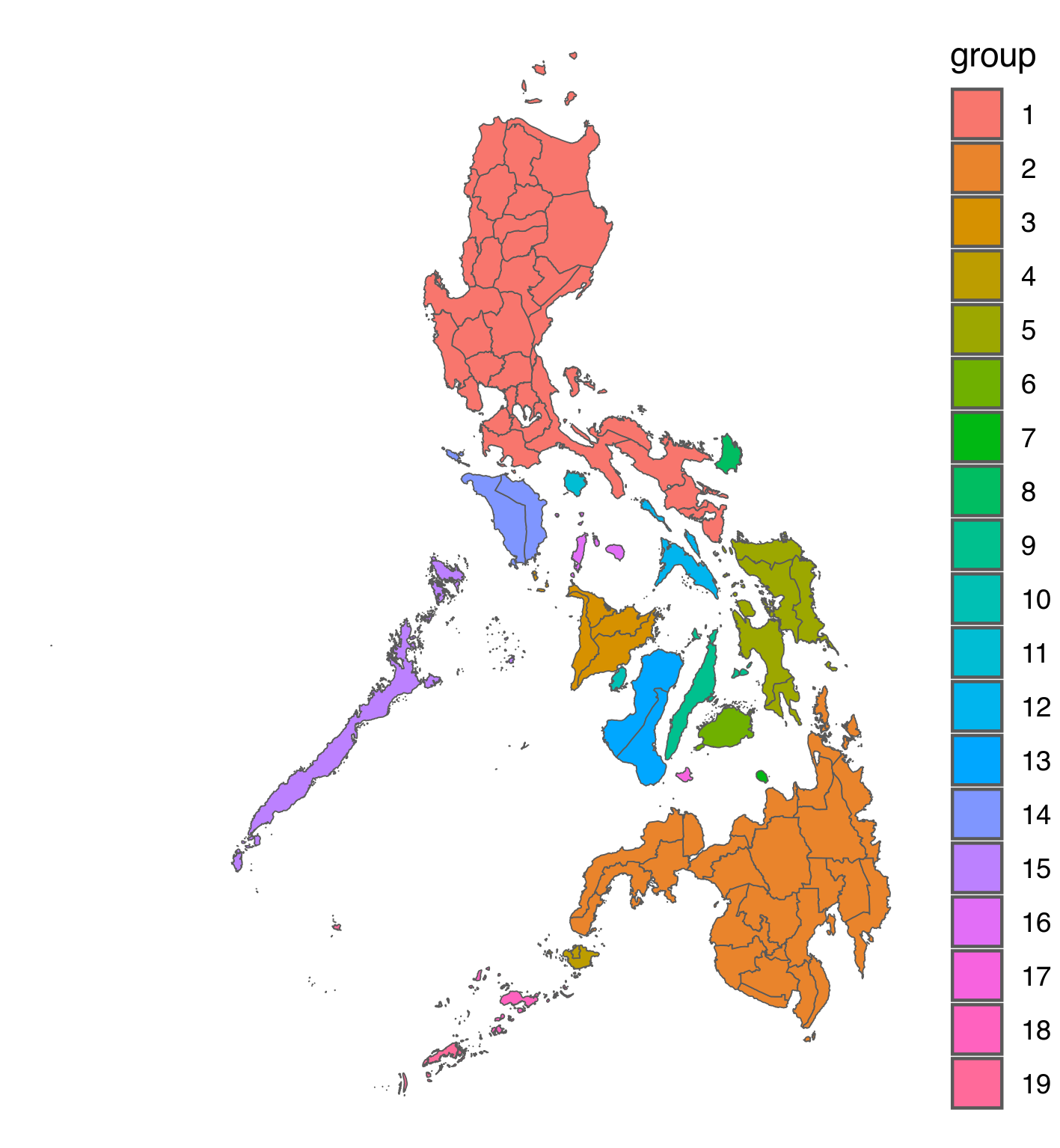}}\hspace{6mm}
    \subfloat[][]
    {\includegraphics[width=0.27\linewidth]{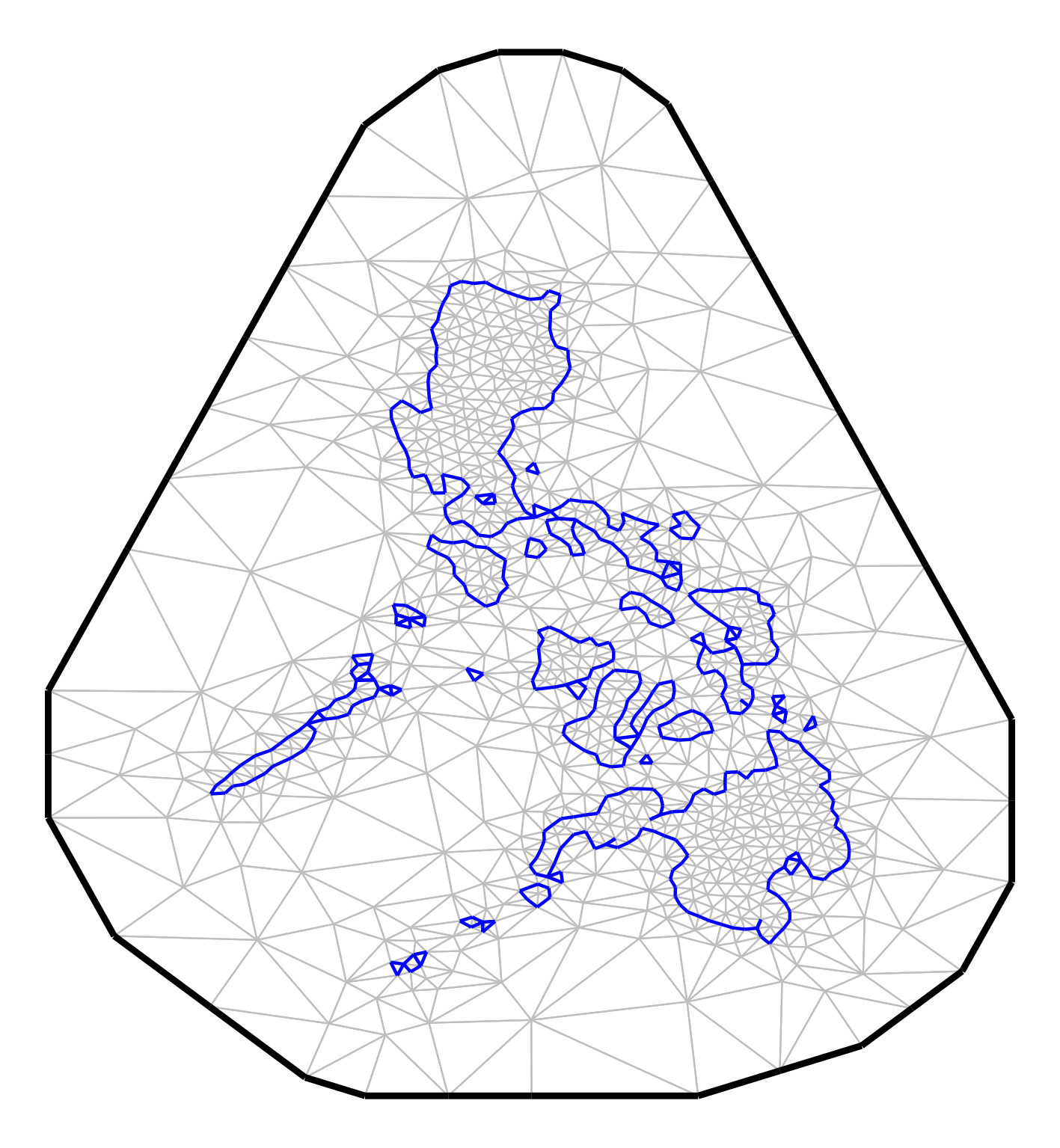}}
    \caption{(a) prediction grid, (b) 19 disconnected graphs for the iCAR model (out of the 19 graphs, 12 are singletons / isolated islands), (c) mesh used for the SPDE approximation}
    \label{fig:PHgraphs}
\end{figure}

An iCAR process is defined with respect to a specific undirected graph, a set of vertices (which here refer to the areas $\{B_1, \ldots, B_{\text{N}}\}$), and the edges (which refer to the set of neighbours). Due the archipelagic structure of the Philippines, some areas or provinces do not have neighbours because they are islands. Specifically, there are 19 disconnected graphs of which 12 are singletons, and the other 7 are connected graphs. These are shown in Figure \ref{fig:PHgraphs}b. When working on disconnected graphs, the precision parameters for each connected graph are not comparable, and the presence of singletons can lead to an improper posterior distribution \citep{freni2018note}.

To circumvent the aforementioned issues, we use the proposed method in \cite{freni2018note}, which is based on the proposed scaling of intrinsic GMRFs in \cite{sorbye2014scaling}. The first step is to define an iCAR model for each connected graph (not including the singletons). The next step is to scale each precision matrix, and to define a common precision parameter, say $\tau$, for all the graphs. For singletons, a standard Gaussian with precision $\tau$ is assumed, i.e., singletons have a non-spatial random effect. This specification means that if a block $B_i$ has a neighbour,  $\text{u}(B_i)$ shrinks to the local mean of the graph where it belongs. On the other hand, if a block $B_i$ does not have a neighbour,  $\text{u}(B_i)$ shrinks to the overall or global mean. Also, $\tau$ operates as the precision parameter for all connected components, i.e., it regulates the degree to which each $\text{u}(B_i)$ shrinks to either the local mean or the global mean. 

In addition to applying a sum to zero constraint, we also add a separate intercept for each connected component as recommended by \cite{freni2018note}. This implies that random effects for each node within a connected component can deviate randomly from the local intercept, just as singletons deviate from the global (overall) intercept. The size of the deviation depends only on $\tau$ and not on the graph structures of the connected components. Neither the sum-to-zero constraint nor a separate intercept is required for singletons.

\subsection{Interaction term $\upsilon(B_i,t)$}\label{subsec:interactionmodel}

We explore four possible specifications for the interaction term $\upsilon(B,t)$ following   \cite{knorr2000bayesian}. These are summarized below:
\begin{enumerate}
    \item Type I interaction assumes that both unstructured effects in space and time interact, i.e., $\upsilon(B_i,t)\overset{\text{iid}}{\sim}\mathcal{N}(0,\sigma_{\upsilon}^2)$. The structure matrix for this interaction is simply the identity matrix, i.e., $\mathbf{R}_{\upsilon}=\mathbb{I}$, so that $\bm{\upsilon} \sim \mathcal{N}\big(\mathbf{0},\sigma^2_{\upsilon}\mathbb{I}\big)$. 
    \item Type II interaction assumes that a structured temporal effect interacts with an unstructured spatial effect. The structure matrix here is the Kronecker product of the structure matrices for the $\text{iid}$ effect and the first-order autoregressive effect in time. This type of interaction implies that each block $B_i$ has its own temporal structure, which is independent of the other blocks/areas.
    \item Type III interaction assumes an interaction between an unstructured temporal effect and a structured spatial effect. The structure matrix is the Kronecker product between the $\text{iid}$ effect in time and the iCAR effect in space. This type of interaction implies that for each time point $t$, there is a spatial structure, which is independent of the other time points. 
    \item Type IV interaction assumes that both structured effects in space and time interact. This type of interaction specifies a temporal (autoregressive) structure for each area which depends on the temporal patterns of the neighbouring areas. 
    \end{enumerate}

\subsection{Specification of priors}\label{subsec:priors}

We assign non-informative priors for the fixed effects, particularly, $\gamma_0\sim\mathcal{N}(0,\infty)$, $\gamma_1\sim\mathcal{N}(0,1000)$, and $\bm{\gamma}_2\sim\mathcal{N}(\bm{0},1000\mathbb{I})$. The precision parameters of the time effects are given the following priors: $\log\Big( 1/\sigma_{\nu}^2\Big)\sim\log \text{Gamma}(1,0.00005)$ and $\log\Big( 1/\sigma_{\zeta}^2\Big)\sim\log \text{Gamma}(1,0.00005) $. The variance parameter for the spatial effect $\psi(B_i)$ is given a penalized complexity (PC) prior, particularly, $\mathbb{P}\Big(1/\sqrt{\tau_{\psi}}>1\Big)=0.01$. PC priors are weakly informative priors which penalize complexity of Gaussian random fields by penalizing departure from a base model \citep{fuglstad2019constructing, simpson2017penalising}. This prior choice assumes a probability of 0.99 of having residual relative risks smaller than 1. The mixing parameter $\phi$ of the spatial effect $\psi(B_i)$ is also given a PC prior, particularly, $\mathbb{P}\Big( \phi<0.5\Big) = 2/3.$ This gives more mass to values $\phi<0.5$. This is conservative since it assumes that the unstructured term accounts for more of the variation in $\psi(B_i)$ \citep{riebler2016intuitive}. The parameters for the spatio-temporal interaction effect $\upsilon(B_i,t)$ are given non-informative priors. For the Type I interaction, we specify $\log\Big(1/\sigma^2_{\upsilon}\Big)\sim\log\text{Gamma}(1,0.00005)$. For the Type II interaction, the variance parameter has the same prior as Type I, but in addition, we specify a log normal prior for the AR parameter $\rho_{\upsilon}$, i.e., $\log\bigg(\dfrac{1+\rho_{\upsilon}}{1-\rho_{\upsilon}}\bigg)\sim\mathcal{N}(0,0.15)$. For the Type III interaction, both the precision parameters of the unstructured time effect and structured time effects are also given the $\log\text{Gamma}(1,0.00005)$ distribution. Finally, for the Type IV interaction, the precision parameters and the AR parameter are also given similar priors as the other interaction types. The prior distributions for the first-stage model parameters are in Section \ref{subsec:priorsfirststage} of the Appendix.

\section{Model Estimation} \label{sec:modelestimation}

\subsection{INLA and SPDE approach} \label{subsec:INLA_SPDE}

We use the integrated nested Laplace approximation (INLA) approach to perform inference in both the climate model and the health model. INLA is an approximate method for Bayesian inference for latent Gaussian models, and has gained popularity in the recent years because of its speed and accuracy \citep{rue2009approximate, van2023new}. 

Suppose that $\mathbf{y}$ denotes the vector of response variables; for the first-stage model, this refers to the climate data, while for the second-stage model, this refers to the observed dengue cases. Suppose that $\bm{x}$ denotes the latent parameters; for the first-stage stations-only model (see Equation \eqref{eq:latentprocess} in the Appendix), this refers to the fixed effects $\beta_0$ and $\bm{\beta}$, and the spatio-temporal effect $\xi(\mathbf{s},t)$. For the first-stage data fusion model, additional latent parameter include the additive bias $\alpha_0(\cdot,t)$ (see Equation \eqref{eq:gsmdata} in the Appendix). For the second-stage model, the latent parameters are the fixed effects $\gamma_0$ and $\bm{\gamma}_1$. Furthermore, suppose that $\bm{\theta}= \begin{pmatrix}
    \bm{\theta}_1 & \bm{\theta}_2
\end{pmatrix}^\intercal$ are the hyperparameters, where $\bm{\theta}_1$ are the hyperparameters linked to the likelihood, e.g., variance/scaling parameters, while $\bm{\theta}_2$ are the hyperparameters linked to the latent parameters. For the first-stage stations-only model, $\bm{\theta}$ includes the measurement error variance $\sigma^2_{e_1}$, the AR parameter $\phi_1$, and the marginal variance and range parameters of the Mat\'ern field $\omega_1(\mathbf{s},t)$ (see Section \ref{subsec:climatemodel} in the Appendix). For the first-stage data fusion model, additional hyperparameters are the marginal variance and range parameter of the Mat\'ern field $\omega_2(\cdot,t)$, the AR parameter $\phi_2$, the multiplicative bias $\alpha_1$, and the measurement error variance $\sigma_{e_2}^2$ (see Equation \eqref{eq:gsmdata} in the Appendix). For the second-stage model, $\bm{\theta}_2$ refers to all parameters linked to the spatio-temporal effect $\varphi(B_i,t)$. 

The Bayesian hierarchical model, for both first-stage and second-stage model, is given as follows:
\begin{equation}
    \begin{aligned}
        \mathbf{y}|\bm{x},\bm{\theta}_1 &\sim \prod_{\forall i} \pi(\text{y}_i|\bm{x},\bm{\theta}_1)\\
        \bm{x}|\bm{\theta}_2&\sim \mathcal{N}\Big(\bm{0}, \mathbf{Q}^{-1}(\bm{\theta}_2)\Big)\\
        \bm{\theta} &\sim \pi(\bm{\theta}),
    \end{aligned}
\end{equation}
which is a latent Gaussian model because of the Gaussian prior on the latent parameters. INLA estimates the posterior marginals via the following nested integrals:
\begin{equation}
    \begin{aligned}
        \pi(x_j|\textbf{y}) &= \int\pi(x_j|\bm{\theta},\textbf{y})\pi(\bm{\theta}|\textbf{y}) \\
        \pi(\theta_j|\textbf{y}) &= \int \pi(\bm{\theta}|\textbf{y})d\bm{\theta}_{-j},
    \end{aligned}
\end{equation}
where $x_j$ and $\theta_j$ refers to a latent parameter and hyperparameter, respectively, and $\bm{\theta}_{-j}$ denotes the vector of all hyperparameters, excluding $\theta_j$;  details  can be found in \cite{rue2009approximate} and \cite{van2023new}. 

To estimate the first-stage models (both the stations-only model and the data fusion model), the stochastic partial differential equation (SPDE) approach is used, which is a computationally  efficient and flexible approach to estimate Gaussian fields of the Mat\'ern class \citep{lindgren2011explicit}. The SPDE approach finds the optimal Gaussian Markov random field (GMRF) to represent a Mat\'ern field, i.e., a representation with local neighbourhood and sparse precision matrix. The Markov property is achieved via a finite element representation which specifies the Mat\'ern field as a linear combination of basis functions defined on a triangulation called mesh; Figure \ref{fig:PHgraphs}c shows the mesh used here. The SPDE representation of the models is presented in Section \ref{subsec:SPDErepresentation} of the Appendix. 

\subsection{Uncertainty propagation}\label{subsec:uncertainty}

A crude two-stage modelling approach is to first evaluate Equation \eqref{eq:blockaverage} using the posterior mean of $x(\mathbf{s},t)$ at the prediction grid spatial locations, and to then plug-in the block-averages into the second-stage model; the formal steps for this are given in Algorithm \ref{alg:plugin}. However, this will potentially underestimate the posterior uncertainty in the second-stage model parameters, since the uncertainty in $\hat{x}(\mathbf{s},t)$ is ignored. In order to account for the uncertainty, we use a resampling approach, which essentially generates several samples from the posterior distribution of first-stage parameters, and then uses each resample as plug-in values to the second-stage model; the formal steps are outlined in Algorithm \ref{alg:resampling}.

\begin{algorithm}
\caption{Implementation of the crude plug-in method}\label{alg:plugin}
    \flushleft
\begin{steps} 
    \item Suppose $\hat{\beta}_0$, $\hat{\bm{\beta}}_1$ and $\{\hat{\tilde{\bm{\xi}}}_{t}\}_{t=1,\ldots,T}$ are posterior means from the first-stage model (see Sections \ref{subsec:climatemodel} and \ref{subsec:SPDErepresentation} of the Appendix). Also, suppose that $\mathbf{B}_{\text{grid}}$ is the projection matrix from the SPDE mesh nodes (Figure \ref{fig:PHgraphs}c) to the point locations of the prediction grid (Figure \ref{fig:PHgraphs}a). We compute the following:
    \begin{equation}\label{eq:alg_plugin}
        \hat{x}(\mathbf{s}^*,t) = \hat{\beta}_0 + \hat{\bm{\beta}}_1^\intercal\bm{z}(\mathbf{s}^*,t) + \mathbf{b}_{\text{s}^*}^\intercal\hat{\tilde{\bm{\xi}}}_{t},
    \end{equation}
    where $\mathbf{b}_{\text{s}^*}$ is the row in $\mathbf{B}_{\text{grid}}$ which corresponds to the spatial location $\mathbf{s}^*$, and $\bm{z}(\mathbf{s}^*,t)$ is the set of covariates for the first-stage model (see Equation \eqref{eq:latentprocess} of the Appendix).
    \item Compute the spatial average given in Equation \eqref{eq:blockaverage} for all $B_i$ and $t$.
    \item Plug in the values of $\hat{x}(B_i,t),\forall B_i, t$, in the second-stage model.
\end{steps}
\end{algorithm}

\begin{algorithm}
\caption{Implementation of the resampling method}\label{alg:resampling}
    \flushleft
    \text{Repeat steps 1--5 for $j=1,2,\ldots,J:$}
\begin{steps}
    \item Generate posterior samples of the latent parameters from the first-stage model (see Sections \ref{subsec:climatemodel} and \ref{subsec:SPDErepresentation} of the Appendix). 
    \item Suppose $\tilde{\beta}_0^{(j)}$, $\tilde{\bm{\beta}}_1^{(j)}$ and $\{\tilde{\tilde{\bm{\xi}}}_{t}^{(j)}\}_{t=1,\ldots\,T}$ are the $j^{\text{th}}$ posterior samples from the first-stage model. We compute the following:
    \begin{equation}
        \tilde{x}^{(j)}(\mathbf{s}^*,t) = \tilde{\beta}_0^{(j)} + \tilde{\bm{\beta}}_1^{{(j)}\intercal}\bm{z}_1(\mathbf{s}^*,t) + \mathbf{b}_{\text{s}^*}^\intercal\tilde{\tilde{\bm{\xi}}}_{t}^{(j)},
    \end{equation}
    where $\mathbf{b}_{\text{s}^*}$ and $\bm{z}(\mathbf{s}^*,t)$ are the same as in Equation \eqref{eq:alg_plugin},
    \item Compute the spatial average given in Equation \eqref{eq:blockaverage} for all $B_i$ and $t$, but here we denote the spatial averages by $\tilde{x}^{(j)}(B_i,t)$, since these are computed using posterior samples of the latent parameters, not the posterior means. 
    \item Plug in the values of $\tilde{x}^{(j)}(B_i,t), \forall B_i,t$, in the second-stage model. 
    \item Store all the posterior estimates, such as $\pi^{(j)}(\gamma_0|\textbf{y})$ and $\pi^{(j)}(\gamma_1|\textbf{y})$. Here, the superscript $(j)$ means that the posteriors are computed using the $j^{\text{th}}$ posterior samples.
    \item Combine all results via model averaging, e.g.,
    \begin{equation}
        \pi(\gamma_0|\textbf{y}) = \dfrac{1}{J}\sum_j^J\pi^{(j)}(\gamma_0|\textbf{y}) \;\;\;\;\;\; \text{and} \;\;\;\;\;\; \pi(\gamma_1|\textbf{y}) = \dfrac{1}{J}\sum_j^J\pi^{(j)}(\gamma_1|\textbf{y}).
    \end{equation}
\end{steps}
\end{algorithm}

When the first-stage model uses data solely from stations, the crude plug-in method and resampling method are straightforwardly implemented using Algorithms \ref{alg:plugin} and \ref{alg:resampling}, respectively. However, there is a slight modification in the steps when the data fusion model is used as the first-stage model. The reason is that the data fusion model is estimated conditional on the multiplicative bias parameter $\alpha_1$ (see Equation \eqref{eq:gsmdata} in the Appendix), which in practice is defined on a grid of values centered on 1 \citep{villejo2025data}. 

Suppose we denote by $\hat{\beta}_0^{(\ell)}$, $\hat{\bm{\beta}}_1^{(\ell)}$ and $\{\hat{\tilde{\bm{\xi}}}_{t}^{(\ell)}\}_{t=1,\ldots,T}$ the posterior means for the (first-stage) model which was fitted conditional on $\alpha_1 = \alpha_1^{(\ell)}$. Here, the first-stage model is indexed by $\ell=1,\ldots,L$, where $L$ is the number of $\alpha_1$ values considered. In addition, suppose that we denote by $w_{\ell}$ the weight for the $\ell^{\text{th}}$ (first-stage) model, i.e., the model fitted conditional on $\alpha_1 = \alpha_1^{(\ell)}$. The weights are computed as:
\begin{equation}
    w_{\ell} = \dfrac{ \pi\Big(\textbf{y}^{(1)}|\alpha_1^{(\ell)}\Big)\pi\Big(\alpha_1^{(\ell)}\Big) }{\sum_{\ell=1}^L \pi\Big(\textbf{y}^{(1)}|\alpha_1^{(\ell)}\Big)\pi\Big(\alpha_1^{(\ell)}\Big)}, \label{eq:BMAweights}
\end{equation}
where $\textbf{y}^{(1)}$ denotes the first-stage data, $\pi\Big(\textbf{y}^{(1)}|\alpha_1^{(\ell)}\Big)$ is the conditional marginal log likelihood given $\alpha_1=\alpha_1^{(\ell)}$, and $\pi\Big(\alpha_1^{(\ell)}\Big)$ is the prior distribution. The spatial averages are then computed as:
\begin{equation}
    \hat{x}(B_i,t)=\dfrac{1}{L}\dfrac{1}{\#B_i}\sum_{\ell=1}^L\sum_{\forall \mathbf{s}^*\in B_i}\hat{x}^{(\ell)}(\mathbf{s}^*,t)w_{\ell} = \dfrac{1}{L}\dfrac{1}{\#B_i}\sum_{\ell=1}^L\sum_{\forall \mathbf{s}^*\in B_i}\Big( \hat{\beta}_0^{(\ell)} + \hat{\bm{\beta}}_1^{(\ell)\intercal}\bm{z}(\mathbf{s}^*,t) + \mathbf{b}_{\text{s}^*}^\intercal\hat{\tilde{\bm{\xi}}}_{t}^{(\ell)} \Big)w_{\ell}. \label{eq:weightedblockaverages}
\end{equation}
Equation \eqref{eq:weightedblockaverages} implies that the spatial average for block $B_i$ and time $t$ is a weighted average of the point predictions from each first-stage model estimated conditional on $\alpha_1=\alpha_1^{(\ell)},\ell=1,\ldots,L$, and where the weights are given in Equation \eqref{eq:BMAweights}. 
 A similar idea is implemented with the resampling method, but the spatial averages are  computed by further averaging all results across the $J$ posterior samples, i.e.,
 \begin{equation*}
    \hat{x}(B_i,t) = \dfrac{1}{J}\dfrac{1}{L}\dfrac{1}{\#B_i}\sum_{j=1}^J\sum_{\ell=1}^L\sum_{\forall \mathbf{s}^*\in B_i}\Big( \tilde{\beta}_0^{(j\ell)} + \tilde{\bm{\beta}}_1^{(j\ell)\intercal}\bm{z}(\mathbf{s}^*,t) + \mathbf{b}_{\text{s}^*}^\intercal\tilde{\tilde{\bm{\xi}}}_{t}^{(j\ell)} \Big)w_{\ell},
\end{equation*}
where $\tilde{\beta}_0^{(j\ell)}$, $\tilde{\beta}_1^{(j\ell)}$, and $\{\tilde{\tilde{\bm{\xi}}}_{t}^{(j\ell)}\}_{t=1,\ldots,T}$ are the $j^{\text{th}}$ posterior samples from the first-stage model fitted conditional on $\alpha_1=\alpha_1^{(\ell)}, \ell=1,\ldots,L$.

\section{Results}\label{sec:results}

This section presents the results for the dengue models, where the input (first-stage) models are the data fusion models. The results for a stations-only model input can be found in the Supplementary Material. Figure \ref{fig:ggpairs_climate} in the Appendix shows the pairwise scatter plot of the three climate variables. The figure shows that relative humidity and log rainfall are positively correlated, while temperature and relative humidity are negatively correlated. On the other hand, temperature and log rainfall are not correlated. Hence, we considered two health models: the first model considers temperature and log rainfall as climate variables (Section \ref{subsubsec_stage2_temprain}), while the second model only has relative humidity as the climate variable (Section \ref{subsubsec_stage2_rh}). We considered the following additional covariates in the model at the province level: population density (\texttt{PopDensity}), and a binary variable which indicates if a time point is during the COVID-19 pandemic (\texttt{covid}).   

\subsection{Temperature and log rainfall}\label{subsubsec_stage2_temprain}

The linear predictor of the health model with temperature and log rainfall as climate covariates is given as follows:
\begin{equation}\label{eq:TempRainModel}
    \begin{aligned}
        \eta\Big(B_i,t\Big) = \gamma_0 + & \gamma_1\widehat{\text{Temperature}}(B_i,t) + \gamma_2\widehat{\text{Temperature}}^2(B_i,t) + \gamma_3\log\widehat{\text{Rain}}(B_i,t) + \gamma_4\text{ClimateType}(B_i,t) \\
    & + \gamma_5\log\widehat{\text{Rain}}(B_i,t)\times\text{ClimateType}(B_i,t)+\gamma_6\text{covid}+\gamma_7\log\text{PopDensity}(B_i,t)+\varphi(B_i,t)
    \end{aligned}
\end{equation}


In Equation \eqref{eq:TempRainModel}, we considered a non-linear effect of temperature, following \cite{xu2020high}. The \texttt{ClimateType} variable is a binary variable which takes a value of `1' for the eastern section of the country, and takes `0' for the western section.  The country's western section has a pronounced dry and wet season, while the eastern part has relatively high rainfall all year round \citep{kintanar_climate,coronas1920climate}. Moreover, we considered an interaction effect between $\log$ rainfall and climate type. This is based on the results in \cite{cawiding2025disentangling}, which showed that the effect of rainfall varies for different regions of the country (see Section \ref{sec:literature}). The values $\widehat{\text{Temperature}}(B_i,t)$ and $\widehat{\text{Rain}}(B_i,t)$ are computed using Equation \eqref{eq:blockaverage}; see also Algorithms \ref{alg:plugin} and \ref{alg:resampling}.

Table \ref{tab:ModelSelectionTempRain} shows the marginal log likelihood (MLik), Watanabe-Akaike Information Criterion (WAIC), and the conditional predictive ordinate (CPO) values for the different models considered. These values are based on the results from the crude plug-in method (Algorithm \ref{alg:plugin}). Results show that Type II interaction model has the highest MLik, the smallest WAIC, and the smallest CPO value as well. Hence, we considered the Type II interaction model for further investigation.

\begin{table}[ht] 
\centering
\scalebox{.9}{\begin{tabular}{|l|rrr|rrr|}
  \hline
 Model & MLik & WAIC & CPO \\ 
  \hline 
  Type I & -7368.66 & 10658.92 & 14182.67\\ 
  Type II & -6814.81 & 10526.76 & 7694.59 \\ 
  Type III & -7337.89 & 10705.02 & 12526.74 \\ 
 Type IV & -11488.96 & 22881.93 & 14366.48  \\ 
   \hline 
\end{tabular}}
\caption{Marginal log likelihood (MLik), WAIC, and $-\sum\log\text{CPO}_i$ for different dengue models with temperature and log rainfall as climate covariates} 
\label{tab:ModelSelectionTempRain}
\end{table}

Table \ref{tab:TempRainFixedEffects} shows the fixed effects estimates. The results show that temperature has a non-linear relationship with dengue. In particular, the higher the temperature, the higher the risk; however, too high temperature leads to a decline in the risk. Moreover, although the main effect of log rainfall is not significant in the model, the interaction between log rainfall and climate type is significant and is negative, with the plug-in approach. The results imply that for a 10\% increase in the amount of rainfall, there is an expected decline in the risks by around 0.43\% for areas in the eastern section of the country, i.e., for areas with uniform amounts of rainfall and relatively wet all year round, log rainfall and dengue are negatively related. This agrees with the results from \cite{cawiding2025disentangling}, which explained that the constant amount of rainfall tends to flush out breeding sites for mosquitoes; thus decreasing the risk of dengue. Note that the resampling approach increased the posterior standard deviation of the coefficient ($\gamma_5$) of this interaction effect, causing it to be no longer significant. Lastly, population density and dengue are positively associated, while the \texttt{covid} binary variable is not significant. For the case with a stations-only climate model as input (see Supplementary Material), the coefficient of log rainfall is significant. The results suggest that for areas in the western section of the country, a 10\% increase in rainfall is associated with an increase in the risks by around $0.09\%$ or $0.07\%$, based on the plug-in method and resampling method, respectively.

\begin{figure}[h!]
    \centering
    \subfloat[][$\gamma_1$, Temperature]
    {\includegraphics[width=0.3\linewidth]{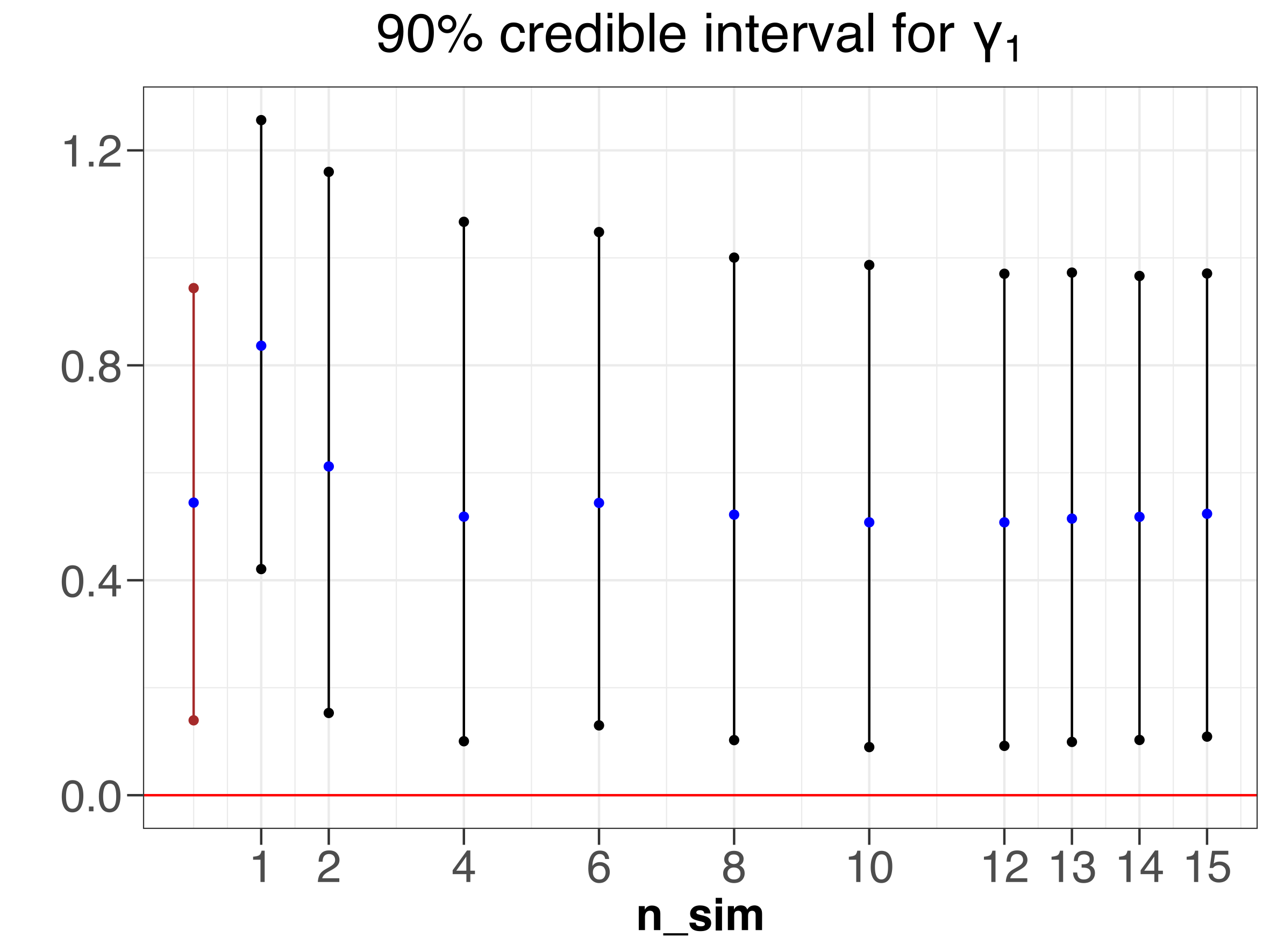}}
    \subfloat[][$\gamma_2$, Temperature$^2$]  {\includegraphics[width=0.3\linewidth]{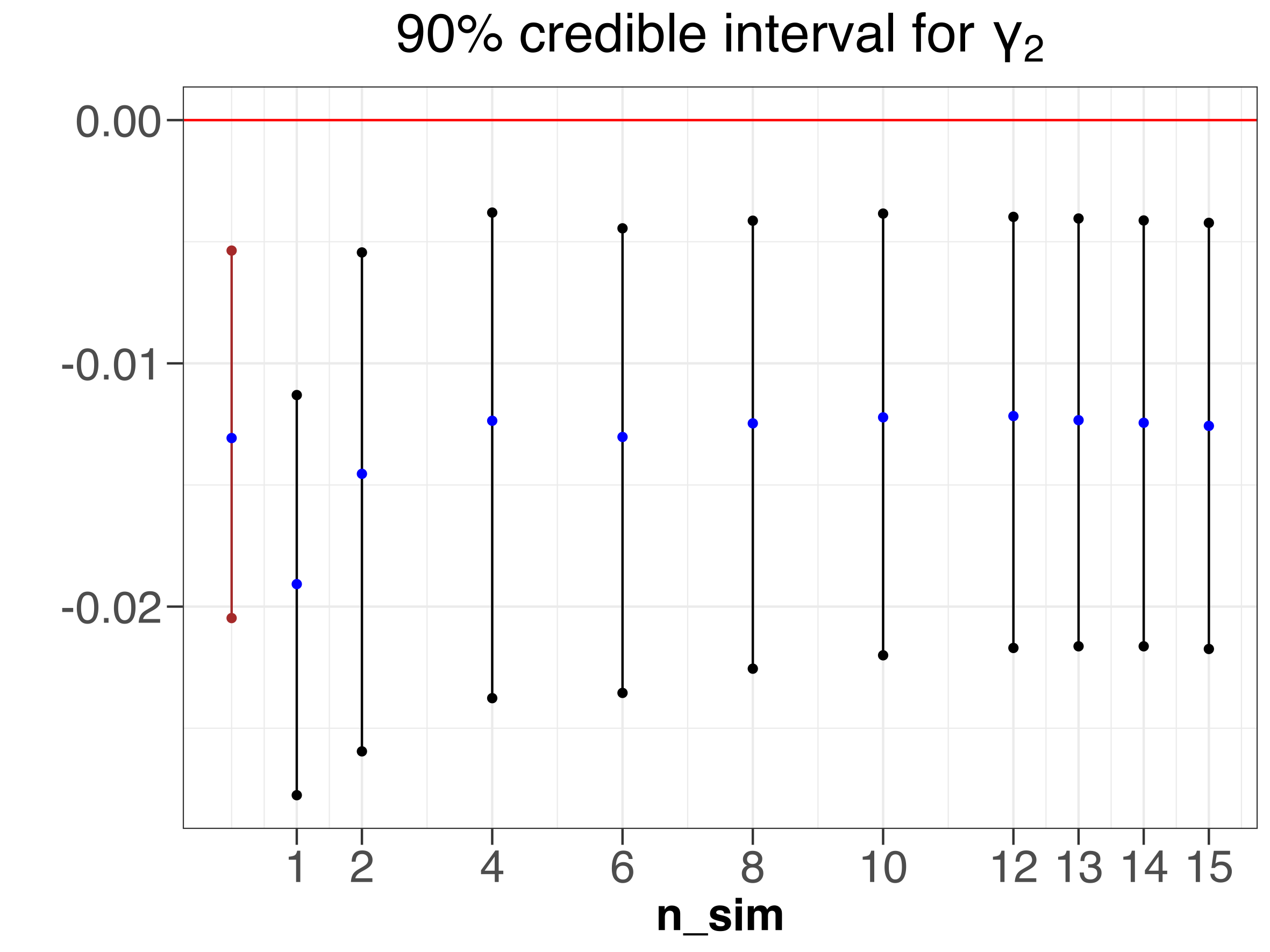}}
    \subfloat[][$\gamma_3$, log Rain $\times$ ClimateType ]  {\includegraphics[width=0.3\linewidth]{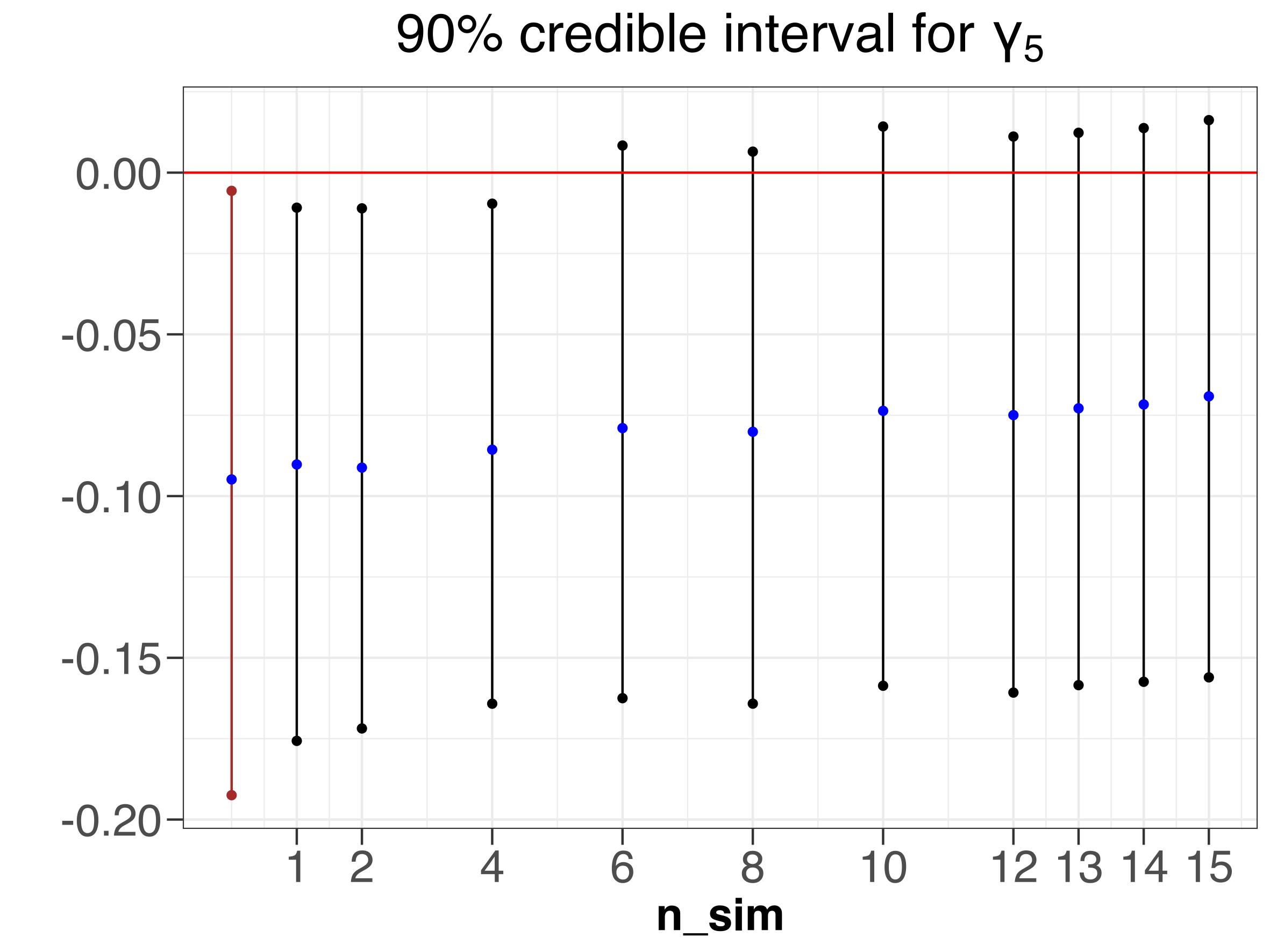}}
    \caption{Plot showing the posterior means and 90\% credible intervals for the following parameters: (a) $\gamma_1$ (b) $\gamma_2$ (c) $\gamma_5$; for the model with temperature and log rainfall as climate covariates. The first vertical line shows the estimates for the plug-in method, while the rest of the lines show the estimates for the resampling method for different number of resamples, from 1 to 15. }
     \label{fig:TempRainGammasDataFusion_risk}
\end{figure}

The results show that the posterior standard deviations from the resampling method are generally larger compared to those of the plug-in method, except for coefficients whose credible intervals (CIs) contain the null value. Figure \ref{fig:TempRainGammasDataFusion_risk} shows a clear comparison of the posterior uncertainties for $\gamma_1$, $\gamma_2$, and $\gamma_5$. We observe an attenuation of the estimated posterior means to the null risk using the resampling method. This is also observed in \citet{lee2017rigorous} and \citet{liu2017incorporating}, where they argue that it is due to the posterior predictive distribution from the first-stage model outweighing the spatio-temporal variation in the data. The same plots for the other covariates are shown in Figure \ref{fig:TempRainOtherGammasDataFusion_risk} of the Appendix.

\begin{table}[!h]
    \centering
    \scalebox{0.9}{\begin{tabular}{|l|rrrr|rrrr|}
  \hline
 &  \multicolumn{4}{c|}{\textbf{Plug-in method}} & \multicolumn{4}{c|}{\textbf{Resampling method}}\\
 Parameter & Mean & SD & P5\% & P95\% & Mean & SD & P5\% & P95\% \\ 
  \hline
\textcolor{brown}{$\gamma_0$, Intercept} & -6.2806 & 3.0463 & -11.4991 & -1.9470 & -6.3940 & 3.3271 & -12.0272 & -1.0479 \\ 
  \textcolor{brown}{$\gamma_1$, Temperature} & 0.5332 & 0.2368 & 0.1545 & 0.9158 & 0.5326 & 0.2665 & 0.1043 & 0.9781  \\ 
  \textcolor{brown}{$\gamma_2$, Temperature$^2$} & -0.0132 & 0.0050 & -0.0206 & -0.0053 & -0.0127 & 0.0053 & -0.0218 & -0.0042 \\
  \textcolor{brown}{$\gamma_3$, log Rain} & -0.0176 & 0.0275 & -0.0645 & 0.0242 & -0.0207 & 0.0239 & -0.0607 & 0.0182 \\ 
  \textcolor{brown}{$\gamma_4$, ClimateType} & 0.3493 & 0.3479 & -0.2184 & 0.9296 & 0.2016 & 0.3578 & -0.3920 & 0.7910  \\ 
  \textcolor{brown}{$\gamma_5$, log Rain $\times$ ClimateType} & -0.0886 & 0.0543 & -0.1786 & -0.0075 & -0.0689 & 0.0528 & -0.1561 & 0.0185 \\ 
  \textcolor{brown}{$\gamma_6$, covid} & -0.1396 & 0.0922 & -0.2879 & 0.0228 & -0.1375 & 0.0915 & -0.2892 & 0.0127 \\
   \textcolor{brown}{$\gamma_7$, log PopDensity} & 0.2124 & 0.0934 & 0.0498 & 0.3590 & 0.1998 & 0.0967 & 0.0404 & 0.3585  \\
   \hline
\end{tabular}}
\caption[]{\label{tab:TempRainFixedEffects}Comparison of estimates of fixed effects between the plug-in method and the resampling method for the dengue model with temperature and log rainfall as climate covariates}
\end{table}

Table \ref{tab:TempRainHyper} in the Appendix shows the estimates of the hyperparameters. Results show that the posterior uncertainty is generally higher for the resampling method compared to the plug-in method. Moreover, the structured effect in time accounts for more variability in the data compared to the unstructured effect. The mixing parameter $\phi$ of the spatial effect is less than 0.5 for the plug-in method, but greater than 0.5 for the resampling method. This suggests that the variability explained by the structured spatial effect is smaller compared to the unstructured effect for the plug-in method, but it is the opposite for the resampling method. Moreover, the time dependence in the interaction effect is strong, since the estimated autoregressive parameter is close to 1.

\begin{figure}[h!]
    \centering
    \subfloat[][Posterior mean]
    {\includegraphics[width=0.45\linewidth]{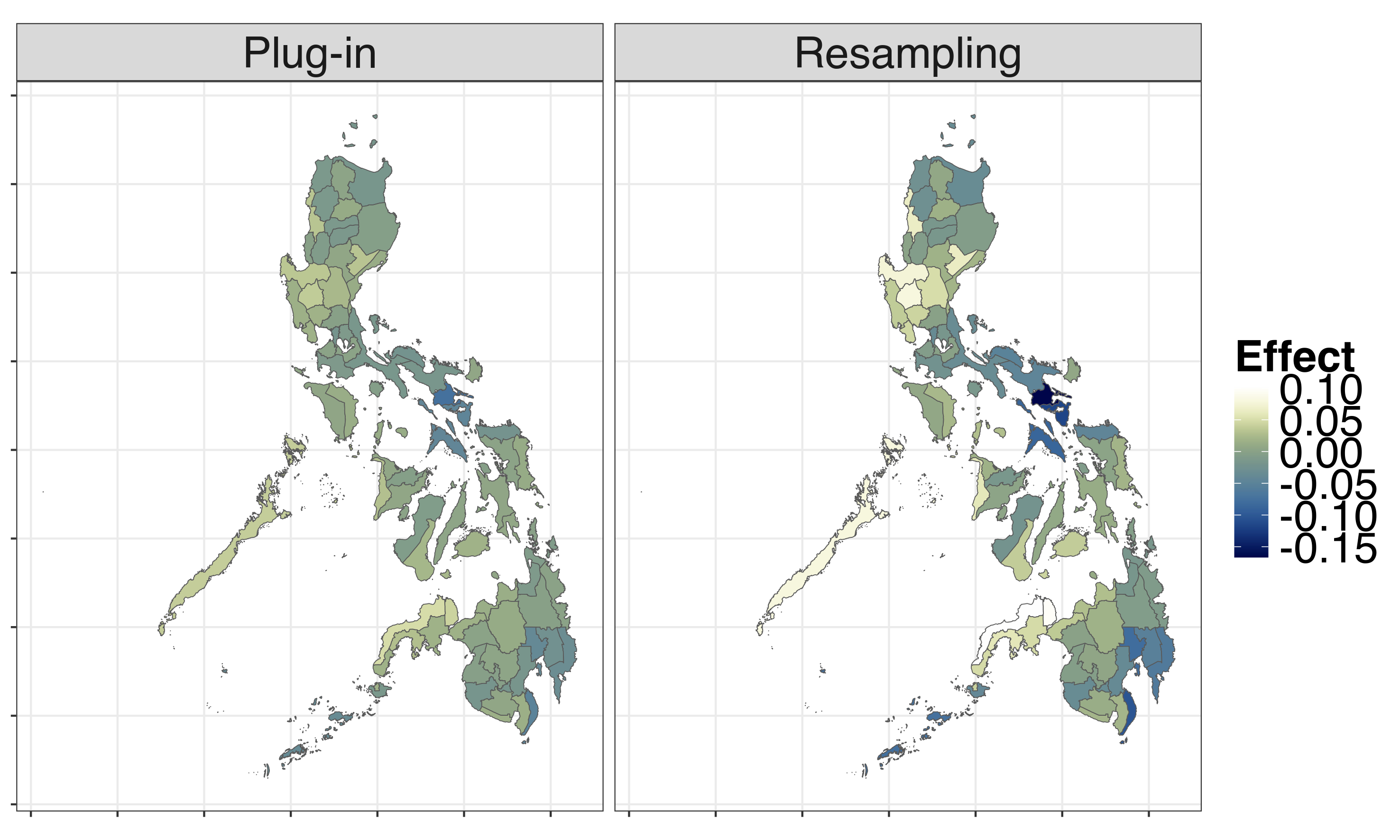}}\vspace{5mm}
    \subfloat[][Posterior standard deviation]{\includegraphics[width=0.45\linewidth]{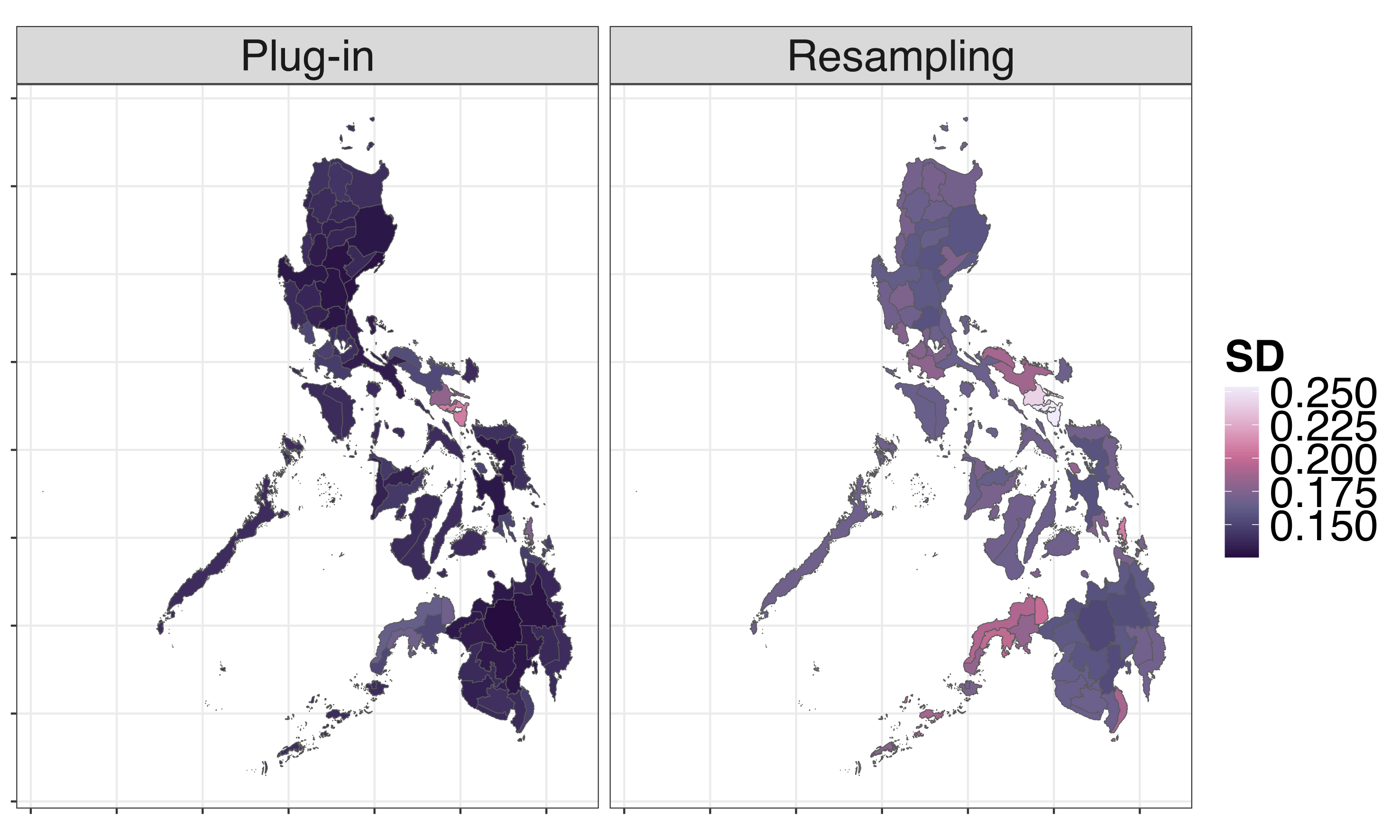}}
    \caption{Comparison of (a) posterior mean and (b) posterior standard deviation of the space effects $\psi(B_i)$ between the plug-in method and resampling method, for the dengue model with temperature and log rainfall as climate covariate}
    \label{fig:TempRainSpaceMeanSD}
\end{figure}

Figure \ref{fig:TempRainSpaceMeanSD}b shows a comparison of the posterior standard deviation of the space effects $\psi(B_i)$ between the plug-in method and resampling method. The figure shows that the resampling method gave higher uncertainty in the spatial effects. The posterior means of $\psi(B_i)$ are shown in Figure \ref{fig:TempRainSpaceMeanSD}a. Note that the posterior means are quite similar between the plug-in method and the resampling method.

\begin{figure}[h!]
    \centering
    \subfloat[][Temperature and log rainfall as covariates]
    {\includegraphics[width=0.45\linewidth]{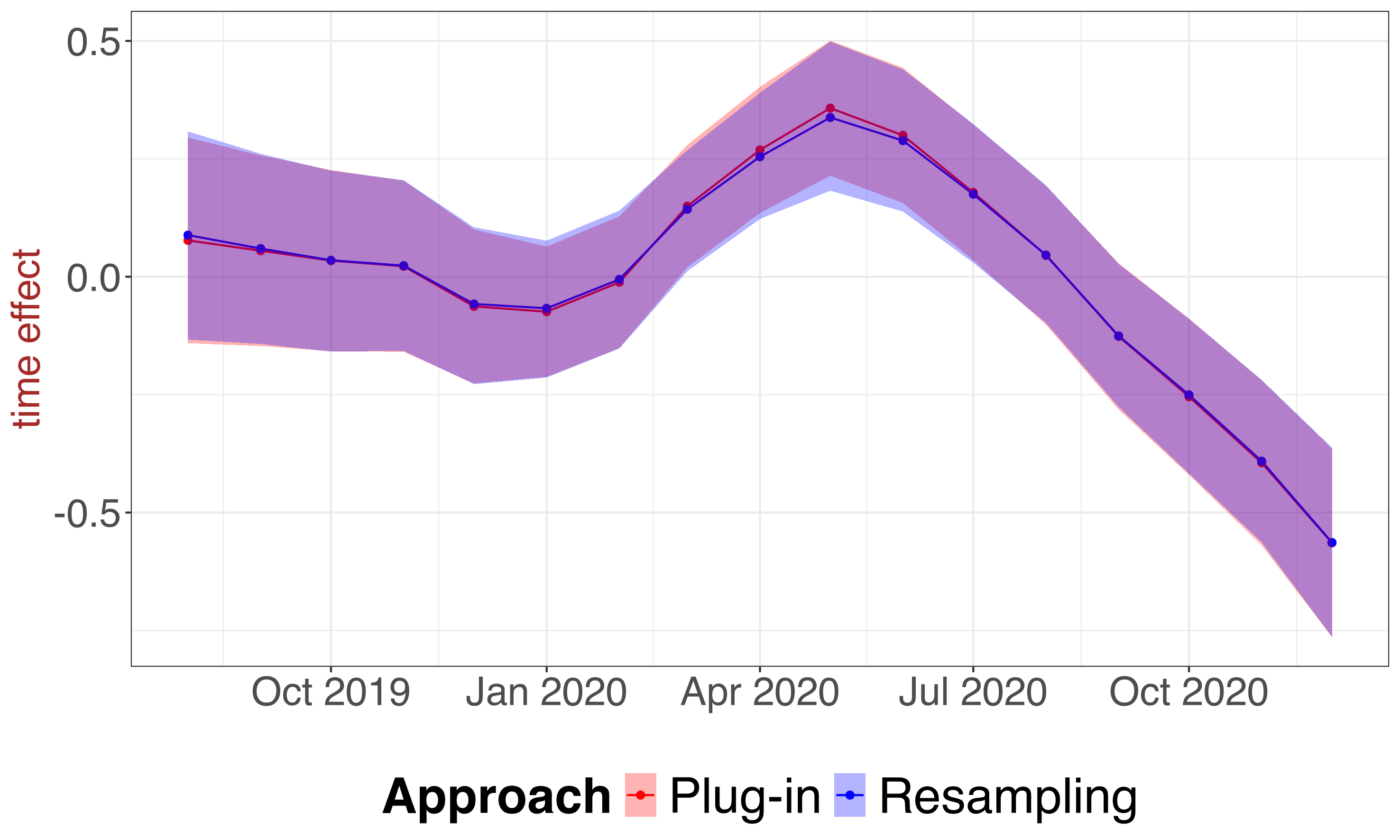}}
    \subfloat[][Relative humidity as covariate]
    {\includegraphics[width=0.45\linewidth]{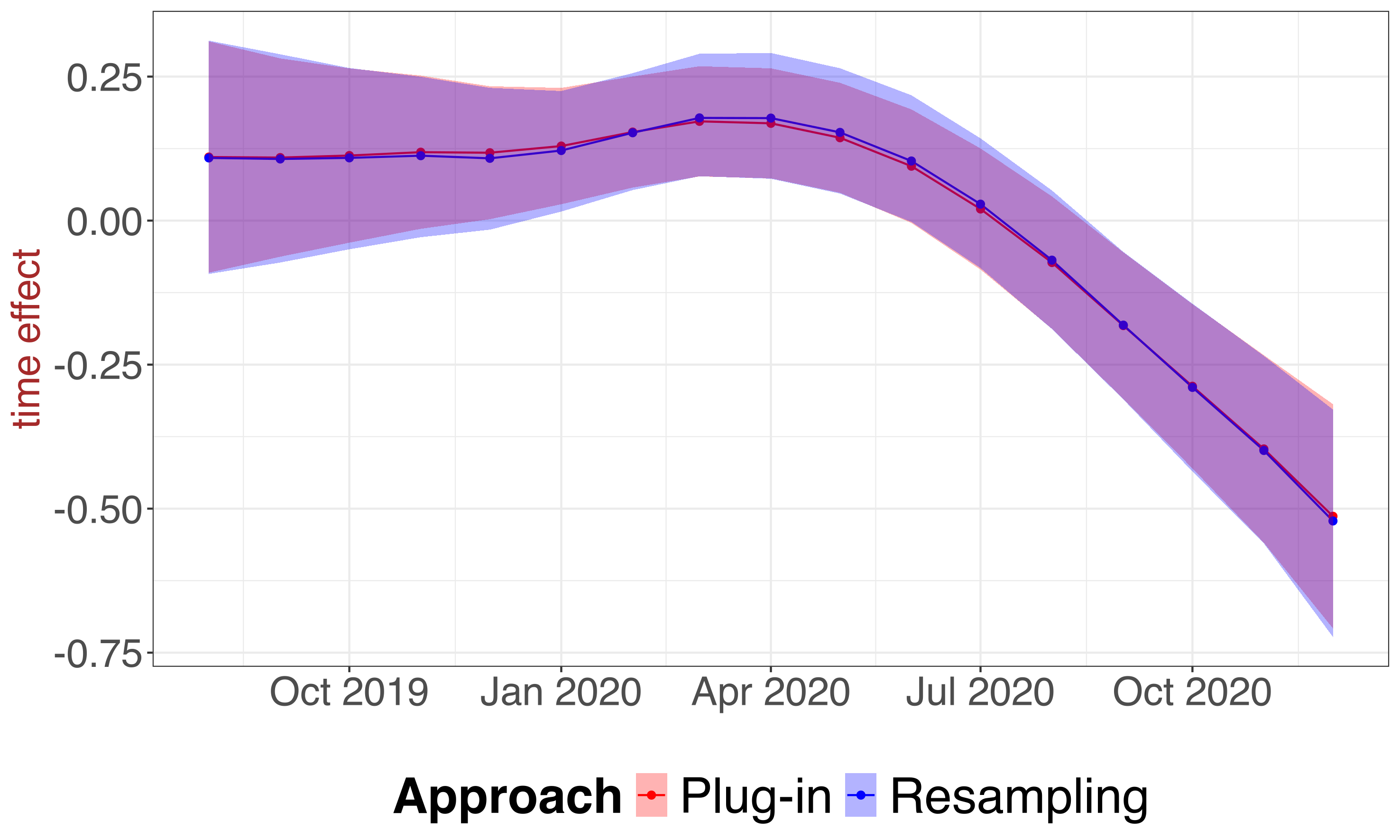}}
    \caption{Plot of the estimated structured time effects $\nu(t)$ with the 95\% credible intervals between the plug-in method and resampling method: (a) temperature and log rain as climate covariates (b) relative humidity as covariate}
    \label{fig:AllTimeEffects_risk}
\end{figure}

Figure \ref{fig:AllTimeEffects_risk}a shows a plot of the estimated structured temporal effects with the corresponding 95\% credible intervals. The plot shows that the posterior uncertainty is very similar between the two methods. Moreover, the plot also shows that the estimated temporal effect is decreasing around April or May 2020, which coincides with the COVID-19 pandemic. This potentially explains why the \texttt{covid} binary variable in the model is not significant, since the drop in the reported dengue cases during this year is already accounted for by the structured temporal effect. 

\begin{figure}[!htbp]
    \centering
    \includegraphics[width=.9\linewidth]{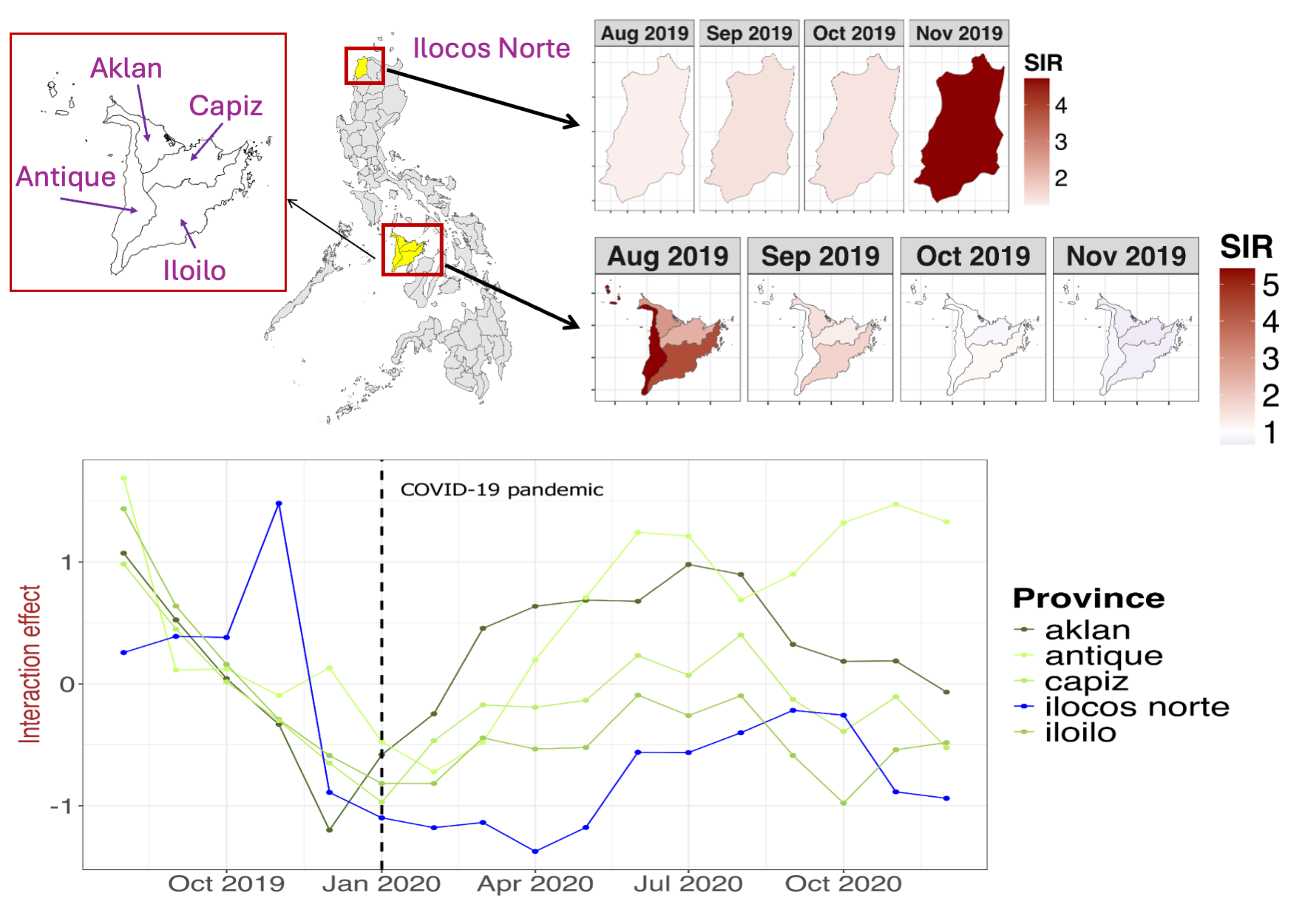}
    \caption{Estimated space-time interaction effect $\upsilon(B_i,t)$ for five provinces. Four of them are contiguous provinces which exhibit the same temporal structure pre-pandemic, and which also agrees with the trend in the SIRs. The fifth province (located in the north) has a decreasing trend in the SIRs for the same time period, and is also accounted for by the space-time effect. The temporal structure during the pandemic varies for the five provinces.}
    \label{fig:TypeII_ineraction}
\end{figure}

Figure \ref{fig:TypeII_ineraction} shows the estimated space-time interaction effect $\nu(B_i,t)$ using the Type II interaction model (see Section \ref{subsec:interactionmodel}) for five provinces. Four of the provinces, which are contiguous and constitute a major island, have decreasing space-time interaction effect pre-pandemic. This also agrees with the trend in the SIRs for the same period (shown in the top portion of Figure \ref{fig:TypeII_ineraction}). For the fifth province considered (Ilocos Norte), which is located in the north, the estimated space time effect shows an increasing trend pre-pandemic, which also agrees with the SIRs. During the COVID-19 pandemic, the estimated effects vary for the 5 provinces, and which shows patterns distinct to each province. This figure confirms the Type II interaction in space and time, which means that in addition to the overall temporal effect, each province exhibits its own temporal structure which is independent of the other provinces.

\subsection{Relative humidity}\label{subsubsec_stage2_rh}

The linear predictor of the health model with relative humidity (RH) as climate covariate has the following form:
\begin{equation}\label{eq:RHModel}
    \begin{aligned}
        \eta\Big(B_i,t\Big) = \gamma_0 + \gamma_1\widehat{\text{RH}}(B_i,t) + & \gamma_2\text{ClimateType}(B_i,t)+ \gamma_3\widehat{\text{RH}}(B_i,t) \times \text{ClimateType}(B_i,t) + \gamma_4\text{covid}+\\
        &\;\;\;\;\;\gamma_5\log\text{PopDensity}(B_i,t)+\varphi(B_i,t) 
    \end{aligned}
\end{equation}

Table \ref{tab:ModelSelectionRH} shows a summary of the metrics to compare the models with different interaction types. The results show that the Type II interaction has the highest marginal log likelihood, the smallest WAIC, and the smallest CPO value. Thus, similar to the model with temperature and log rainfall as climate covariates, we also consider the model with Type II interaction for further investigation. 

\begin{table}[!h] 
\centering
\scalebox{.9}{\begin{tabular}{|l|rrr|rrr|}
  \hline
 Model & MLik & WAIC & CPO \\ 
  \hline 
  Type I & -7363.05 & 10652.39 & 14247.44 \\ 
  Type II  & -6808.63 & 10522.15 & 7756.46  \\ 
  Type III & -7326.95 & 10701.90 & 12542.34 \\ 
 Type IV & -12091.47 & 23698.59 & 14558.67  \\ 
   \hline 
\end{tabular}}
\caption{Marginal log likelihood (MLik), WAIC, and $-\sum\log\text{CPO}_i$ for different dengue models with relative humidity as climate covariate} 
\label{tab:ModelSelectionRH}
\end{table}

Table \ref{tab:RHFixedEffects_risk} shows the estimates of the fixed effects. The main effect of relative humidity is significant and positive. The interaction between relative humidity and climate type is also significant and negative, both for the plug-in approach and the resampling approach. This is the same relationship that log rainfall has with dengue, which is expected since relative humidity and log rainfall are positively correlated. For areas in the eastern section of the country, a one-unit increase in RH is associated with a 1.84\% or 1.40\% decline in the risks, based on the plug-in method and the resampling method, respectively. On the other hand, for areas in the western section of the country, a one-unit increase in RH is associated with a 1.59\% or 1.56\% increase in risks, based on the plug-in method and the resampling method, respectively. Moreover, both population density and the \texttt{covid} variable are  not significant. The non-significance of the \texttt{covid} variable variable is potentially due the temporal random effect accounting for the decline in the dengue risks (see Figure \ref{fig:AllTimeEffects_risk}b). The posterior standard deviations in the coefficients are generally higher for the resampling approach compared to the plug-in method. A comparison of the 90\% credible intervals, similar to Figure \ref{fig:TempRainGammasDataFusion_risk}, is shown in Figure \ref{fig:RHOtherGammasDataFusion_risk} of the Appendix.

\begin{table}[!h]
    \centering
    \scalebox{0.9}{\begin{tabular}{|l|rrrr|rrrr|}
  \hline
 & \multicolumn{4}{c|}{\textbf{Plug-in method}} & \multicolumn{4}{c|}{\textbf{Resampling method}}\\
 Parameter & Mean & SD & P5\% & P95\% & Mean & SD & P5\% & P95\% \\ 
  \hline
\textcolor{brown}{$\gamma_0$, Intercept} & -2.2360 & 0.8622 & -3.6905 & -0.8305 & -2.1124 & 0.8681 & -3.5365 & -0.6733   \\
  \textcolor{brown}{$\gamma_1$, RH} & 0.0170 & 0.0071 & 0.0039 & 0.0294 & 0.0155 & 0.0073 & 0.0035 & 0.0276  \\
  \textcolor{brown}{$\gamma_2$, ClimateType} & 2.5001 & 1.4705 & 0.5529 & 5.0257 & 2.2384 & 1.4407 & -0.1160 & 4.5985  \\
  \textcolor{brown}{$\gamma_3$, RH $\times$ ClimateType} & -0.0356 & 0.0163 & -0.0616 & -0.0072 & -0.0296 & 0.0167 & -0.0577 & -0.0021  \\
  \textcolor{brown}{$\gamma_4$, covid} & -0.0668 & 0.0632 & -0.1589 & 0.0497 & -0.0645 & 0.0680 & -0.1777 & 0.0479  \\ 
  \textcolor{brown}{$\gamma_5$, log PopDensity} & 0.0953 & 0.0896 & -0.0452 & 0.2347 & 0.1035 & 0.0928 & -0.0487 & 0.2582  \\ 
   \hline
\end{tabular}}
\caption[]{\label{tab:RHFixedEffects_risk}Comparison of estimates of fixed effects between the plug-in method and the resampling method for the dengue model with relative humidity as climate covariate.}
\end{table}

Table \ref{tab:RHHyper} in the Appendix shows the estimated hyperparameters. The results show that the posterior standard deviations for the hyperparameters are significantly higher for the resampling method. Figure \ref{fig:RHSpaceMeanSD}b in the Appendix shows a comparison of the posterior SD of the spatial effects $\psi(B_i)$. Here, it is also apparent that the posterior standard deviation from the resampling method is higher compared to the plug-in method. The posterior means of $\psi(B_i)$ are provided in Figure \ref{fig:RHSpaceMeanSD}a in the Appendix, which shows that the estimated posterior means between the plug-in method and resampling method are similar. Finally, Figure \ref{fig:AllTimeEffects_risk}b shows a comparison of the posterior means and 95\% credible intervals for the structured time effect $\nu(t)$. 

\subsection{Estimated risks}

\begin{figure}[t]
    \centering
    \captionsetup{justification=centering}
    \subfloat[][Plug-in method]
    {\includegraphics[width=0.25\linewidth]{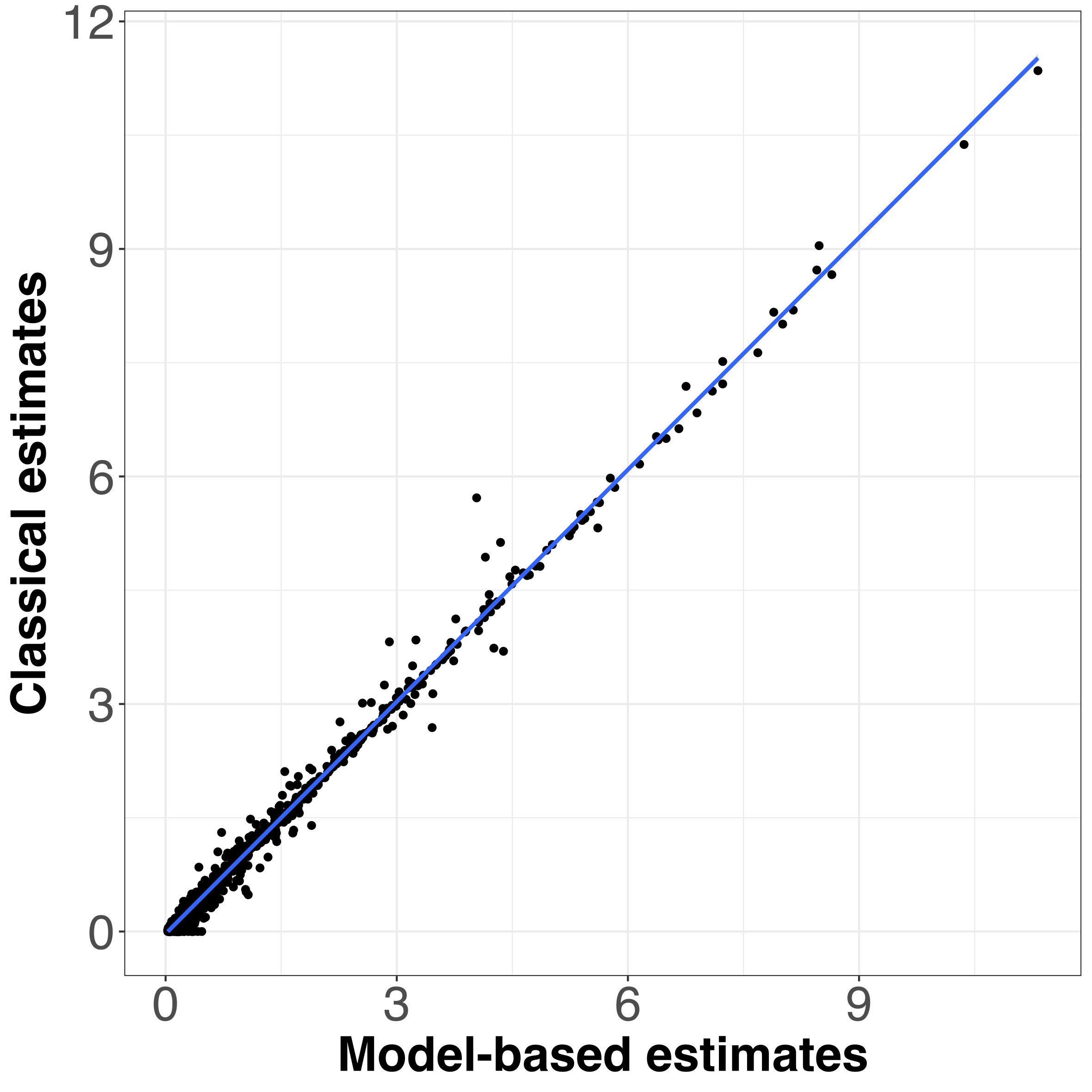}}\hspace{10mm}
    \subfloat[][Resampling method]
    {\includegraphics[width=0.25\linewidth]{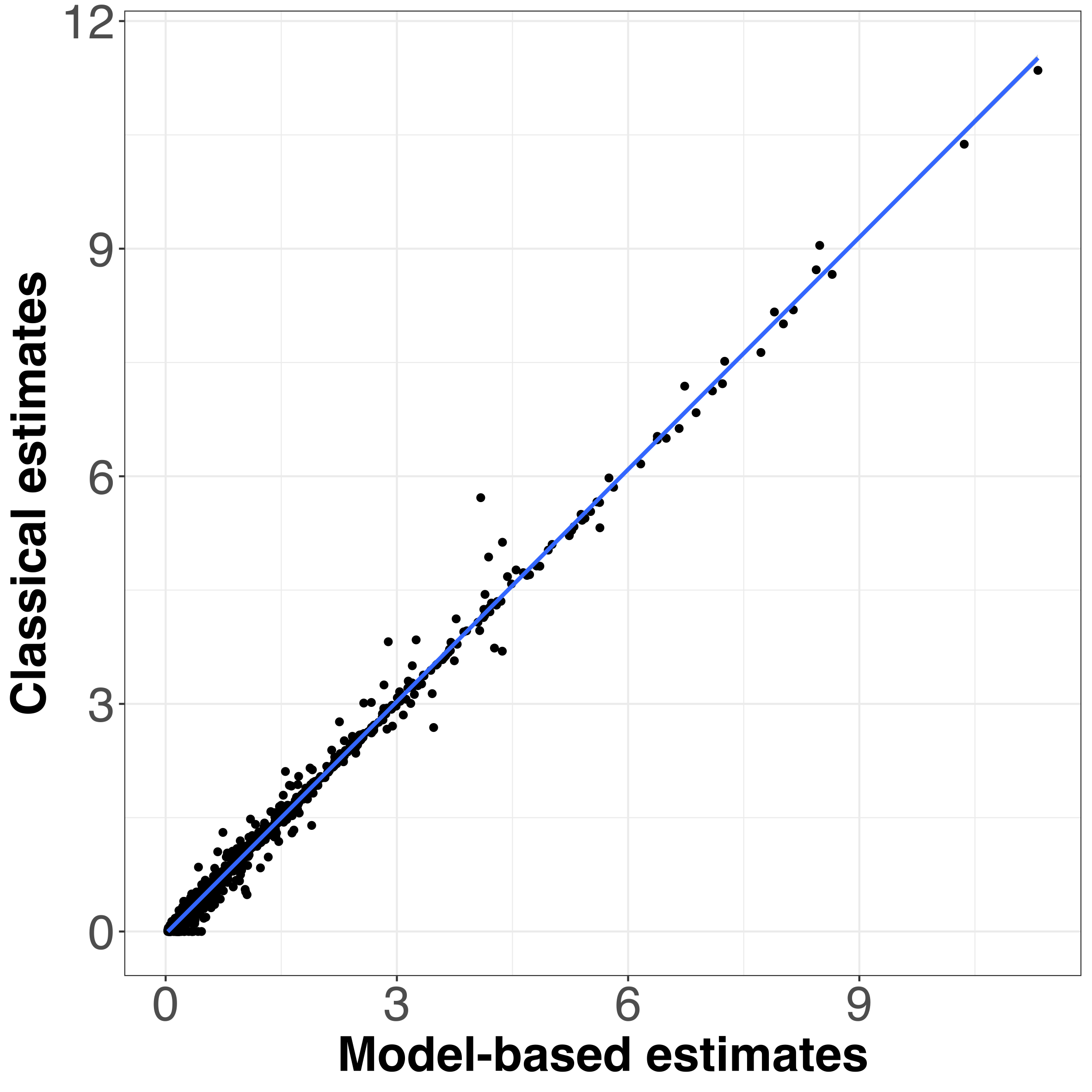}}
    \caption{Comparison of classical SIR estimates and model-based SIR estimates from the health model with temperature and log rainfall as climate covariates: (a) plug-in method (b) resampling method}
    \label{fig:TempRainScatterplotSIR}
\end{figure}

Figure \ref{fig:TempRainScatterplotSIR} shows a comparison of the observed SIRs, which are viewed as classical estimates of risks, versus the model-based estimates $\hat{\lambda}(B_i,t)$ using Equation \eqref{eq:TempRainModel}, i.e., the model with temperature and log rainfall as climate covariates. The figure shows an agreement between the classical estimates and model-based estimates, from both the plug-in method and resampling method. Figure \ref{fig:RHScatterplotSIR} in the Appendix shows the same scatter plots, but from the health model with relative humidity as the climate covariate. It also shows a general agreement between the classical estimates and the model-based estimates of SIR.

\begin{figure}[h!]
    \centering
    \subfloat[][Estimated block-level risks $\hat{\lambda}(B_i,t)$]
    {\includegraphics[width=0.9\linewidth]{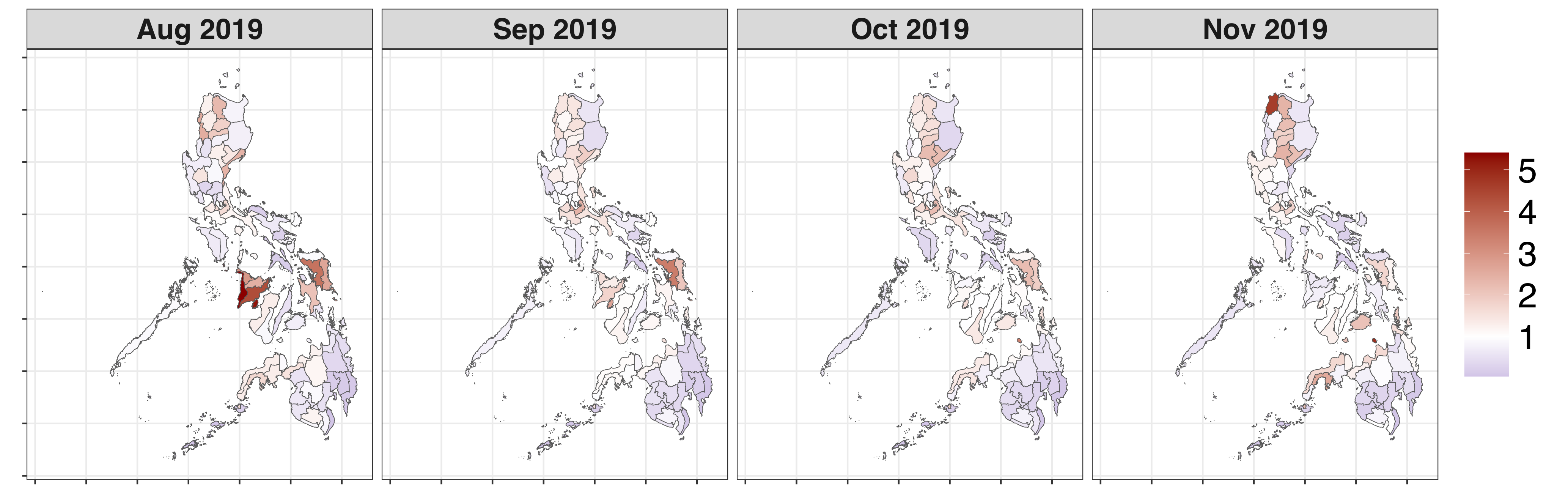}}
    
    \subfloat[][Posterior SD of $\hat{\lambda}(B_i,t)$]
    {\includegraphics[width=0.9\linewidth]{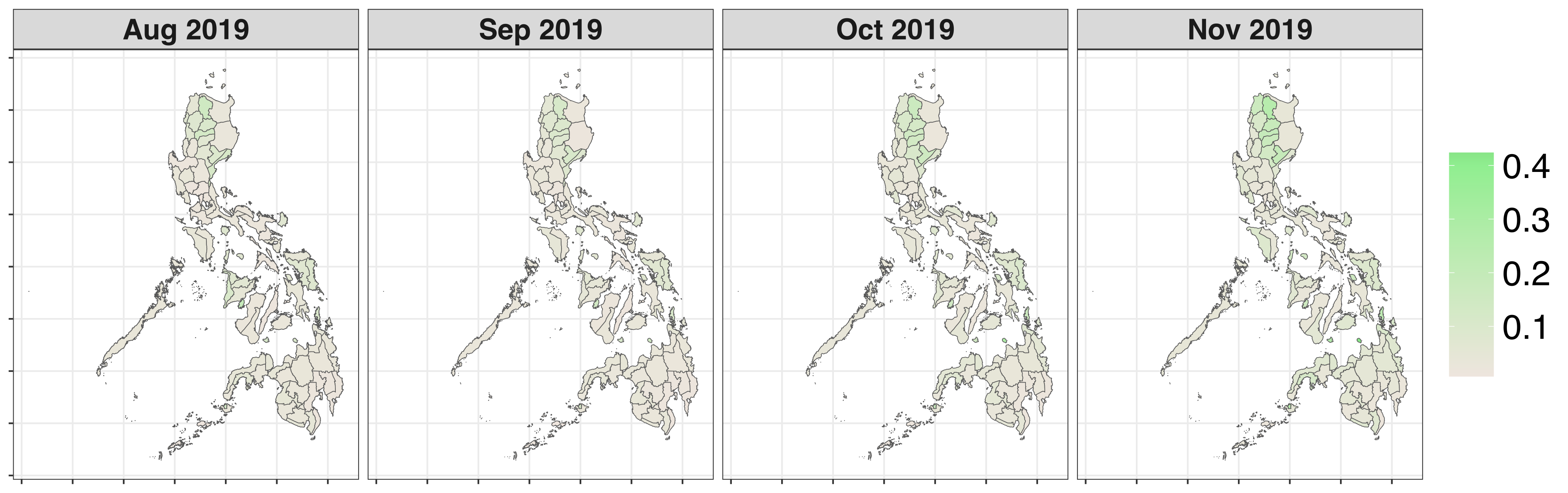}}

    \subfloat[][Probability of exceedance $\mathbb{P}\big(\lambda(B_i,t) > 1 \big)$]
    {\includegraphics[width=0.9\linewidth]{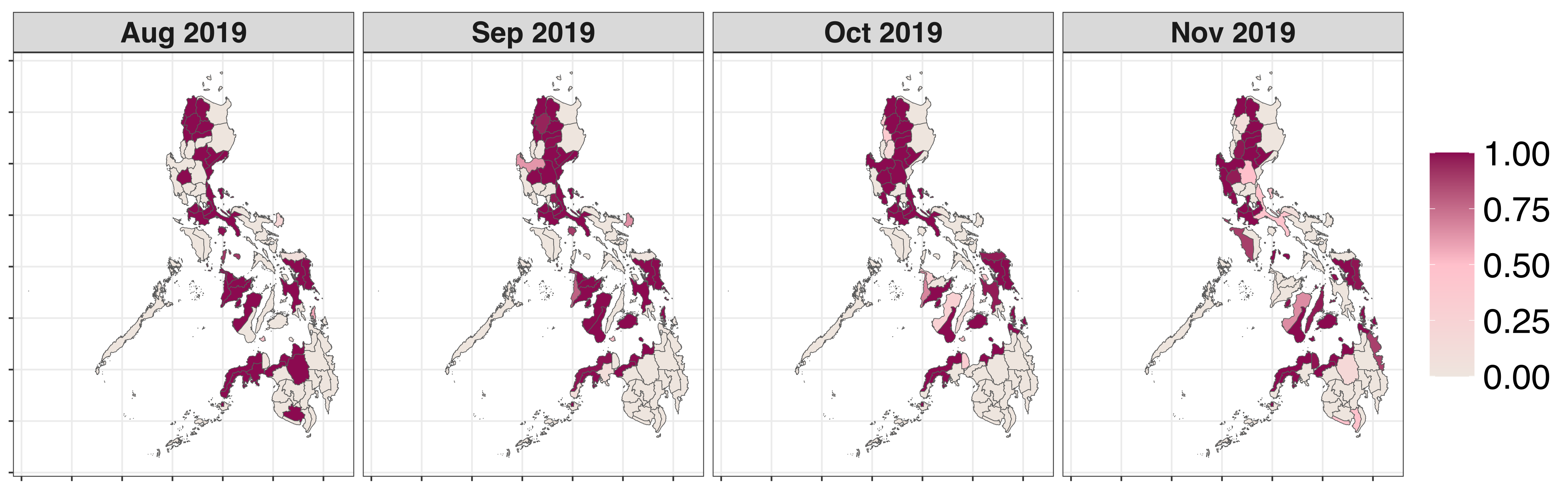}}
    
    \caption{Posterior risk estimates based on the model with temperature and log rainfall as climate covariates and the resampling approach: (a) estimated block-level risks $\hat{\lambda}(B_i,t)$ for August 2019 to November 2019, (b) corresponding posterior uncertainties of $\hat{\lambda}(B_i,t)$, (c) probability of exceedance in the risks $\mathbb{P}\big(\lambda(B_i,t) > 1 \big)$}
    \label{fig:posteriorestimates_blocks}
\end{figure}

Figure \ref{fig:posteriorestimates_blocks}a shows the estimated risks $\hat{\lambda}(B_i,t)$ for the months August to November 2019 using the resampling method and the health model with temperature and log rainfall as climate covariates. This agrees with Figure \ref{fig:DengueSIRsAugtoNov2019}, which shows the classical SIR estimates for the same months. Figure \ref{fig:SIRsDataFusionInput} in the Appendix shows a comparison between the plug-in method and resampling method on $\hat{\lambda}(B_i,t)$. It shows that both approaches have equivalent estimates, which is expected based on Figure \ref{fig:TempRainScatterplotSIR}.  

Moreover, Figure \ref{fig:posteriorestimates_blocks}b shows the corresponding posterior uncertainties in the estimated risks in Figure\ref{fig:posteriorestimates_blocks}a. Areas with high estimated risks appear to have high uncertainties. A comparison of the posterior uncertainties between the plug-in methd and resampling method is shown in Figure \ref{fig:SIRsDataFusionInputSD} in the Appendix. It shows that there is no difference in the posterior uncertainty between the plug-in method and the resampling method. To investigate this further, we look at the variance-covariance structure of the different components of the linear predictor (Equation \ref{eq:TempRainModel}) across the different resamples. In particular, for each posterior sample, we first compute the posterior variance-covariance matrix of the model components. We then average the values for all resamples. Equation \eqref{eq:varcov_fixed_temprain_datafusion} in the Appendix shows the variance-covariance matrix for the fixed effects (across resamples) of the linear predictor in Equation \eqref{eq:TempRainModel}. Note that most of the covariances are negative. Moreover, Equation \eqref{eq:varcov_random_temprain_datafusion} in the Appendix shows the variance-covariance structure for the random effects in the linear predictor. The results show that most covariances are close to zero. Finally, Equation \eqref{eq:crosscov_temprain_datafusion} in the Appendix shows the cross-covariance between the fixed and random effects in the linear predictor, which shows an equal mix of positive and negative linear association between the components. Since most of the pairs of components in the linear predictor in Equation \eqref{eq:TempRainModel} are negatively correlated across the resamples, then this potentially explains why the posterior uncertainty in the dengue risks in the resampling method is similar to the plug-in method. Although the resampling method generally gives higher uncertainty for individual components of the linear predictor, the uncertainty in a linear combination of these components can be washed away because of the latent correlation structure.

Figure \ref{fig:posteriorestimates_blocks}c shows the probability that the dengue risks $\lambda(B_i,t)$ exceed 1, i.e., $\mathbb{P}\Big( \lambda(B_i,t) > 1 \Big)$ for the same months. Note that most of the areas with an estimated probability of exceedence equal to 1 are the same areas badly hit by dengue during the dengue epidemic in the country \citep{CNNepidemic}. A comparison of the estimates between the plug-in and resampling methods are shown in Figure \ref{fig:ProbExceedanceDataFusion} in the Appendix. Both approaches give the same estimates.


\begin{figure}[!h]
    \centering
    \includegraphics[width=0.8\linewidth]{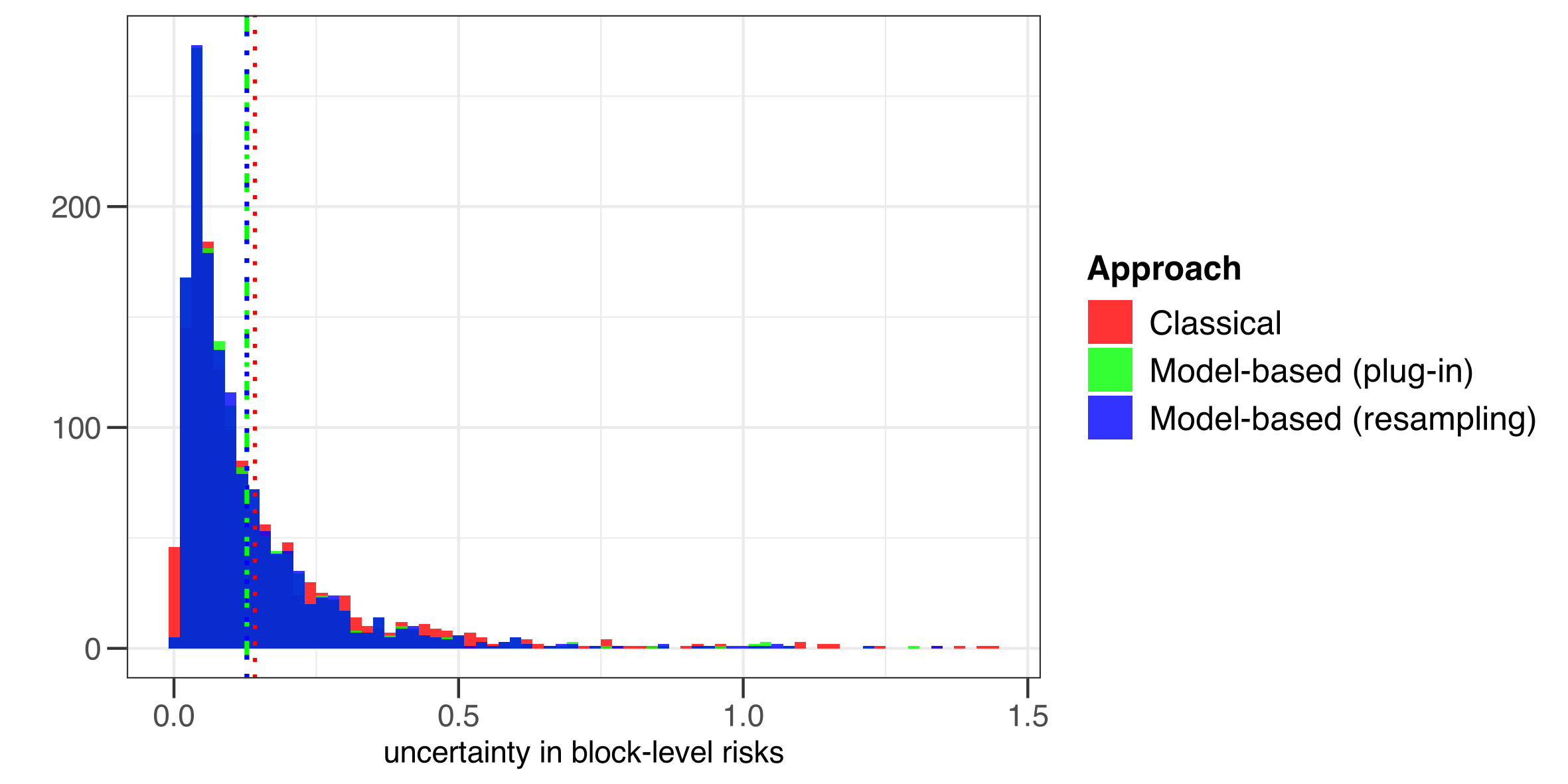}
    \caption{Comparison of the posterior standard deviations in the estimated risks $\hat{\lambda(B_i,t)}$ between three approaches: classical approach based on the asymptotic (Gaussian) distribution of the SIR, model-based estimates from the plug-in approach, model-based estimates from the resampling approach. The model here has temperature and log rainfall as climate covariates. The broken lines are the means of the values for each approach.}
    \label{fig:compareSDs_classicalvsmodelbased}
\end{figure}

Finally, Figure \ref{fig:compareSDs_classicalvsmodelbased} shows a comparison of the posterior standard deviations in the estimated risks $\hat{\lambda(B_i,t)}$ between three approaches: a classical approach based on the asymptotic (Gaussian) distribution of the SIR, model-based estimates from the plug-in approach, and model-based estimates from the resampling approach. Note that there are relatively higher uncertainty values for the classical approach, while the estimates from the model-based plug-in and resampling approaches are almost indistinguishable. On average, the classical approach has higher uncertainty estimates than the model-based approaches (see the broken lines in Figure \ref{fig:compareSDs_classicalvsmodelbased}).


\section{Conclusions}\label{sec:conclusions}

The main aim of this work is to provide additional evidence on the relationship between climate variables and dengue incidence, particularly for the Philippines. In this work, we propose a generalized linear mixed model, where we specify the climate covariates as fixed effects in the health model, and then specify both structured and unstructured random effects in space and time, including their interaction, in order to account for extra variability in the data unexplained by the climate variables. We use the integrated nested Laplace approximation (INLA) approach \citep{rue2009approximate, van2023new} to do model inference. The Bayesian modelling framework that this work employed has not been used thus far based on currently published work which links climate and dengue in the Philippines (see Section \ref{sec:intro}).  Moreover, most of the existing work only looks at certain regions of the country; while this work looks at the data for the entire country.  

The association between climate and dengue has long been established in the literature \citep{mcmichael2003climate,hales2002potential, naish2014climate}. Our results agree with existing studies on the link between climate and dengue. We have shown that temperature is positively related with dengue, although there is a non-linear relationship. In particular, for very high temperature values, the association becomes negative \citep{liu2023effect}, which is explained by the fact that excessively high temperature can shorten the lifespan of mosquitoes and reduce their population size \citep{myer2020mapping}. Moreover, our results show that rainfall has varying effects on dengue, depending on the spatial location of an area, which we defined based on the climate types of the country. In particular, we segmented the country into the eastern and western section. In the western section, rainfall and dengue are positively related, while in the eastern section, the relationship is negative. The eastern section of the country has a low variation in the amount of rainfall and is relatively wet all year round. This phenomenon tends to wash away the breeding sites of mosquitoes, thus explaining its negative relationship with dengue. On the other hand, for the western section, episodes of wet and dry conditions are more pronounced. Phenomena of sporadic rainfall during dry conditions create more breeding sites for mosquitoes, which enhances dengue transmission. This agrees with the results in \cite{cawiding2025disentangling}, who also looked into this spatially varying effect of rainfall on dengue in the Philippines. Relative humidity, which is highly correlated with rainfall, also exhibits the same association with dengue. 

The spatial and temporal effects in the model capture extra variation in the data unexplained by the climate variables and other covariates. When it comes to the spatial effect, most of the variability in space is explained by the conditional autoregressive component, i.e., there are remaining spatial correlations in the data after accounting for the covariate effects. Similarly, for the temporal effect, the structured (random walk) time effect accounts for more of the variability in the data compared to the unstructured effect. In fact, there is a clear decline in the estimated temporal effects during 2020, which is the start of the COVID-19 pandemic, and also a year of very low number of reported dengue cases. We specified a \texttt{covid} binary variable in the model, which turned out to be not significant, most likely since the information is already captured by the random walk effect in time. Moreover, an interaction effect between space and time is shown to be important in the model. In particular, the interaction specifies that each province has its own temporal structure that is independent of the other provinces. A more advanced specification of the random effects is to use the variance partitioning approach proposed in \cite{franco2022variance}. Essentially, it specifies a global (space-time) precision parameter, which is partitioned into the main effect (combined space and time) and the interaction effect via a mixing parameter. The variance explained by the main effect is further partitioned into the space and time effects via another mixing parameter. This kind of specification enhances model interpretability, and also allows an intuitive prior specification. 

Our work used a two-stage modelling approach to link climate and dengue. The first stage fits the climate models, and then the second stage fits the health model for dengue using the climate predictions from the first stage as input. This is a practical framework for doing analysis since the climate models we use are complex, especially the data fusion models. Also, we used climate predictions from existing data fusion models for meteorological data in the Philippines proposed in \cite{villejo2025data}. Thus, a joint modelling framework, i.e., fitting both climate model and health model, is unnecessary. More important reasons for not pursuing a joint modelling framework are as follows. Firstly, our work explores different epidemiological models, e.g., different  random effects specifications or covariate effects; thus, a joint modelling framework is not practical. Another reason is to avoid any feedback effects, i.e., when the health data distorts the climate models, and which usually happens when data from the first stage is sparse \citep{wakefield2006health, shaddick2002modelling, gryparis2009measurement}. Also,  the physical process specifies a clear one-directional relationship between climate and dengue, i.e., climate affects dengue but not the other way around. In other words, the choice of a two-stage modelling framework is guided by both practical and scientific considerations. 

The use of a two-stage approach requires proper propagation of the uncertainty from the first-stage model to the second-stage model. In this work, we used the posterior sampling approach. This approach considers different realizations from the estimated posteriors in the first stage, specifically the climate predictions which we input to the second-stage model. This implies a set of estimated health models,  one for each posterior sample. The final posterior estimates for the second-stage model parameters are then combined using model averaging. In this work, the posterior uncertainties were stable after around 12 resamples. The results show that the posterior means of the model parameters and disease risks computed using the plug-in and resampling approach are equivalent. The posterior standard deviations in the parameter estimates are generally higher for the resampling approach compared to the plug-in approach, but the difference is not substantial. A methodological innovation in this area is to consider an uncertainty propagation approach which does not require resampling from the posteriors and which does not require fitting the second-stage model several times. 

In this work, we computed the block-level estimates of the first-stage latent process $x(\mathbf{s},t)$, which are then used as a covariate in the second-stage model. A future work related to the current problem of  linking a point-referenced covariate and an areal response variable is to use a new model specification which defines a latent intensity field in the second stage. The predictor expression in the second stage would take the following form: $\log\Big(\mu(B,t)\Big)=\log\Big(\bigintssss_B \mu(\mathbf{s},t)d\mathbf{s}\Big) = \log\Big(\bigintssss_B \exp\Big\{ \gamma_0+\gamma_1x(\mathbf{s},t)\Big\}d\mathbf{s}\Big)$, where $\mu(B,t)$ is the mean of the Poisson count for block $B$ at time $t$. This assumes that the mean count at block $B$ at time $t$ is an aggregation of an intensity field $\mu(\mathbf{s},t)$ over $B$, and where $\mu(\mathbf{s},t)$ is non-linearly related to the first-stage latent process $x(\mathbf{s},t)$.  Whereas the model we used in this work specifies the predictor expression in the second-stage as linear with respect to $x(\mathbf{s},t)$, the new specification is a highly non-linear model. This type of model is straightforward to implement using the \texttt{inlabru} library, which extends the class of models that can be fitted using INLA, specifically models which are non-linear in the latent parameters \citep{bachl2019inlabru, lindgren2024inlabru}. 

There are several important indicators of dengue which we did not incorporate in this work's epidemiological models. One important index is the Southern Oscillation Index (SOI), which is an indicator of \textit{El Ni\~no} and \textit{La Ni\~na} episodes. The former is an episode of above-average temperature levels, while the latter implies colder and wetter conditions. A positive SOI is associated with much warmer and wetter conditions than the average, which is ideal for breeding of mosquitoes \citep{mcmichael2003climate}. SOI is shown to be an important indicator of dengue transmission, but the magnitude of its effects could vary for different countries \citep{hales1999nino, hales1996dengue, mcmichael2003climate}. Another extension of the model is to incorporate lagged effects of the climate variables \citep{carvajal2018machine, cruz2024current}, which are also shown to be significant indicators of dengue transmission. This can be pursued in a future work, but we think that this should be used on data with higher time resolution, such as considering weekly cases and daily records of the climate indicators. Future models should also consider vector abundance and biological characteristics of pathogens \citep{murphy2022climate}. Finally, social factors and economic factors are also important indicators in the model. Examples of social factors are human behaviour, such as water storage practices, and land use, such as irrigation/forest clearance/livestock and agricultual practices. Moreover, some economic factors are poverty, population displacement/travel, housing, urbanization, and public health infrastructure \citep{mcmichael2003climate}. There is a complex interaction among the aforementioned factors and the transmission of infectious diseases \citep{foster2001ipcc}, which is something that should be carefully considered for future work.




\bibliographystyle{apalike}
\bibliography{example}

\begin{thebibliography}{}

\bibitem[Abdullah et~al., 2022]{abdullah2022association}
Abdullah, N. A. M.~H., Dom, N.~C., Salleh, S.~A., Salim, H., and Precha, N. (2022).
\newblock The association between dengue case and climate: A systematic review and meta-analysis.
\newblock {\em One Health}, page 100452.

\bibitem[Adin et~al., 2023]{adin2023automatic}
Adin, A., Krainski, E., Lenzi, A., Liu, Z., Mart{\'\i}nez-Minaya, J., and Rue, H. (2023).
\newblock Automatic cross-validation in structured models: {Is} it time to leave out leave-one-out?
\newblock {\em arXiv preprint arXiv:2311.17100}.

\bibitem[Bachl et~al., 2019]{bachl2019inlabru}
Bachl, F.~E., Lindgren, F., Borchers, D.~L., and Illian, J.~B. (2019).
\newblock inlabru: an r package for bayesian spatial modelling from ecological survey data.
\newblock {\em Methods in Ecology and Evolution}, 10(6):760--766.

\bibitem[Bakka et~al., 2018]{bakka2018spatial}
Bakka, H., Rue, H., Fuglstad, G.-A., Riebler, A., Bolin, D., Illian, J., Krainski, E., Simpson, D., and Lindgren, F. (2018).
\newblock Spatial modeling with r-inla: A review.
\newblock {\em Wiley Interdisciplinary Reviews: Computational Statistics}, 10(6):e1443.

\bibitem[Bauer et~al., 2015]{bauer2015quiet}
Bauer, P., Thorpe, A., and Brunet, G. (2015).
\newblock The quiet revolution of numerical weather prediction.
\newblock {\em Nature}, 525(7567):47--55.

\bibitem[BBC, 2019]{BBCepidemic}
BBC (2019).
\newblock Philippines declares dengue epidemic as deaths surge.
\newblock \url{https://www.bbc.co.uk/news/world-asia-49249144#:~:text=Western%20Visayas%20had%20the%20most,levels%20for%20three%20consecutive%20weeks.}
\newblock Accessed: 2025-17-02.

\bibitem[Besag et~al., 1991]{besag1991bayesian}
Besag, J., York, J., and Molli{\'e}, A. (1991).
\newblock Bayesian image restoration, with two applications in spatial statistics.
\newblock {\em Annals of the institute of statistical mathematics}, 43:1--20.

\bibitem[Blangiardo and Cameletti, 2015]{blangiardo2015spatial}
Blangiardo, M. and Cameletti, M. (2015).
\newblock {\em Spatial and spatio-temporal Bayesian models with R-INLA}.
\newblock John Wiley \& Sons.

\bibitem[Blangiardo et~al., 2016]{blangiardo2016two}
Blangiardo, M., Finazzi, F., and Cameletti, M. (2016).
\newblock Two-stage bayesian model to evaluate the effect of air pollution on chronic respiratory diseases using drug prescriptions.
\newblock {\em Spatial and spatio-temporal epidemiology}, 18:1--12.

\bibitem[Buczak et~al., 2014]{buczak2014prediction}
Buczak, A.~L., Baugher, B., Babin, S.~M., Ramac-Thomas, L.~C., Guven, E., Elbert, Y., Koshute, P.~T., Velasco, J. M.~S., Roque~Jr, V.~G., Tayag, E.~A., et~al. (2014).
\newblock Prediction of high incidence of dengue in the philippines.
\newblock {\em PLoS neglected tropical diseases}, 8(4):e2771.

\bibitem[Cameletti et~al., 2019]{cameletti2019bayesian}
Cameletti, M., Gomez-Rubio, V., and Blangiardo, M. (2019).
\newblock Bayesian modelling for spatially misaligned health and air pollution data through the inla-spde approach.
\newblock {\em Spatial Statistics}, 31:100353.

\bibitem[Cameletti et~al., 2013]{cameletti2013spatio}
Cameletti, M., Lindgren, F., Simpson, D., and Rue, H. (2013).
\newblock Spatio-temporal modeling of particulate matter concentration through the spde approach.
\newblock {\em AStA Advances in Statistical Analysis}, 97:109--131.

\bibitem[Carvajal et~al., 2018]{carvajal2018machine}
Carvajal, T.~M., Viacrusis, K.~M., Hernandez, L. F.~T., Ho, H.~T., Amalin, D.~M., and Watanabe, K. (2018).
\newblock Machine learning methods reveal the temporal pattern of dengue incidence using meteorological factors in metropolitan manila, philippines.
\newblock {\em BMC infectious diseases}, 18:1--15.

\bibitem[Cawiding et~al., 2025]{cawiding2025disentangling}
Cawiding, O.~R., Jeon, S., Tubera-Panes, D., de~los Reyes~V, A.~A., and Kim, J.~K. (2025).
\newblock Disentangling climate’s dual role in dengue dynamics: A multiregion causal analysis study.
\newblock {\em Science Advances}, 11(7):eadq1901.

\bibitem[CDCP, 2025]{USCDCP}
CDCP (2025).
\newblock Areas with risk of dengue.
\newblock \url{https://www.cdc.gov/dengue/areas-with-risk/index.html#:~:text=Dengue%20is%20a%20common%20disease,illness%20in%20areas%20with%20risk.}
\newblock Accessed: 2025-03-12.

\bibitem[Col{\'o}n-Gonz{\'a}lez et~al., 2021]{colon2021projecting}
Col{\'o}n-Gonz{\'a}lez, F.~J., Sewe, M.~O., Tompkins, A.~M., Sj{\"o}din, H., Casallas, A., Rockl{\"o}v, J., Caminade, C., and Lowe, R. (2021).
\newblock Projecting the risk of mosquito-borne diseases in a warmer and more populated world: a multi-model, multi-scenario intercomparison modelling study.
\newblock {\em The Lancet Planetary Health}, 5(7):e404--e414.

\bibitem[Coronas, 1920]{coronas1920climate}
Coronas, J. (1920).
\newblock {\em The Climate and Weather of the Philippines, 1903-1918, by Rev. Jos{\'e} Coronas. SJ, Chief, Meteorological Division, Weather Bureau, Manila Observatory}.
\newblock Manila,: Bureau of Printing.

\bibitem[Couper et~al., 2021]{couper2021will}
Couper, L.~I., Farner, J.~E., Caldwell, J.~M., Childs, M.~L., Harris, M.~J., Kirk, D.~G., Nova, N., Shocket, M., Skinner, E.~B., Uricchio, L.~H., et~al. (2021).
\newblock How will mosquitoes adapt to climate warming?
\newblock {\em Elife}, 10:e69630.

\bibitem[Cruz et~al., 2024]{cruz2024current}
Cruz, E.~I., Salazar, F.~V., Aguila, A. M.~A., Villaruel-Jagmis, M.~V., Ramos, J., and Paul, R.~E. (2024).
\newblock Current and lagged associations of meteorological variables and aedes mosquito indices with dengue incidence in the philippines.
\newblock {\em PLOS Neglected Tropical Diseases}, 18(7):e0011603.

\bibitem[Dulay et~al., 2013]{dulay2013climate}
Dulay, A. V.~S., Bautista, J.~R., and Teves, F.~G. (2013).
\newblock Climate change and incidence of dengue fever (df) and dengue hemorrhagic fever (dhf) in iligan city, lanao del norte, philippines.
\newblock {\em Internasional Research Journal of Biological Sciences}, 2(7):37--41.

\bibitem[Duque-Lee et~al., 2020]{duque2020correlation}
Duque-Lee, C.~D., Yu, A. K.~D., Ytienza, S. I.~E., Yu, A. M.~D., Yu, V. C.~S., Wangkay, K. A.~K., Wong, M. A.~R., Zhang, E. M.~T., Yumul, W.~D., Zipagan, Z. M.~R., et~al. (2020).
\newblock Correlation between incidence of dengue and climatic factors in the philippines: An ecological study.
\newblock {\em Health Sciences Journal}, 9(2):1--1.

\bibitem[ECDC, 2023]{ECDPC_factsheet}
ECDC (2023).
\newblock Factsheet for health professionals about dengue.
\newblock \url{https://www.ecdc.europa.eu/en/dengue-fever/facts}.
\newblock Accessed: 2025-02-27.

\bibitem[ECDC, 2024]{ECDC_infog}
ECDC (2024).
\newblock An emerging threat: mosquito-borne diseases in europe.

\bibitem[eClinicalMedicine, 2024]{DengueEClinicalMed}
eClinicalMedicine (2024).
\newblock Dengue as a growing global health concern.
\newblock {\em eClinicalMedicine}, 77.

\bibitem[Edillo et~al., 2022]{edillo2022temperature}
Edillo, F., Ymbong, R.~R., Bolneo, A.~A., Hernandez, R.~J., Fuentes, B.~L., Cortes, G., Cabrera, J., Lazaro, J.~E., and Sakuntabhai, A. (2022).
\newblock Temperature, season, and latitude influence development-related phenotypes of philippine aedes aegypti (linnaeus): Implications for dengue control amidst global warming.
\newblock {\em Parasites \& Vectors}, 15(1):74.

\bibitem[Edillo et~al., 2024]{edillo2024detecting}
Edillo, F., Ymbong, R.~R., Navarro, A.~O., Cabahug, M.~M., and Saavedra, K. (2024).
\newblock Detecting the impacts of humidity, rainfall, temperature, and season on chikungunya, dengue and zika viruses in aedes albopictus mosquitoes from selected sites in cebu city, philippines.
\newblock {\em Virology Journal}, 21(1):42.

\bibitem[Edillo et~al., 2015]{edillo2015economic}
Edillo, F.~E., Halasa, Y.~A., Largo, F.~M., Erasmo, J. N.~V., Amoin, N.~B., Alera, M. T.~P., Yoon, I.-K., Alcantara, A.~C., and Shepard, D.~S. (2015).
\newblock Economic cost and burden of dengue in the philippines.
\newblock {\em The American journal of tropical medicine and hygiene}, 92(2):360.

\bibitem[Ewing et~al., 2016]{ewing2016modelling}
Ewing, D.~A., Cobbold, C.~A., Purse, B., Nunn, M., and White, S.~M. (2016).
\newblock Modelling the effect of temperature on the seasonal population dynamics of temperate mosquitoes.
\newblock {\em Journal of theoretical biology}, 400:65--79.

\bibitem[Focks et~al., 1995]{focks1995simulation}
Focks, D.~A., Daniels, E., Haile, D.~G., Keesling, J.~E., et~al. (1995).
\newblock A simulation model of the epidemiology of urban dengue fever: literature analysis, model development, preliminary validation, and samples of simulation results.
\newblock {\em American journal of tropical medicine and hygiene}, 53(5):489--506.

\bibitem[Forlani et~al., 2020]{forlani2020joint}
Forlani, C., Bhatt, S., Cameletti, M., Krainski, E., and Blangiardo, M. (2020).
\newblock A joint bayesian space--time model to integrate spatially misaligned air pollution data in r-inla.
\newblock {\em Environmetrics}, 31(8):e2644.

\bibitem[Foster, 2001]{foster2001ipcc}
Foster, B. (2001).
\newblock Ipcc third assessment report.
\newblock {\em The Scientific Basis: Geneva, Switzerland}.

\bibitem[Francisco et~al., 2021]{francisco2021dengue}
Francisco, M.~E., Carvajal, T.~M., Ryo, M., Nukazawa, K., Amalin, D.~M., and Watanabe, K. (2021).
\newblock Dengue disease dynamics are modulated by the combined influences of precipitation and landscape: A machine learning approach.
\newblock {\em Science of The Total Environment}, 792:148406.

\bibitem[Franco-Villoria et~al., 2022]{franco2022variance}
Franco-Villoria, M., Ventrucci, M., and Rue, H. (2022).
\newblock Variance partitioning in spatio-temporal disease mapping models.
\newblock {\em Statistical Methods in Medical Research}, 31(8):1566--1578.

\bibitem[Freni-Sterrantino et~al., 2018]{freni2018note}
Freni-Sterrantino, A., Ventrucci, M., and Rue, H. (2018).
\newblock A note on intrinsic conditional autoregressive models for disconnected graphs.
\newblock {\em Spatial and spatio-temporal epidemiology}, 26:25--34.

\bibitem[Fuentes and Raftery, 2001]{fuentes2001model}
Fuentes, M. and Raftery, A.~E. (2001).
\newblock Model validation and spatial interpolation by combining observations with outputs from numerical models via bayesian melding.
\newblock Technical report, WASHINGTON UNIV SEATTLE DEPT OF STATISTICS.

\bibitem[Fuentes and Raftery, 2005]{fuentes2005model}
Fuentes, M. and Raftery, A.~E. (2005).
\newblock Model evaluation and spatial interpolation by bayesian combination of observations with outputs from numerical models.
\newblock {\em Biometrics}, 61(1):36--45.

\bibitem[Fuglstad et~al., 2019]{fuglstad2019constructing}
Fuglstad, G.-A., Simpson, D., Lindgren, F., and Rue, H. (2019).
\newblock Constructing priors that penalize the complexity of gaussian random fields.
\newblock {\em Journal of the American Statistical Association}, 114(525):445--452.

\bibitem[Gelfand et~al., 2010]{gelfand2010handbook}
Gelfand, A.~E., Diggle, P., Guttorp, P., and Fuentes, M. (2010).
\newblock {\em Handbook of spatial statistics}.
\newblock CRC press.

\bibitem[Gettelman et~al., 2022]{gettelman2022future}
Gettelman, A., Geer, A.~J., Forbes, R.~M., Carmichael, G.~R., Feingold, G., Posselt, D.~J., Stephens, G.~L., van~den Heever, S.~C., Varble, A.~C., and Zuidema, P. (2022).
\newblock The future of {Earth} system prediction: {Advances} in model-data fusion.
\newblock {\em Science Advances}, 8(14):eabn3488.

\bibitem[G{\'o}mez-Rubio et~al., 2020]{gomez2020bayesian}
G{\'o}mez-Rubio, V., Bivand, R.~S., and Rue, H. (2020).
\newblock Bayesian model averaging with the integrated nested laplace approximation.
\newblock {\em Econometrics}, 8(2):23.

\bibitem[Gryparis et~al., 2009]{gryparis2009measurement}
Gryparis, A., Paciorek, C.~J., Zeka, A., Schwartz, J., and Coull, B.~A. (2009).
\newblock Measurement error caused by spatial misalignment in environmental epidemiology.
\newblock {\em Biostatistics}, 10(2):258--274.

\bibitem[Gubler et~al., 2014]{gubler2014dengue}
Gubler, D.~J., Ooi, E.~E., Vasudevan, S., and Farrar, J. (2014).
\newblock {\em Dengue and dengue hemorrhagic fever}.
\newblock CABI.

\bibitem[Gubler et~al., 2001]{gubler2001climate}
Gubler, D.~J., Reiter, P., Ebi, K.~L., Yap, W., Nasci, R., and Patz, J.~A. (2001).
\newblock Climate variability and change in the united states: potential impacts on vector-and rodent-borne diseases.
\newblock {\em Environmental health perspectives}, 109(suppl 2):223--233.

\bibitem[Hales et~al., 2002]{hales2002potential}
Hales, S., De~Wet, N., Maindonald, J., and Woodward, A. (2002).
\newblock Potential effect of population and climate changes on global distribution of dengue fever: an empirical model.
\newblock {\em The Lancet}, 360(9336):830--834.

\bibitem[Hales et~al., 1999]{hales1999nino}
Hales, S., Weinstein, P., Souares, Y., and Woodward, A. (1999).
\newblock El ni{\~n}o and the dynamics of vectorborne disease transmission.
\newblock {\em Environmental Health Perspectives}, 107(2):99--102.

\bibitem[Hales et~al., 1996]{hales1996dengue}
Hales, S., Weinstein, P., and Woodward, A. (1996).
\newblock Dengue fever epidemics in the south pacific: driven by el nino southern oscillation?
\newblock {\em The Lancet}, 348(9042):1664--1665.

\bibitem[Iguchi et~al., 2018]{iguchi2018meteorological}
Iguchi, J.~A., Seposo, X.~T., and Honda, Y. (2018).
\newblock Meteorological factors affecting dengue incidence in davao, philippines.
\newblock {\em BMC public health}, 18:1--10.

\bibitem[Kintanar, 1984]{kintanar_climate}
Kintanar, R.~L. (1984).
\newblock {\em Climate of the {Philippines}}.
\newblock Philippine Atmospheric, Geophysical and Astronomical Services Administration (PAGASA).

\bibitem[Knorr-Held, 2000]{knorr2000bayesian}
Knorr-Held, L. (2000).
\newblock Bayesian modelling of inseparable space-time variation in disease risk.
\newblock {\em Statistics in medicine}, 19(17-18):2555--2567.

\bibitem[Lawson et~al., 2016]{lawson2016handbook}
Lawson, A.~B., Banerjee, S., Haining, R.~P., and Ugarte, M.~D. (2016).
\newblock {\em Handbook of spatial epidemiology}.
\newblock CRC press.

\bibitem[Lee et~al., 2017]{lee2017rigorous}
Lee, D., Mukhopadhyay, S., Rushworth, A., and Sahu, S.~K. (2017).
\newblock A rigorous statistical framework for spatio-temporal pollution prediction and estimation of its long-term impact on health.
\newblock {\em Biostatistics}, 18(2):370--385.

\bibitem[Lindgren et~al., 2024]{lindgren2024inlabru}
Lindgren, F., Bachl, F., Illian, J., Suen, M.~H., Rue, H., and Seaton, A.~E. (2024).
\newblock inlabru: software for fitting latent gaussian models with non-linear predictors.
\newblock {\em arXiv preprint arXiv:2407.00791}.

\bibitem[Lindgren and Rue, 2015]{lindgren2015bayesian}
Lindgren, F. and Rue, H. (2015).
\newblock Bayesian spatial modelling with r-inla.
\newblock {\em Journal of statistical software}, 63:1--25.

\bibitem[Lindgren et~al., 2011]{lindgren2011explicit}
Lindgren, F., Rue, H., and Lindstr{\"o}m, J. (2011).
\newblock An explicit link between gaussian fields and gaussian markov random fields: the stochastic partial differential equation approach.
\newblock {\em Journal of the Royal Statistical Society: Series B (Statistical Methodology)}, 73(4):423--498.

\bibitem[Liu et~al., 2017]{liu2017incorporating}
Liu, Y., Shaddick, G., and Zidek, J.~V. (2017).
\newblock Incorporating high-dimensional exposure modelling into studies of air pollution and health.
\newblock {\em Statistics in Biosciences}, 9:559--581.

\bibitem[Liu et~al., 2023]{liu2023effect}
Liu, Z., Zhang, Q., Li, L., He, J., Guo, J., Wang, Z., Huang, Y., Xi, Z., Yuan, F., Li, Y., et~al. (2023).
\newblock The effect of temperature on dengue virus transmission by aedes mosquitoes.
\newblock {\em Frontiers in cellular and infection microbiology}, 13:1242173.

\bibitem[Macdonald.~G, 1957]{macdonald1957epidemiology}
Macdonald.~G, M.~G. (1957).
\newblock {\em The epidemiology and control of malaria.}
\newblock Oxford University Press.

\bibitem[Marigmen and Addawe, 2022a]{marigmen2022climatic}
Marigmen, J. L.~D. and Addawe, R.~C. (2022a).
\newblock Climatic influences on dengue incidence in baguio city, philippines: A multiple linear regression approach.
\newblock In {\em AIP Conference Proceedings}, volume 2472. AIP Publishing.

\bibitem[Marigmen and Addawe, 2022b]{marigmen2022forecasting}
Marigmen, J. L. D.~C. and Addawe, R.~C. (2022b).
\newblock Forecasting and on the influence of climatic factors on rising dengue incidence in baguio city, philippines.
\newblock {\em Journal of Computational Innovation and Analytics (JCIA)}, 1(01):43--68.

\bibitem[McMichael, 2003]{mcmichael2003climate}
McMichael, A.~J. (2003).
\newblock {\em Climate change and human health: risks and responses}.
\newblock World Health Organization.

\bibitem[MetOffice, 2024]{UKMetRH}
MetOffice (2024).
\newblock Why is humidity important.

\bibitem[Murphy et~al., 2022]{murphy2022climate}
Murphy, A.~K., Salazar, F.~V., Bonsato, R., Uy, G., Ebol, A.~P., Boholst, R.~P., Davis, C., Frentiu, F.~D., Bambrick, H., Devine, G.~J., et~al. (2022).
\newblock Climate variability and aede s vector indices in the southern philippines: An empirical analysis.
\newblock {\em PLOS Neglected Tropical Diseases}, 16(6):e0010478.

\bibitem[Murphy and Nathanson, 1994]{murphy1994emergence}
Murphy, F.~A. and Nathanson, N. (1994).
\newblock The emergence of new virus diseases: an overview.
\newblock In {\em Seminars in Virology}, volume~5, pages 87--102. Elsevier.

\bibitem[Murray et~al., 2013]{murray2013epidemiology}
Murray, N. E.~A., Quam, M.~B., and Wilder-Smith, A. (2013).
\newblock Epidemiology of dengue: past, present and future prospects.
\newblock {\em Clinical epidemiology}, pages 299--309.

\bibitem[Myer et~al., 2020]{myer2020mapping}
Myer, M.~H., Fizer, C.~M., Mcpherson, K.~R., Neale, A.~C., Pilant, A.~N., Rodriguez, A., Whung, P.-Y., and Johnston, J.~M. (2020).
\newblock Mapping aedes aegypti (diptera: Culicidae) and aedes albopictus vector mosquito distribution in brownsville, tx.
\newblock {\em Journal of medical entomology}, 57(1):231--240.

\bibitem[Naish et~al., 2014]{naish2014climate}
Naish, S., Dale, P., Mackenzie, J.~S., McBride, J., Mengersen, K., and Tong, S. (2014).
\newblock Climate change and dengue: a critical and systematic review of quantitative modelling approaches.
\newblock {\em BMC infectious diseases}, 14(1):1--14.

\bibitem[Ong et~al., 2022]{ong2022perspectives}
Ong, E.~P., Obeles, A. J.~T., Ong, B. A.~G., and Tantengco, O. A.~G. (2022).
\newblock Perspectives and lessons from the philippines’ decades-long battle with dengue.
\newblock {\em The Lancet Regional Health--Western Pacific}, 24.

\bibitem[PAGASA, 2023]{PHClimate_PAGASA}
PAGASA (2023).
\newblock Climate of the philippines.
\newblock \url{https://www.pagasa.dost.gov.ph/information/climate-philippines}.
\newblock Accessed: 2023-06-12.

\bibitem[Patz et~al., 2000]{patz2000potential}
Patz, J.~A., McGeehin, M.~A., Bernard, S.~M., Ebi, K.~L., Epstein, P.~R., Grambsch, A., Gubler, D.~J., Reither, P., Romieu, I., Rose, J.~B., et~al. (2000).
\newblock The potential health impacts of climate variability and change for the united states: executive summary of the report of the health sector of the us national assessment.
\newblock {\em Environmental health perspectives}, 108(4):367--376.

\bibitem[Pineda-Cortel et~al., 2019]{pineda2019modeling}
Pineda-Cortel, M. R.~B., Clemente, B.~M., and Nga, P. T.~T. (2019).
\newblock Modeling and predicting dengue fever cases in key regions of the philippines using remote sensing data.
\newblock {\em Asian Pacific Journal of Tropical Medicine}, 12(2):60--66.

\bibitem[Promprou et~al., 2005]{promprou2005climatic}
Promprou, S., Jaroensutasinee, M., and Jaroensutasinee, K. (2005).
\newblock Climatic factors affecting dengue haemorrhagic fever incidence in southern thailand.
\newblock {\em Bulletin of the World Health Organization}.

\bibitem[Riebler et~al., 2016]{riebler2016intuitive}
Riebler, A., S{\o}rbye, S.~H., Simpson, D., and Rue, H. (2016).
\newblock An intuitive bayesian spatial model for disease mapping that accounts for scaling.
\newblock {\em Statistical methods in medical research}, 25(4):1145--1165.

\bibitem[Rigau-P{\'e}rez et~al., 1998]{rigau1998dengue}
Rigau-P{\'e}rez, J.~G., Clark, G.~G., Gubler, D.~J., Reiter, P., Sanders, E.~J., and Vorndam, A.~V. (1998).
\newblock Dengue and dengue haemorrhagic fever.
\newblock {\em The lancet}, 352(9132):971--977.

\bibitem[Rue et~al., 2009]{rue2009approximate}
Rue, H., Martino, S., and Chopin, N. (2009).
\newblock Approximate bayesian inference for latent gaussian models by using integrated nested laplace approximations.
\newblock {\em Journal of the royal statistical society: Series b (statistical methodology)}, 71(2):319--392.

\bibitem[Ryan et~al., 2019]{ryan2019global}
Ryan, S.~J., Carlson, C.~J., Mordecai, E.~A., and Johnson, L.~R. (2019).
\newblock Global expansion and redistribution of aedes-borne virus transmission risk with climate change.
\newblock {\em PLoS neglected tropical diseases}, 13(3):e0007213.

\bibitem[Schr{\"o}dle and Held, 2011]{schrodle2011spatio}
Schr{\"o}dle, B. and Held, L. (2011).
\newblock Spatio-temporal disease mapping using inla.
\newblock {\em Environmetrics}, 22(6):725--734.

\bibitem[Seposo et~al., 2023]{seposo2023socio}
Seposo, X., Valenzuela, S., and Apostol, G.~L. (2023).
\newblock Socio-economic factors and its influence on the association between temperature and dengue incidence in 61 provinces of the philippines, 2010--2019.
\newblock {\em PLOS Neglected Tropical Diseases}, 17(10):e0011700.

\bibitem[Seposo et~al., 2024]{seposo2024projecting}
Seposo, X., Valenzuela, S., Apostol, G. L.~C., Wangkay, K.~A., Lao, P.~E., and Enriquez, A.~B. (2024).
\newblock Projecting temperature-related dengue burden in the philippines under various socioeconomic pathway scenarios.
\newblock {\em Frontiers in Public Health}, 12:1420457.

\bibitem[Seposo, 2021]{seposo2021dengue}
Seposo, X.~T. (2021).
\newblock Dengue at the time of covid-19 in the philippines.
\newblock {\em Western Pacific Surveillance and Response Journal: WPSAR}, 12(2):38.

\bibitem[Shaddick and Wakefield, 2002]{shaddick2002modelling}
Shaddick, G. and Wakefield, J. (2002).
\newblock Modelling daily multivariate pollutant data at multiple sites.
\newblock {\em Journal of the Royal Statistical Society Series C: Applied Statistics}, 51(3):351--372.

\bibitem[Simpson et~al., 2017]{simpson2017penalising}
Simpson, D., Rue, H., Riebler, A., Martins, T.~G., and S{\o}rbye, S.~H. (2017).
\newblock Penalising model component complexity: {A} principled, practical approach to constructing priors.
\newblock {\em Statistical Science}.

\bibitem[S{\o}rbye and Rue, 2014]{sorbye2014scaling}
S{\o}rbye, S.~H. and Rue, H. (2014).
\newblock Scaling intrinsic gaussian markov random field priors in spatial modelling.
\newblock {\em Spatial Statistics}, 8:39--51.

\bibitem[Stoddard et~al., 2013]{stoddard2013house}
Stoddard, S.~T., Forshey, B.~M., Morrison, A.~C., Paz-Soldan, V.~A., Vazquez-Prokopec, G.~M., Astete, H., Reiner~Jr, R.~C., Vilcarromero, S., Elder, J.~P., Halsey, E.~S., et~al. (2013).
\newblock House-to-house human movement drives dengue virus transmission.
\newblock {\em Proceedings of the National Academy of Sciences}, 110(3):994--999.

\bibitem[Su, 2008]{su2008correlation}
Su, G. L.~S. (2008).
\newblock Correlation of climatic factors and dengue incidence in metro manila, philippines.
\newblock {\em AMBIO: A Journal of the Human Environment}, 37(4):292--294.

\bibitem[Subido and Aniversario, 2022]{subido2022correlation}
Subido, M.~E. and Aniversario, I.~S. (2022).
\newblock A correlation study between dengue incidence and climatological factors in the philippines.
\newblock {\em Asian Res J Math}, 18:110--119.

\bibitem[Sumi et~al., 2017]{sumi2017effect}
Sumi, A., Telan, E., Chagan-Yasutan, H., Piolo, M., Hattori, T., and Kobayashi, N. (2017).
\newblock Effect of temperature, relative humidity and rainfall on dengue fever and leptospirosis infections in manila, the philippines.
\newblock {\em Epidemiology \& Infection}, 145(1):78--86.

\bibitem[Szpiro et~al., 2011]{szpiro2011efficient}
Szpiro, A.~A., Sheppard, L., and Lumley, T. (2011).
\newblock Efficient measurement error correction with spatially misaligned data.
\newblock {\em Biostatistics}, 12(4):610--623.

\bibitem[Undurraga et~al., 2017]{undurraga2017disease}
Undurraga, E.~A., Edillo, F.~E., Erasmo, J. N.~V., Alera, M. T.~P., Yoon, I.-K., Largo, F.~M., and Shepard, D.~S. (2017).
\newblock Disease burden of dengue in the philippines: adjusting for underreporting by comparing active and passive dengue surveillance in punta princesa, cebu city.
\newblock {\em The American journal of tropical medicine and hygiene}, 96(4):887.

\bibitem[Undurraga et~al., 2013]{undurraga2013use}
Undurraga, E.~A., Halasa, Y.~A., and Shepard, D.~S. (2013).
\newblock Use of expansion factors to estimate the burden of dengue in southeast asia: a systematic analysis.
\newblock {\em PLoS neglected tropical diseases}, 7(2):e2056.

\bibitem[Van~Niekerk et~al., 2023]{van2023new}
Van~Niekerk, J., Krainski, E., Rustand, D., and Rue, H. (2023).
\newblock A new avenue for bayesian inference with inla.
\newblock {\em Computational Statistics \& Data Analysis}, 181:107692.

\bibitem[Villejo et~al., 2023]{villejo2023data}
Villejo, S.~J., Illian, J.~B., and Swallow, B. (2023).
\newblock Data fusion in a two-stage spatio-temporal model using the inla-spde approach.
\newblock {\em Spatial Statistics}, 54:100744.

\bibitem[Villejo et~al., 2025]{villejo2025data}
Villejo, S.~J., Martino, S., Lindgren, F., and Illian, J.~B. (2025).
\newblock A data fusion model for meteorological data using the inla-spde method.
\newblock {\em Journal of the Royal Statistical Society: Series C (Applied Statistics)}.

\bibitem[Wakefield and Shaddick, 2006]{wakefield2006health}
Wakefield, J. and Shaddick, G. (2006).
\newblock Health-exposure modeling and the ecological fallacy.
\newblock {\em Biostatistics}, 7(3):438--455.

\bibitem[Waller, 2004]{waller2004applied}
Waller, L. (2004).
\newblock Applied spatial statistics for public health data.
\newblock {\em Willey \& Sons}.

\bibitem[Waller and Carlin, 2010]{waller2010disease}
Waller, L.~A. and Carlin, B.~P. (2010).
\newblock Disease mapping.
\newblock {\em Chapman \& Hall/CRC handbooks of modern statistical methods}, 2010:217.

\bibitem[WHO, 2023a]{WHODengueSituation}
WHO (2023a).
\newblock Dengue - global situation.
\newblock \url{https://www.who.int/emergencies/disease-outbreak-news/item/2023-DON498}.
\newblock Accessed: 2025-20-02.

\bibitem[WHO, 2023b]{WHO_NTD}
WHO (2023b).
\newblock Neglected tropical diseases.
\newblock \url{https://www.who.int/health-topics/neglected-tropical-diseases#tab=tab_1}.
\newblock Accessed: 2025-03-01.

\bibitem[Xu et~al., 2017]{xu2017climate}
Xu, L., Stige, L.~C., Chan, K.-S., Zhou, J., Yang, J., Sang, S., Wang, M., Yang, Z., Yan, Z., Jiang, T., et~al. (2017).
\newblock Climate variation drives dengue dynamics.
\newblock {\em Proceedings of the National Academy of Sciences}, 114(1):113--118.

\bibitem[Xu et~al., 2020]{xu2020high}
Xu, Z., Bambrick, H., Yakob, L., Devine, G., Frentiu, F.~D., Salazar, F.~V., Bonsato, R., and Hu, W. (2020).
\newblock High relative humidity might trigger the occurrence of the second seasonal peak of dengue in the philippines.
\newblock {\em Science of The Total Environment}, 708:134849.

\bibitem[Yeung and Faidell, 2019]{CNNepidemic}
Yeung, J. and Faidell, S. (2019).
\newblock Philippines declares a national dengue epidemic after 622 deaths.
\newblock \url{https://edition.cnn.com/2019/08/07/health/philippines-dengue-epidemic-intl-hnk/index.html}.
\newblock Accessed: 2025-17-02.

\bibitem[Zhu et~al., 2003]{zhu2003hierarchical}
Zhu, L., Carlin, B.~P., and Gelfand, A.~E. (2003).
\newblock Hierarchical regression with misaligned spatial data: relating ambient ozone and pediatric asthma er visits in atlanta.
\newblock {\em Environmetrics: The official journal of the International Environmetrics Society}, 14(5):537--557.

\end{thebibliography}





\newpage


\section{Appendix}

\subsection{First-stage model -- climate model}\label{subsec:climatemodel}

Let $x(\mathbf{s},t)$ be the true (unknown) value of the climate variable at a spatial location $\mathbf{s}$ and time $t$. We assume:
\begin{equation}
  \begin{aligned}
  \color{black} x(\mathbf{s},t) &= \beta_0 + \bm{\beta}^\intercal \bm{z}(\mathbf{s},t)+\xi(\mathbf{s},t)  \\
    &\xi(\mathbf{s},t) = \phi_{1} \xi(\mathbf{s},t-1) + \omega_1(\mathbf{s},t), \;\;\; |\phi_1|<1, \label{eq:latentprocess} 
  \end{aligned}
\end{equation}

where $\beta_0$ is an intercept, and $\bm{z}(\mathbf{s},t)$ is a set of known covariates,  typically topographical features such as elevation, with linear effect $\bm{\beta}$, and  $\xi(\mathbf{s},t)$ is a spatio-temporal random effect, which evolves in time as an autoregressive (AR) process of order 1, where $\phi_1$ is the AR parameter. Moreover, $\omega_1(\mathbf{s},t)$ is a Gaussian innovation process with the Mat\'ern covariance function. Innovations $\omega_1(\mathbf{s},t)$ are temporally independent, i.e,
\begin{align}
&\text{Cov}\Big(\omega_1(\mathbf{s}_i,t),\omega_1(\mathbf{s}_j,u)\Big) = \begin{cases} 
            0 & t \neq u \\
  			\Sigma_{i,j}& t = u
  		\end{cases} \label{eq:materncovariance1}\\
    &\Sigma_{i,j}= \frac{\sigma_{1}^2}{2^{\nu_1-1}\Gamma(\nu)}\Big(\kappa_1\norm{\mathbf{s}_i-\mathbf{s}_j}\Big)^{\nu_1}K_{\nu_1}\Big(\kappa_1\norm{\mathbf{s}_i-\mathbf{s}_j}\Big), \label{eq:materncovariance2}
\end{align}
where $\sigma_{1}^2$ is the marginal variance, and $\kappa_1$ is a scaling parameter linked to the spatial range. The mean-square differentiability parameter $\nu_1$ is fixed equal to 1, as it is often poorly identified  \citep{lindgren2011explicit}. Moreover, $\norm{\cdot}$ is the Euclidean distance in $\mathbb{R}^2$ between two locations  $\mathbf{s}_i$ and $\mathbf{s}_j$, and $K_{\nu_1}(\cdot)$ is the modified Bessel function of the second kind and order $\nu_1>0$. We assume that $\xi(\mathbf{s},t)$ follows its stationary distribution at $t=1$, i.e., $\xi(\mathbf{s},1)\sim N\Big(0,\sigma^2_{1}/(1-\phi_1^2)\Big)$. 

We assume that the observed value from station $\mathbf{s}_i$ at time $t$, denoted by $\text{w}_1(\mathbf{s}_i,t), i=1,\ldots,\text{n}_{\text{w}_1}, t=1,\ldots,T$, follows the classical error model:
\begin{equation}
\text{w}_1(\mathbf{s}_i,t) = x(\mathbf{s}_i,t) + e_1(\mathbf{s}_i,t), \;\;\; e_1(\mathbf{s}_i,t)\overset{\text{iid}}{\sim}\mathcal{N}(0,\sigma_{e_1}^2). \label{eq:stationsmodel}
\end{equation}
The iid assumption in the random noise $e(\mathbf{s}_i,t)$ is justified as weather stations are far away from each other and operate independently of each other.

In the data fusion model, in addition to data from weather stations, outputs from the Global Spectral Model (GSM) are incorporated. Such data have potentially both an additive and a multiplicative bias.
Let $\text{w}_2(\mathbf{g}_j,t)$ be the forecast value at the grid cell with centroid $\mathbf{g}_j$ at time $t$, $j=1,\ldots,\text{n}_{\text{w}_2}, t=1,\ldots,T$. We assume that:
\begin{equation}\label{eq:gsmdata}
  \begin{aligned}
\text{w}_2(\mathbf{g}_j,t) &= \alpha_0(\mathbf{g}_j,t) + \alpha_1 x(\mathbf{g}_j,t) + e_2(\mathbf{g}_j,t), \\
  &\alpha_0(\mathbf{g}_j,t) = \phi_2\alpha_0(\mathbf{g}_j,t-1) + \omega_2(\mathbf{g}_j,t),
  \end{aligned}
\end{equation}
where $\alpha_0(\mathbf{g}_j,t)$ is a spatially-structured additive bias which varies in time following an AR1 model, where $|\phi_2|<1$ is the AR parameter. The term $\omega_2(\mathbf{g}_j,t)$ is another time-independent Gaussian process with the Mat\'ern covariance function, similar to Equations \eqref{eq:materncovariance1} and \eqref{eq:materncovariance2}, and with its own set of parameters, namely, $\sigma^2_2$ and $\kappa_2$. The parameter $\alpha_1$ is a multiplicative bias parameter and assumed to be constant in space and time. Finally, the $e_2(\mathbf{g}_j,t)$ is a random noise, i.e., $e_2(\mathbf{g}_j,t)\overset{\text{iid}}{\sim}\mathcal{N}\Big(0,\sigma_{e_2}^2\Big)$. 

The data fusion model jointly fits Equations \eqref{eq:latentprocess}, \eqref{eq:stationsmodel}, and \eqref{eq:gsmdata}, and assumes that both weather stations and GSM output data are error-prone realizations of the same latent process, $x(\mathbf{s},t)$. We assume that weather stations data are  more accurate, while the GSM outcomes have better time and space coverage but might be  biased. Details of the climate data fusion model are discussed in \cite{villejo2025data}, which highlights, using leave-group-out cross-validation \citep{adin2023automatic}, that the data fusion model yields better predictions than the stations-only model and a regression calibration model. 

\subsection{Specification of priors for first-stage model}\label{subsec:priorsfirststage}

For the fixed effects of the first-stage model, we have $\beta_0\sim\mathcal{N}(0,\infty)$ and $\bm{\beta}\sim\mathcal{N}(\bm{0},1000\mathbb{I})$. We assign penalized-complexity (PC) priors for the Mat\'ern parameters. This is expressed in terms of probability statements, particularly $\mathbb{P}(\sigma_{1}<\sigma_{1_0})=0.5$ and $\mathbb{P}(\rho_{1}>300)=0.5$, where $\sigma_{1}$ and $\rho_{1}$ are the marginal standard deviation and range parameter of the Mat\'ern field $\omega_1(\mathbf{s},t)$. The value 300 km is one-third of the maximum distance in the spatial domain, while the value of $\sigma_{1_0}$ is the empirical standard deviation of the data. For the AR parameter $\phi_1$, we have $\log\bigg(\dfrac{1+\phi_1}{1-\phi_1}\bigg)\sim\mathcal{N}(0,0.15)$. For the measurement error variance $\sigma_{e_1}^2$, we also assign a PC prior, particularly, $\mathbb{P}(\sigma_{e_1}<1)=0.5$. For the data fusion model, we have three additional parameters: $\sigma_{2}, \rho_{2}$ and $\phi_2$, which are the parameters of the Mat\'ern field $\omega_2(\mathbf{s},t)$. The prior choice for these models are discussed in \cite{villejo2025data}. 

\subsection{Predictor expressions}

The following are the predictor expressions for the climate models:
\begin{equation}
\small
    \begin{aligned}
        \text{Temperature}(\mathbf{s},t) &= \beta_0 + \beta_1 \log\Big(\text{Elevation}(\mathbf{s},t)\Big) + \beta_2\text{Cool}(\mathbf{s},t) + \beta_3\text{ClimateType}(\mathbf{s},t) + \xi(\mathbf{s},t)\\
        \log\text{RH}(\mathbf{s},t) &= \beta_0 + \beta_1\log\Big(\text{Temperature}(\mathbf{s},t)\Big) + \beta_2\log\Big(\text{Temperature}(\mathbf{s},t)\Big)^2 + \beta_3\log\Big(\text{Elevation}(\mathbf{s},t) \Big) \\ &\;\;\;\;\;\;\;\;\;\;\;\;\;\;\;\;\;\;\;\;\;\;\;+ \beta_4\text{ClimateType}(\mathbf{s},t) + \xi(\mathbf{s},t) \\
        \log\Big(\text{Rain}(\mathbf{s},t)+1\Big) &= \beta_0 + \beta_1\log\Big(\text{Temperature}(\mathbf{s},t)\Big) + \beta_2\log\Big(\text{Temperature}(\mathbf{s},t)\Big)^2 + \beta_3\log\Big(\text{Season}(\mathbf{s},t) \Big) \\
        &\;\;\;\;\;\;\;\;\;\;\;\;\;\;\;\;\;\;\;\;\;\;\; + \beta_4\text{ClimateType}(\mathbf{s},t) + \beta_5\text{Season}(\mathbf{s},t)\times \text{ClimateType} (\mathbf{s},t) + \xi(\mathbf{s},t)\label{eq:climatemodels}
    \end{aligned}
\end{equation}
The climate models in Equation \eqref{eq:climatemodels} are based on the data fusion models proposed in \cite{villejo2025data}. The \texttt{Cool} variable is a binary variable which takes a value of 1 during the months December to February, and 0 otherwise. The \texttt{ClimateType} variable is also another binary variable which takes a value of `1' for the eastern section of the country, and takes `0' for the western section. These variables are based on the fact that most of the country's western section has a pronounced dry and wet seasons, while the eastern part has relatively high rainfall all year round \citep{kintanar_climate,coronas1920climate}. The \texttt{Season} variable is also a binary variable which takes a value of 1 for June to November, characterized as a wet period, and takes value of 0 for the other months, which is typically a dry period. 

\subsection{SPDE representation}\label{subsec:SPDErepresentation}

The basis function representation of the field $\omega_{\text{h}}(\mathbf{s},t), \text{h}=1,2$, is given by $\omega_{\text{h}}(\mathbf{s},t) = \sum_{k=1}^K\vartheta_k(\mathbf{s})\chi_{\text{h}kt}$, where $\{\vartheta_k(\mathbf{s})\}$ are basis functions chosen to be piecewise linear on each triangle,  $\{\chi_{\text{h}kt}\}$ are Gaussian weights, and $K$ is the number of vertices or nodes in the mesh. This solution is completely specified by the precision matrix of the Gaussian weights $\{\chi_{\text{h}kt}\}$, which is a sparse matrix \citep{lindgren2011explicit}. Thus, for each time point $t$, the Mat\'ern vectors $\bm{\omega}_{1t} = \begin{pmatrix}
    \omega_{\text{1}}(\mathbf{s}_1,t) & \cdots & \omega_{\text{1}}(\mathbf{s}_{\text{n}_{\text{w}_1}},t)
\end{pmatrix}$ and $\bm{\omega}_{2t} = \begin{pmatrix}
    \omega_{\text{2}}(\mathbf{g}_1,t) & \cdots & \omega_{\text{2}}(\mathbf{g}_{\text{n}_{\text{w}_2}},t)
\end{pmatrix}$ are represented through the GMRFs $\tilde{\bm{\omega}}_{1t} \sim \mathcal{N}\Big(\bm{0}, \mathbf{Q}^{\inv}_{S_1}\Big)$ and $\tilde{\bm{\omega}}_{2t} \sim \mathcal{N}\Big(\bm{0}, \mathbf{Q}^{\inv}_{S_2}\Big)$, where $\mathbf{Q}^{\inv}_{S_1}$ and $\mathbf{Q}^{\inv}_{S_2}$ are sparse precision matrices. The spatio-temporal field $\bm{\xi}_t$ in Equation \eqref{eq:latentprocess} and the additive bias field $\bm{\alpha}_{0t}$ in Equation \eqref{eq:gsmdata} can be written as follows:
\begin{equation}
    \begin{aligned}
        \tilde{\bm{\xi}}_t &= \phi_1 \tilde{\bm{\xi}}_{t-1} + \tilde{\bm{\omega}}_{1t}, \;\;\; \tilde{\bm{\omega}}_{1t} \sim \mathcal{N}\Big(\bm{0}, \mathbf{Q}^{\inv}_{S_1}\Big) \\
        \tilde{\bm{\alpha}}_{0t} &= \phi_2  \tilde{\bm{\alpha}}_{0,t-1} + \tilde{\bm{\omega}}_{2t}, \;\;\; \tilde{\bm{\omega}}_{2t} \sim \mathcal{N}\Big(\bm{0}, \mathbf{Q}^{\inv}_{S_2}\Big)
    \end{aligned},
\end{equation}
for $t=1,\ldots,T$, and with $\tilde{\bm{\xi}}_1\sim\mathcal{N}\Big(\bm{0},\mathbf{Q}^{\inv}_{S_1}/(1-\phi_1^2) \Big)$ and $\tilde{\bm{\alpha}}_{01}\sim\mathcal{N}\Big(\bm{0},\mathbf{Q}^{\inv}_{S_2}/(1-\phi_2^2) \Big)$. 

This implies that the first-stage observation models are  specified as:
\begin{equation}
\begin{aligned}
    \mathbf{w}_{1t} &= \beta_0\mathbf{1} + \bm{Z}_{t}\bm{\beta} + \mathbf{B_1}\tilde{\bm{\xi}}_t + \sigma^2_{e_1}\mathbb{I}_{\text{n}_{\text{w}_1}} \\ 
    \mathbf{w}_{2t} &= \mathbf{B}_2\tilde{\bm{\alpha}_{0t}} + \alpha_1 \Big( \beta_0\mathbf{1} + \bm{Z}_{1t}\bm{\beta} + \mathbf{B_1}\tilde{\bm{\xi}}_t  \Big) + \sigma^2_{e_2}\mathbb{I}_{\text{n}_{\text{w}_2}} \label{eq:SPDErepresentation}
\end{aligned},
\end{equation}
where $\mathbf{B_1}$ and $\mathbf{B_2}$ are appropriate projection matrices, which map the values of the fields from the mesh to the spatial location of the observed data $\mathbf{w}_{1t}$  and $\mathbf{w}_{2t}$, respectively. The SPDE representation of the observation models in Equation \eqref{eq:SPDErepresentation} are latent Gaussian models; thus, inference can be performed using INLA. Note that in the INLA framework, the Gaussian weights $\{\chi_{\text{h}kt}\}$ are considered latent parameters. For the data fusion model, which fits a joint model for $\mathbf{w}_{1t}$  and $\mathbf{w}_{2t}$, a Bayesian model averaging (BMA) approach with INLA is performed. This necessitates specifying a grid of values for the bias parameter $\alpha_1$ and then fitting the joint model conditional on $\alpha_1$. The final marginal posterior estimates and other posterior quantities of interest are then computed by model averaging; see \cite{villejo2025data} for details, while the original idea of BMA with INLA is discussed in \cite{gomez2020bayesian}.

\begin{figure}[H]
    \centering
    \includegraphics[width=0.55\linewidth]{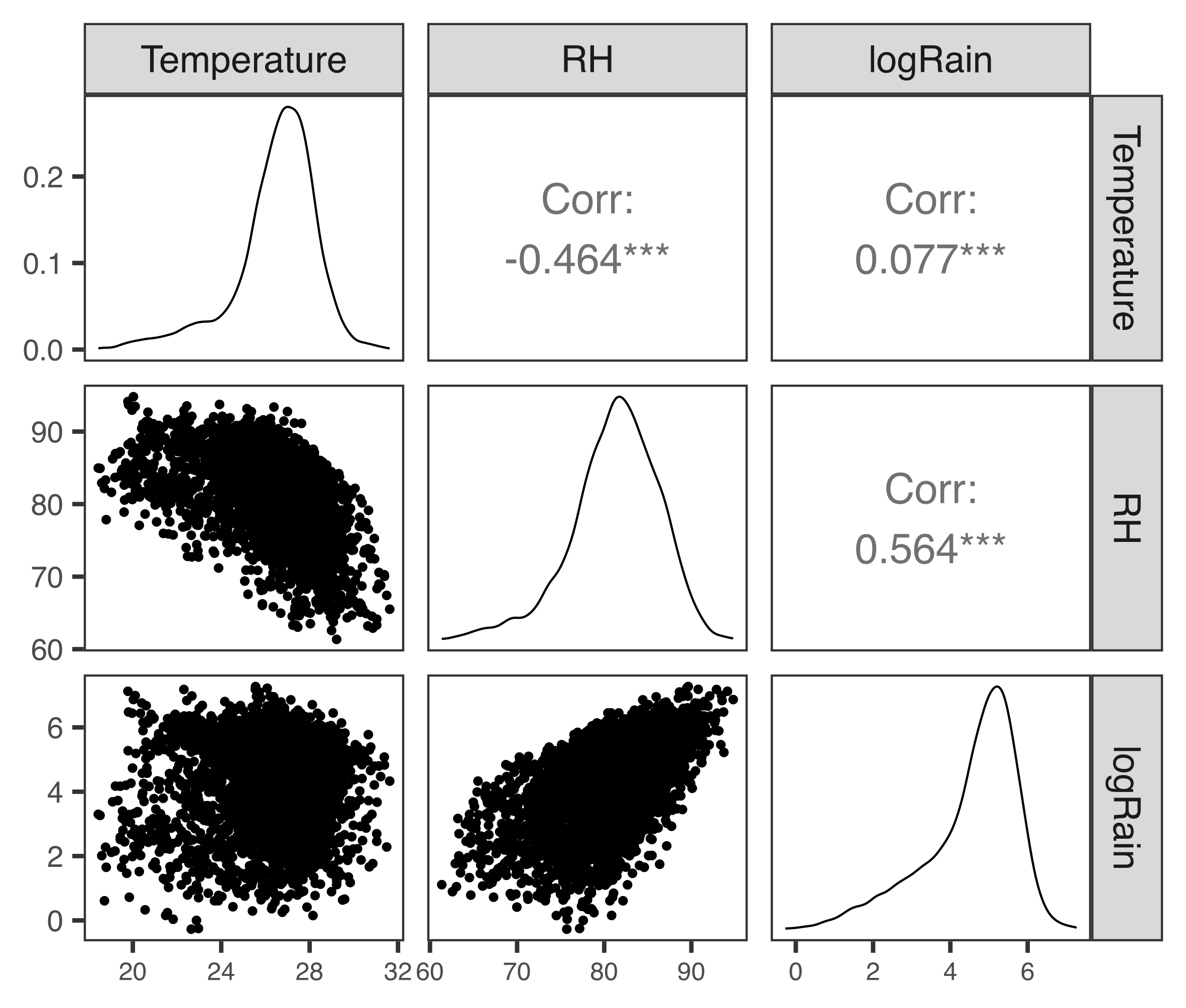}
    \caption{Pairwise correlation among the block-level estimates of the climate variables: temperature, relative humidity, and log rainfall}
    \label{fig:ggpairs_climate}
\end{figure}

\begin{table}[!h]
    \centering
    \begin{tabular}{|l|rrrr|rrrr|}
  \hline
 & \multicolumn{4}{c|}{\textbf{Plug-in method}} & \multicolumn{4}{c|}{\textbf{Resampling method}}\\
 Parameter & Mean & SD & P2.5\% & P97.5\% & Mean & SD & P2.5\% & P97.5\% \\ 
  \hline
\textcolor{brown}{$\sigma^2_{\nu}$, RW2 time} & 0.0084 & 0.0091 & 0.0001 & 0.0326 & 0.0726 & 0.1020 & 0.0060 & 0.3063 \\ 
  \textcolor{brown}{$\sigma^2_{\zeta}$, iid time} & 0.0005 & 0.0015 & 0.0000 & 0.0032 & 0.0003 & 0.0009 & 0.0000 & 0.0021 \\ 
  \textcolor{brown}{$\sigma^2_{\psi}$, space} & 0.0357 & 0.0471 & 0.0003 & 0.1657 & 0.0400 & 0.1142 & 0.0000 & 0.3363 \\
  \textcolor{brown}{$\phi$} & 0.2420 & 0.2420 & 0.0031 & 0.8415 & 0.6801 & 0.4677 & 0.0072 & 1.3654 \\ 
  \textcolor{brown}{$\sigma^2_{\upsilon}$, interaction} & 1.1442 & 0.1304 & 0.9534 & 1.4548 & 0.7028 & 0.4623 & 0.0075 & 1.3630 \\ 
  \textcolor{brown}{$\rho$} & 0.9070 & 0.0103 & 0.8900 & 0.9289  & 0.9018 & 0.0131 & 0.8751 & 0.9257  \\ 
   \hline
\end{tabular}
\caption{Comparison of hyperparameter estimates between the plug-in method and the resampling method for the dengue model with temperature and log rainfall as the climate covariate}
\label{tab:TempRainHyper}
\end{table}

\begin{table}[]
    \centering
    \begin{tabular}{|l|rrrr|rrrr|}
  \hline
 & \multicolumn{4}{c|}{\textbf{Plug-in method}} & \multicolumn{4}{c|}{\textbf{Resampling method}}\\
 Parameter & Mean & SD & P2.5\% & P97.5\% & Mean & SD & P2.5\% & P97.5\% \\ 
  \hline
\textcolor{brown}{$\sigma^2_{\nu}$, RW2 time} & 0.0127 & 0.0212 & 0.0007 & 0.0627 & 0.0185 & 0.1266 & 0.0003 & 0.0737  \\ 
  \textcolor{brown}{$\sigma^2_{\zeta}$, iid time} & 0.0002 & 0.0003 & 0.0000 & 0.0009 & 0.0003 & 0.0009 & 0.0000 & 0.0016  \\ 
  \textcolor{brown}{$\sigma^2_{\psi}$, space} & 0.0000 & 0.0000 & 0.0000 & 0.0000 & 0.0608 & 0.2589 & 0.0000 & 0.5053   \\
  \textcolor{brown}{$\phi$} & 0.1850 & 0.2130 & 0.0017 & 0.7741 & 0.7091 & 0.4780 & 0.0110 & 1.3783 \\ 
  \textcolor{brown}{$\sigma^2_{\upsilon}$, interaction} & 1.1861 & 0.1435 & 0.9487 & 1.5098 & 0.7326 & 0.4779 & 0.0117 & 1.3872
  \\ 
  \textcolor{brown}{$\rho$} & 0.9061 & 0.0116 & 0.8831 & 0.9284 & 0.9040 & 0.0120 & 0.8783 & 0.9253 \\ 
   \hline
\end{tabular}
\caption{Comparison of hyperparameter estimates between the plug-in method and the resampling method for the dengue model with relative humidity as the climate covariate}
\label{tab:RHHyper}
\end{table}

\begin{figure}[h!]
    \centering
    \subfloat[][$\gamma_0$, Intercept]
    {\includegraphics[width=0.4\linewidth]{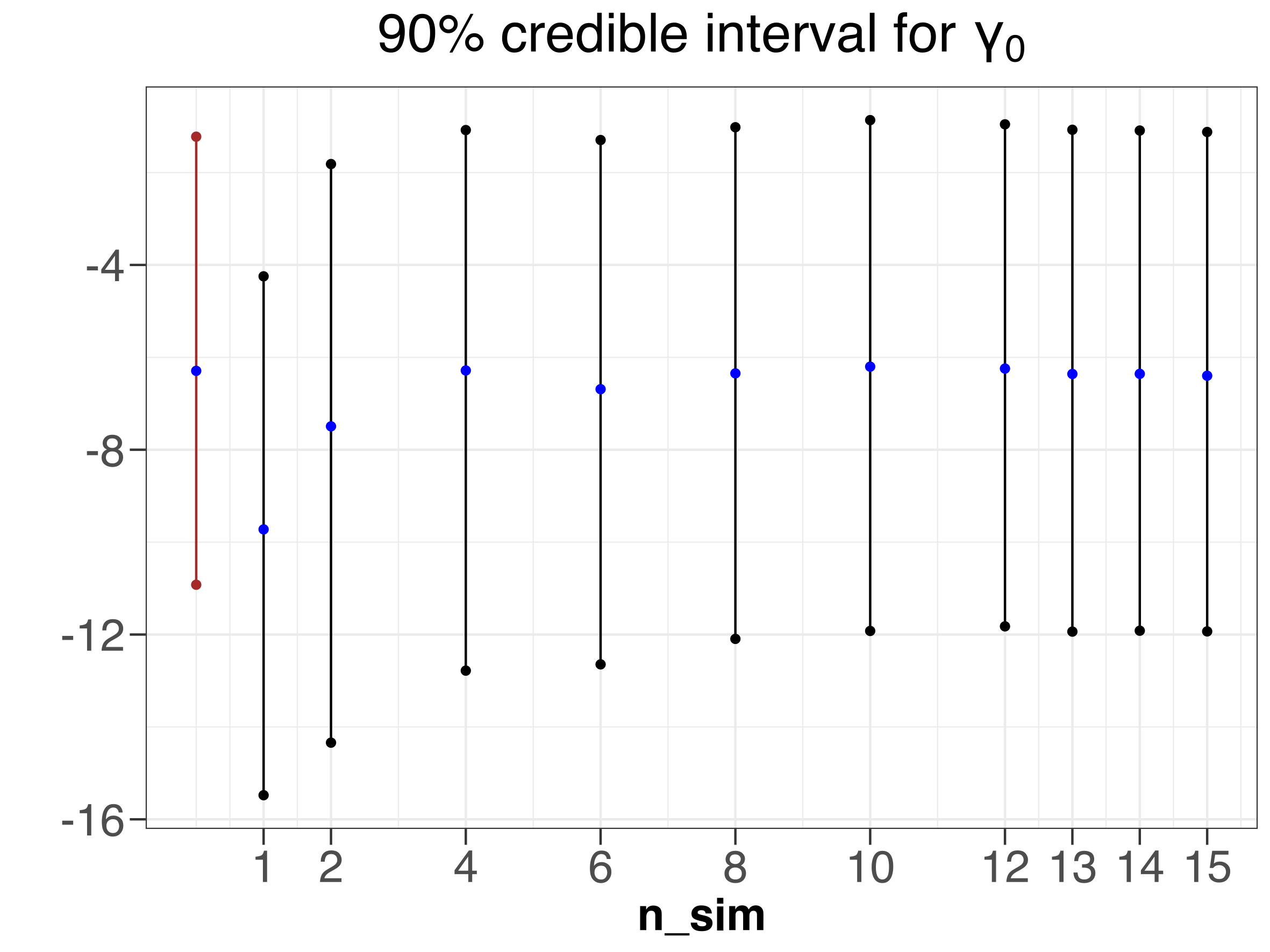}}
    \subfloat[][$\gamma_3$, log Rain]
    {\includegraphics[width=0.4\linewidth]{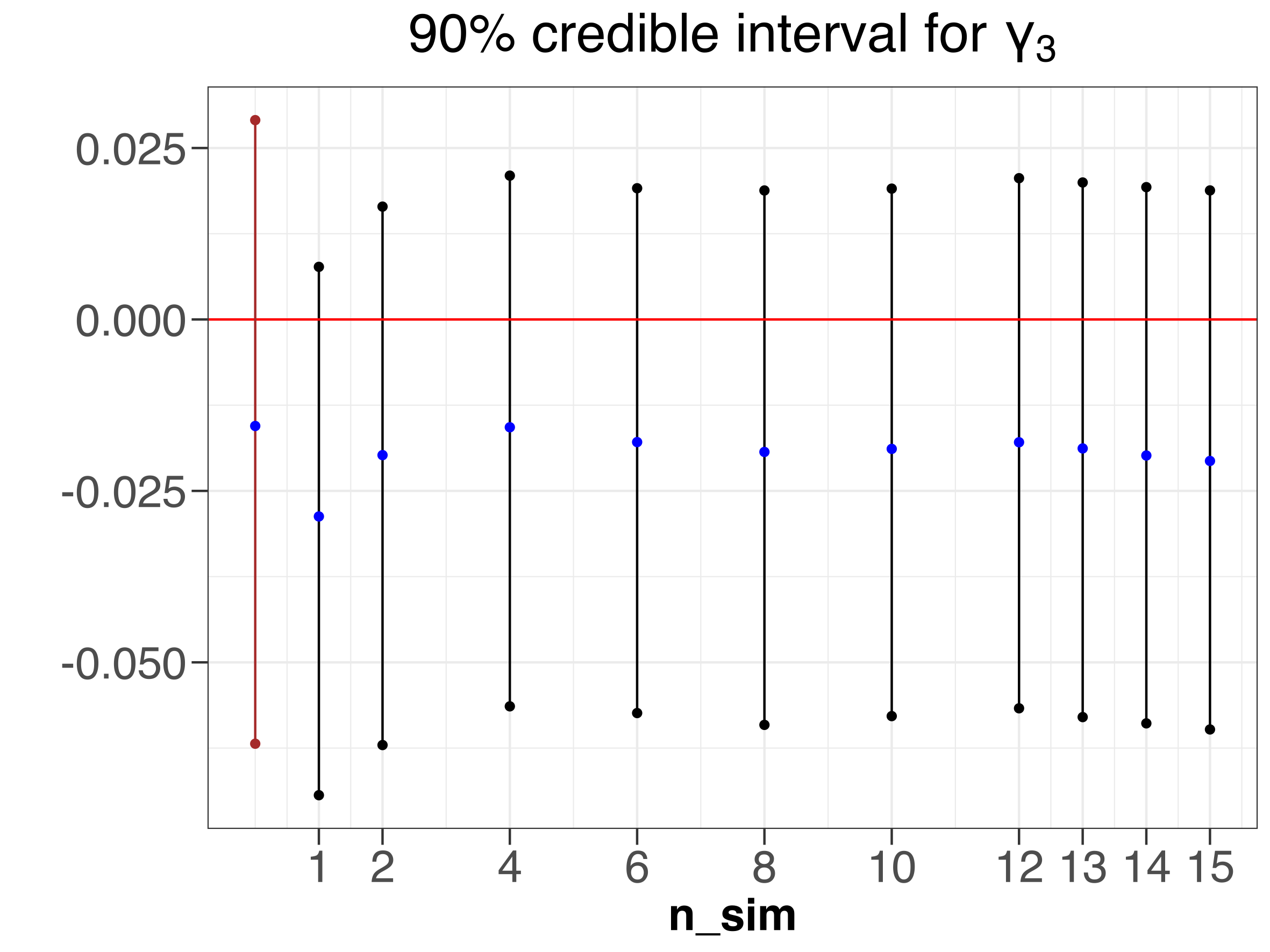}}

    \subfloat[][$\gamma_4$, Climate Type]
    {\includegraphics[width=0.4\linewidth]{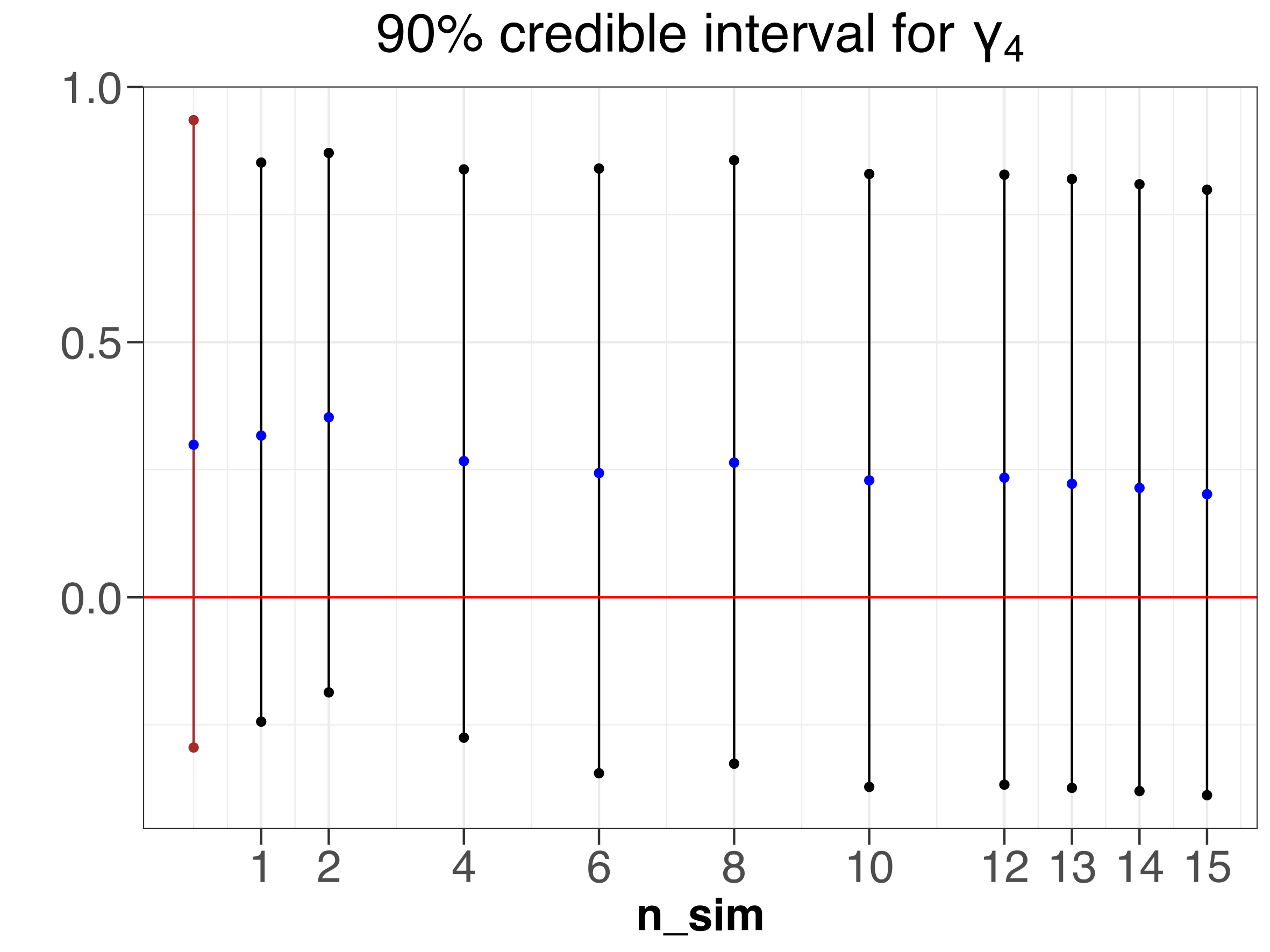}}
    \subfloat[][$\gamma_6$, covid]
    {\includegraphics[width=0.4\linewidth]{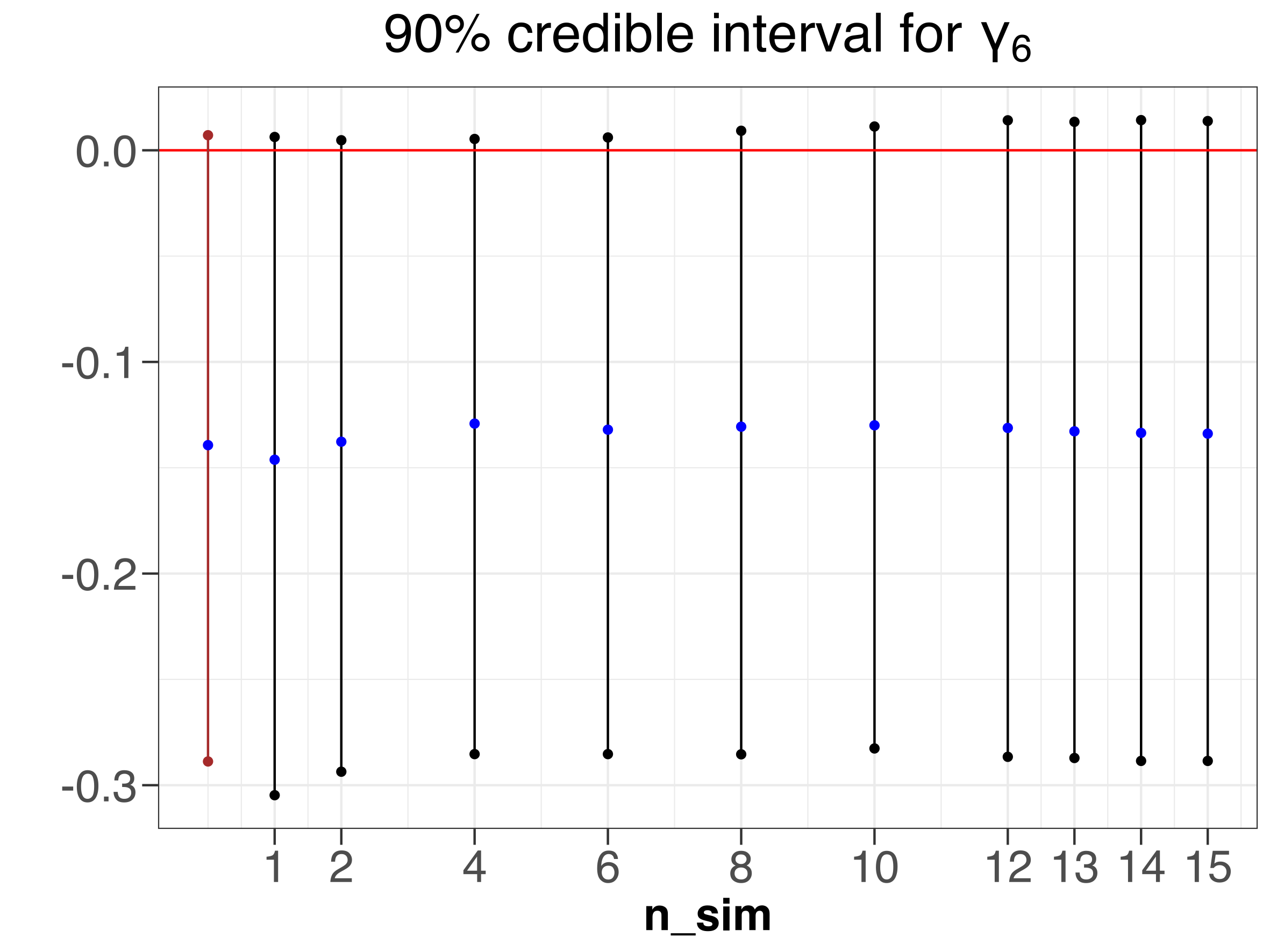}}

    \subfloat[][$\gamma_7$, log Population Density]
    {\includegraphics[width=0.4\linewidth]{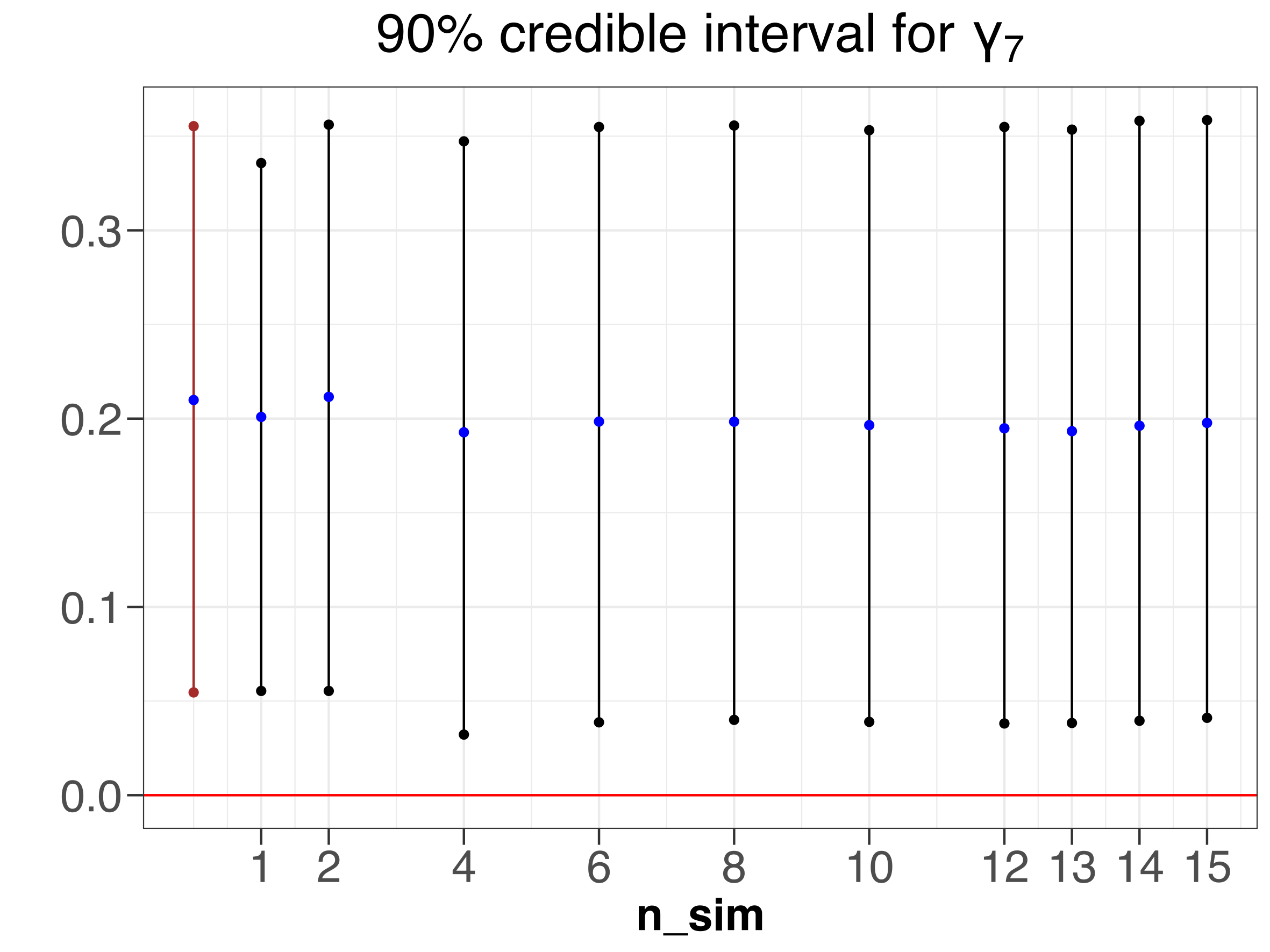}}
    
    \caption{Plots showing the posterior means and 90\% credible intervals of the fixed effects (except $\gamma_1$, $\gamma_2$, and $\gamma_5$) for the dengue model with temperature and log rainfall as covariates. The first vertical line shows the estimates for the plug-in method, while the rest of the lines show the estimates for the resampling method for different number of resamples, from 1 to 15. }
    \label{fig:TempRainOtherGammasDataFusion_risk}
\end{figure}

\begin{figure}[h!]
    \centering
    \subfloat[][$\gamma_0$, Intercept]
    {\includegraphics[width=0.4\linewidth]{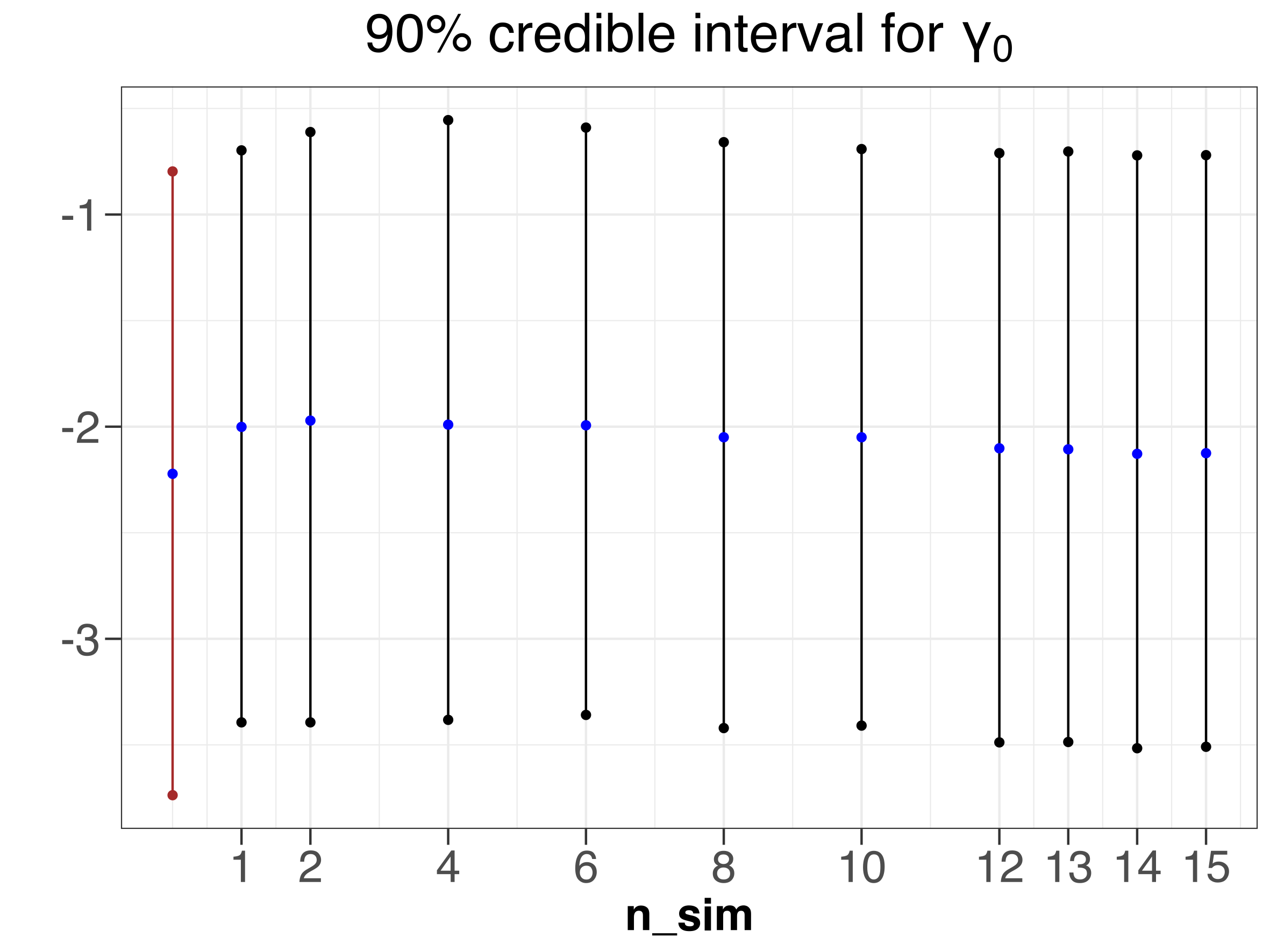}}
    \subfloat[][$\gamma_1$, Relative humidity]
    {\includegraphics[width=0.4\linewidth]{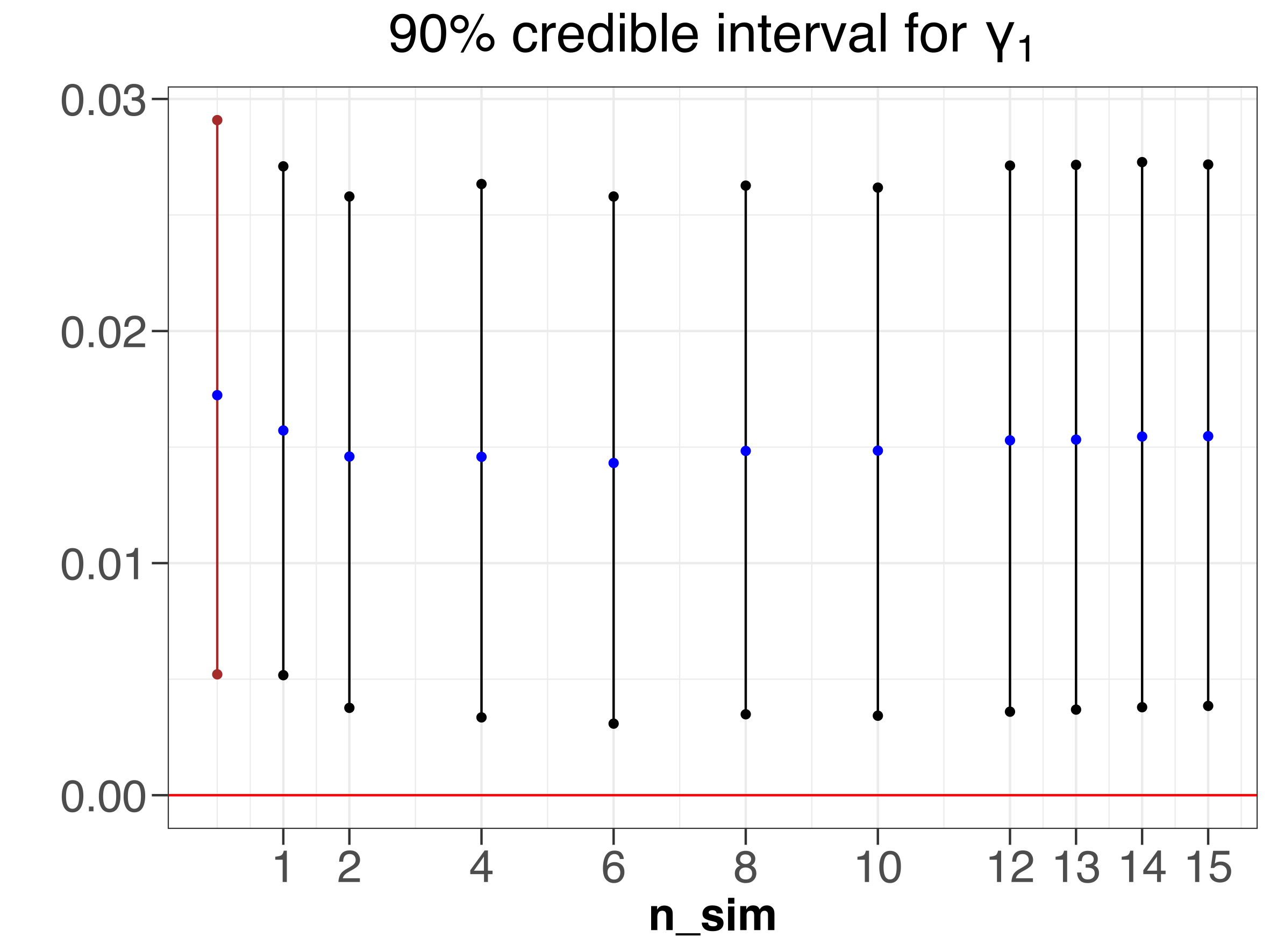}}

    \subfloat[][$\gamma_2$, Climate Type]
    {\includegraphics[width=0.4\linewidth]{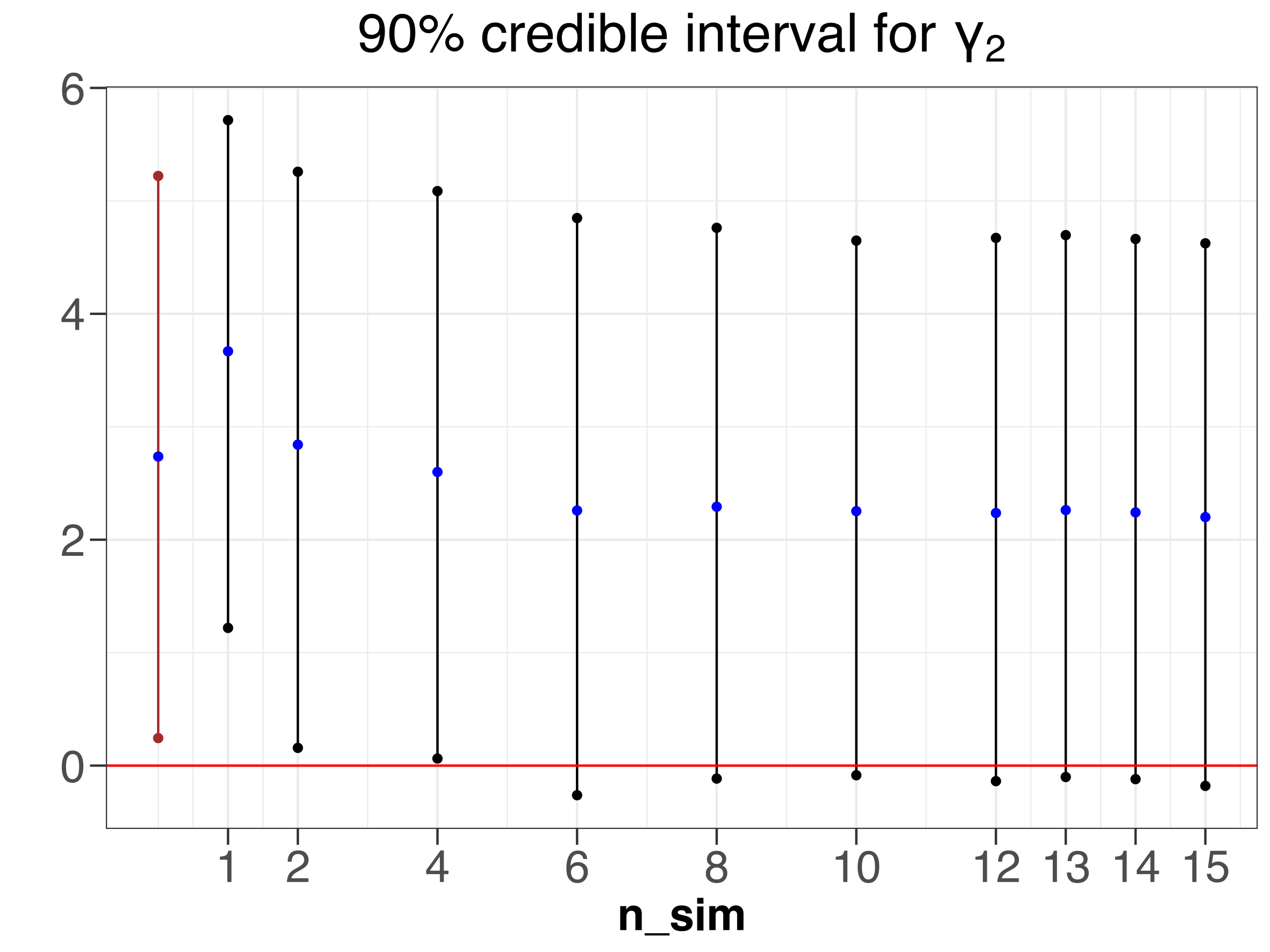}}
    \subfloat[][$\gamma_3$, RH $\times$ Climate Type]
    {\includegraphics[width=0.4\linewidth]{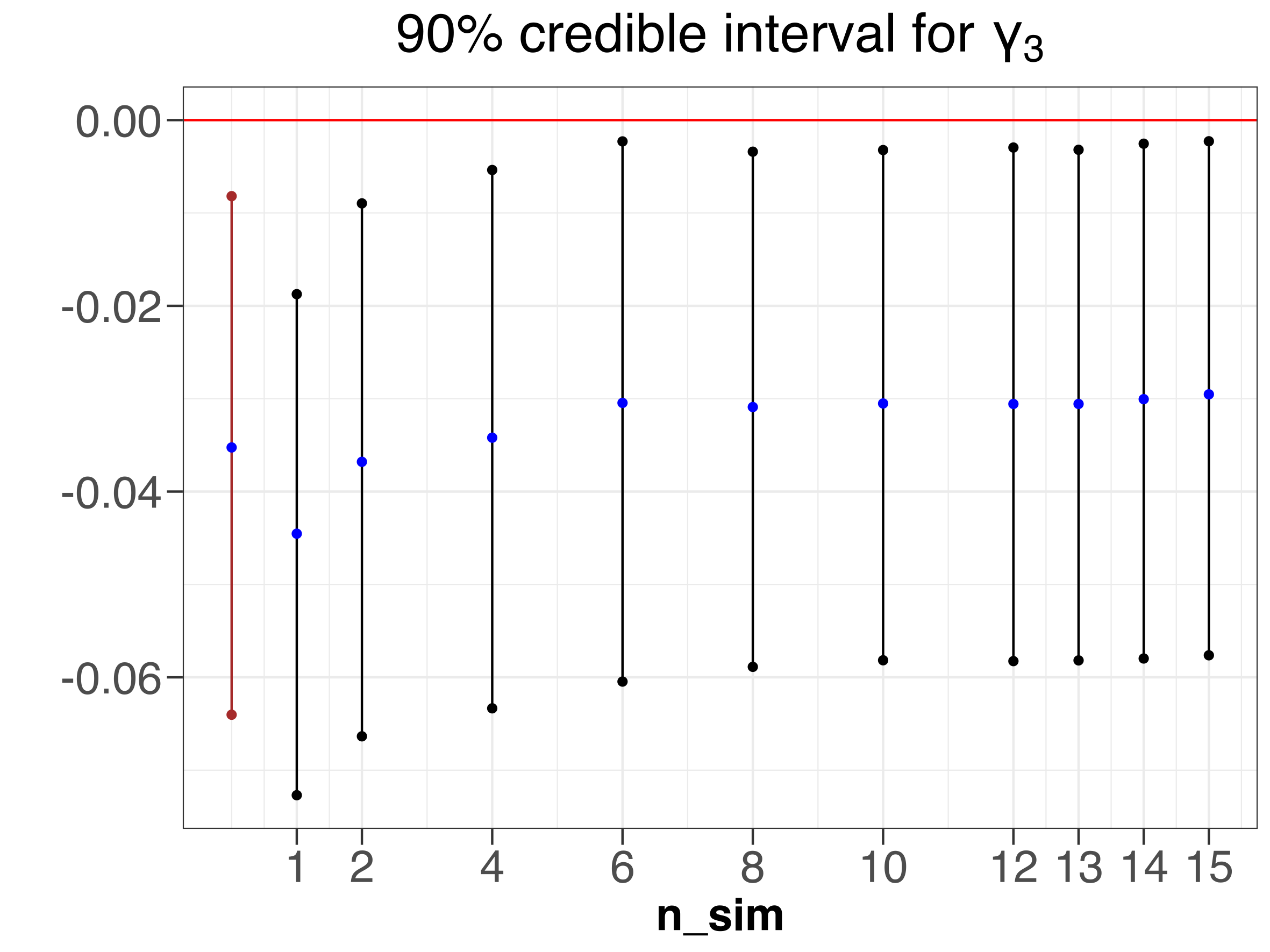}}

    \subfloat[][$\gamma_4$, covid]
    {\includegraphics[width=0.4\linewidth]{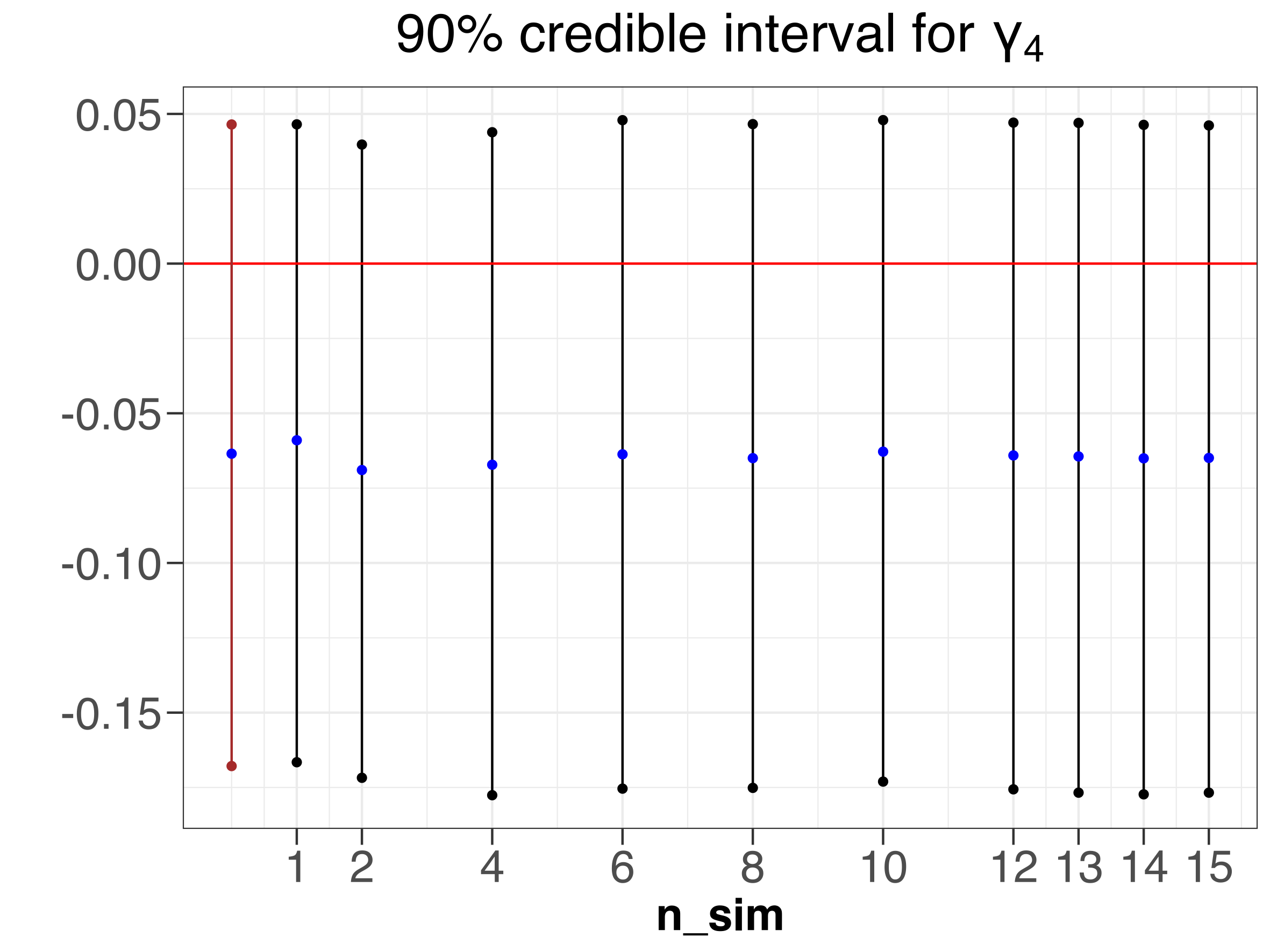}}
    \subfloat[][$\gamma_5$, log Population Density]
    {\includegraphics[width=0.4\linewidth]{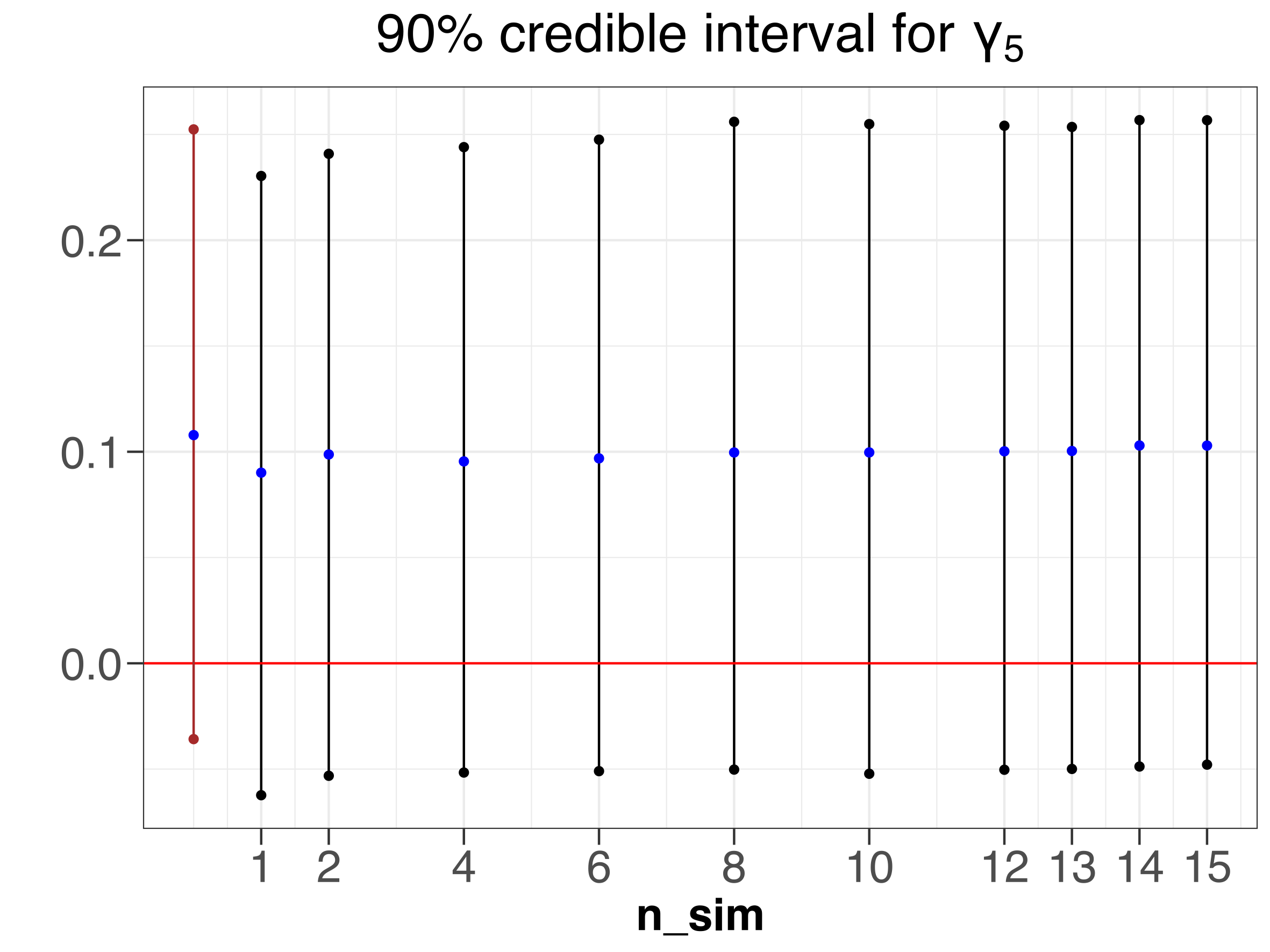}}
    \caption{Plots showing the posterior means and 90\% credible intervals of the fixed effects for the dengue model with relative humidity as climate covariate. The first vertical line shows the estimates for the plug-in method, while the rest of the lines show the estimates for the resampling method for different number of resamples, from 1 to 15. }
    \label{fig:RHOtherGammasDataFusion_risk}
\end{figure}

\begin{figure}[h!]
    \centering
    \subfloat[][Posterior mean]
    {\includegraphics[width=0.4\linewidth]{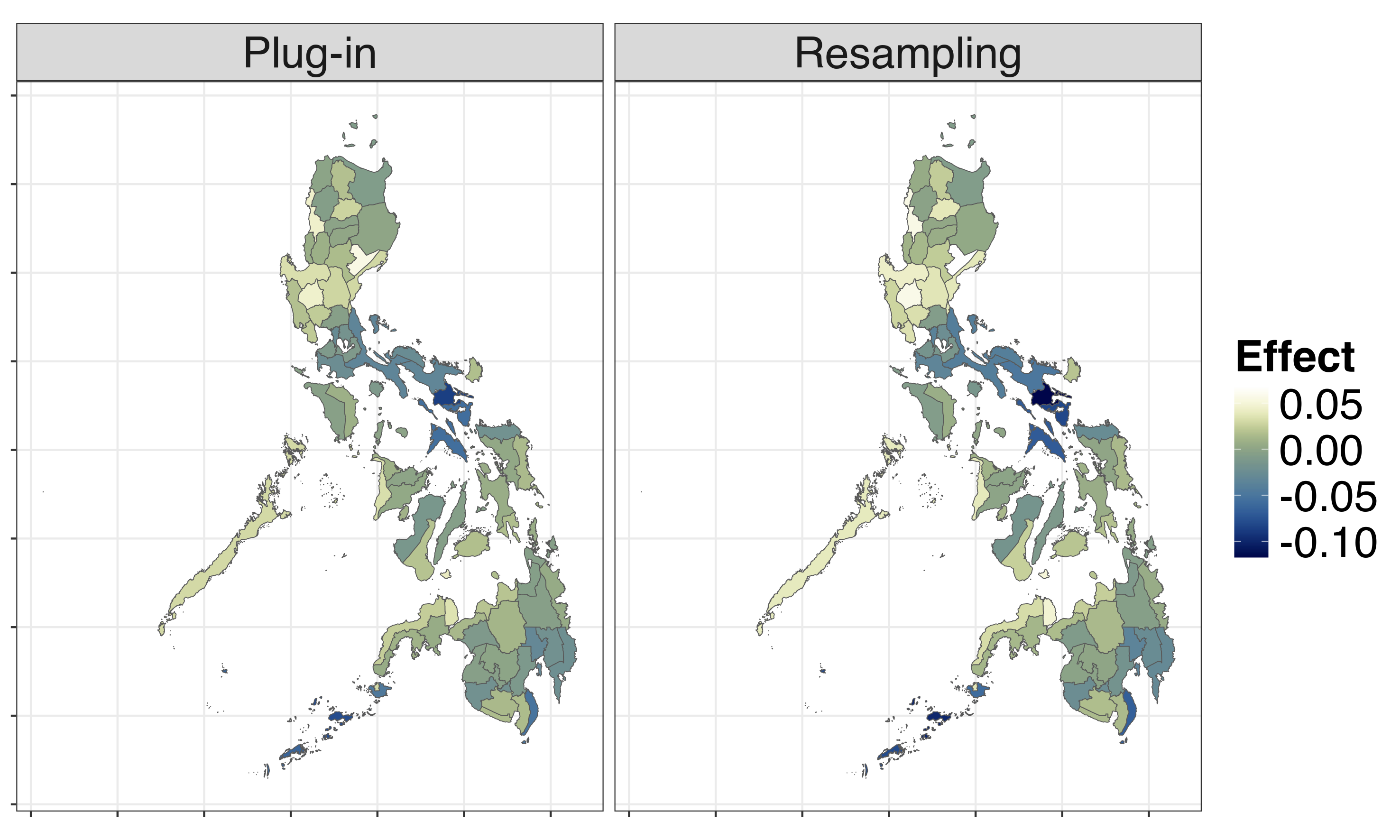}}
    \subfloat[][Posterior standard deviation]
    {\includegraphics[width=0.4\linewidth]{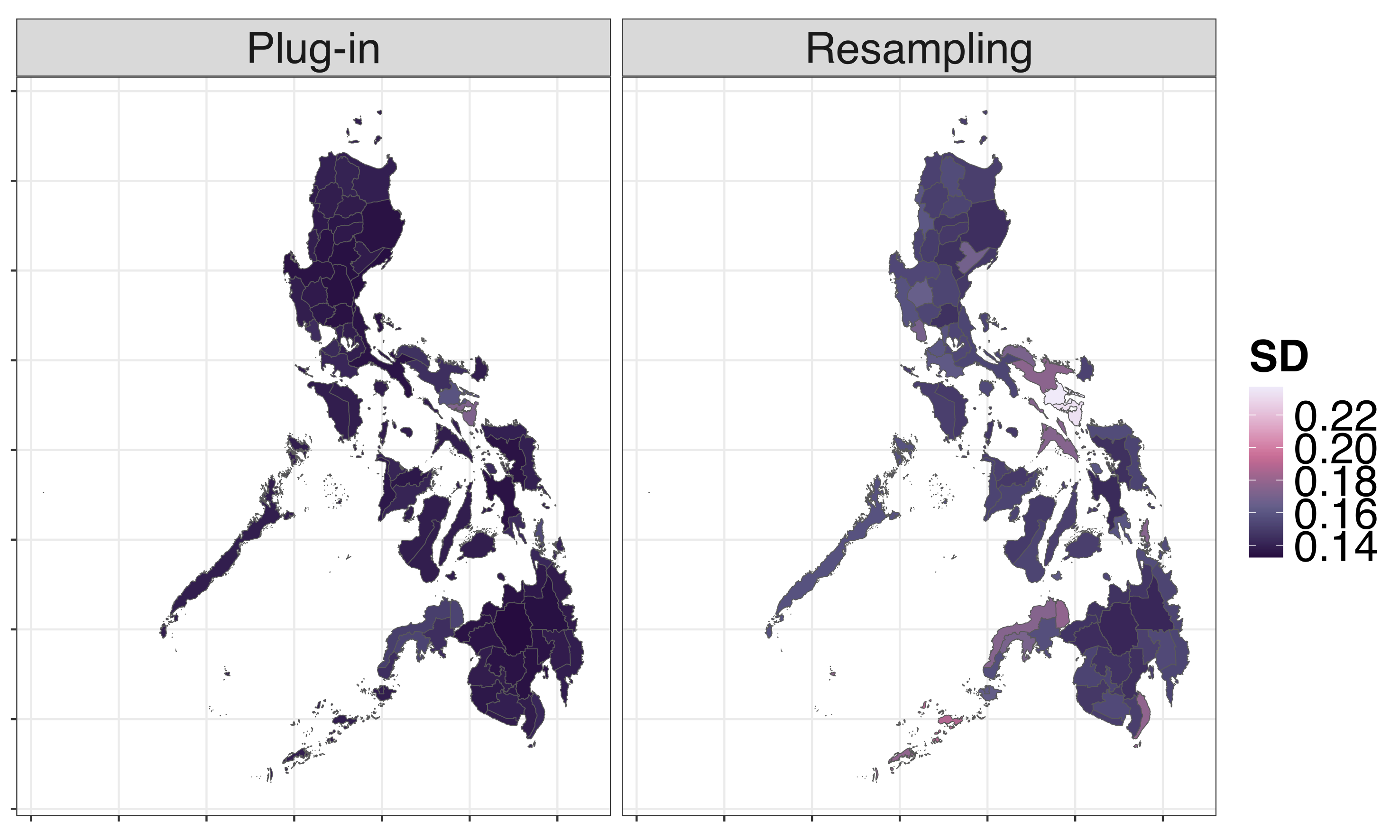}}
    \caption{Comparison of (a) posterior mean and (b) posterior standard deviation of the space effects $\psi(B_i)$ between the plug-in method and resampling method, for the model with relative humidity as climate covariate}
    \label{fig:RHSpaceMeanSD}
\end{figure}



\begin{figure}[h!]
    \centering
    \captionsetup{justification=centering}
    \subfloat[][Plug-in method]
    {\includegraphics[width=0.25\linewidth]{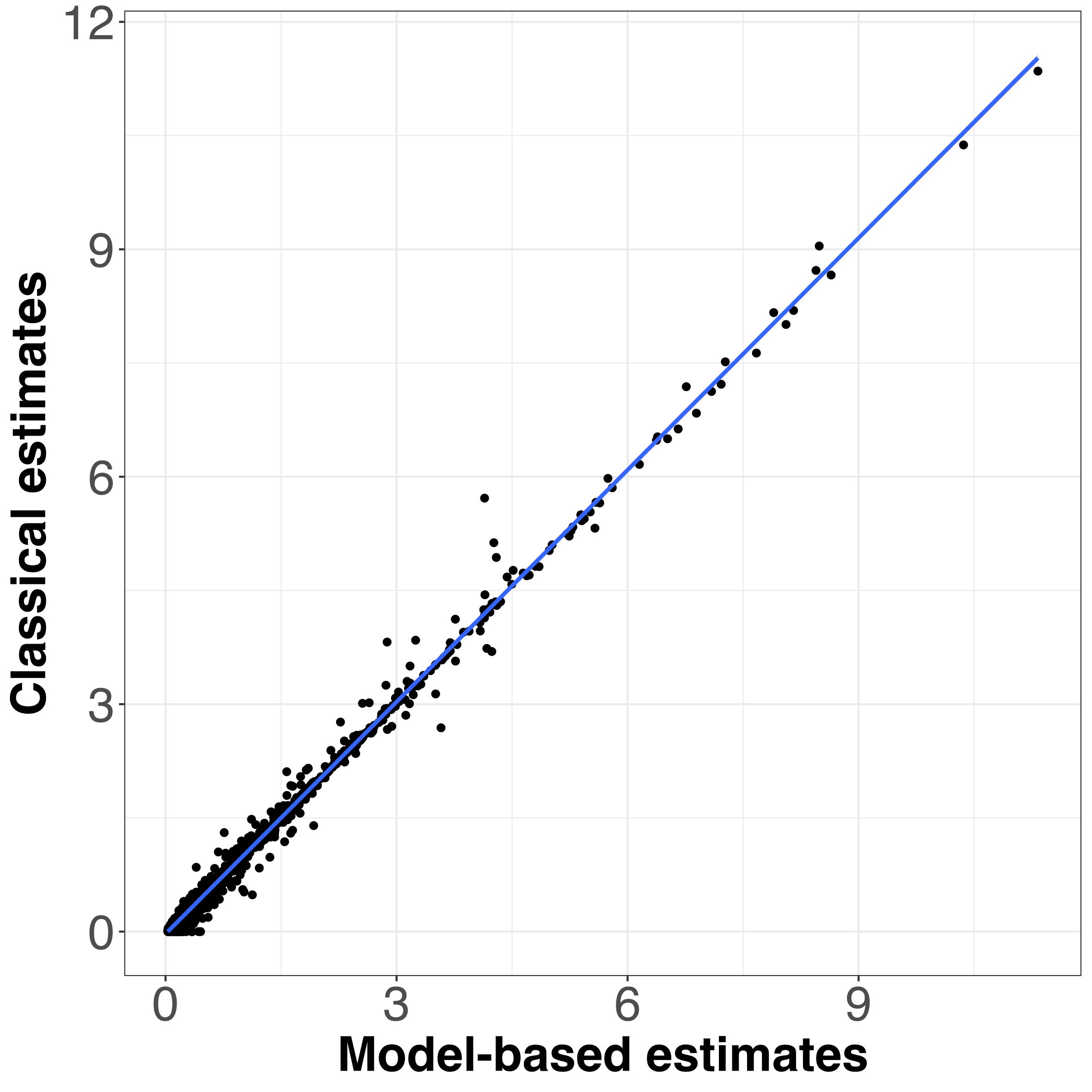}}\hspace{10mm}
    \subfloat[][Resampling method]
    {\includegraphics[width=0.25\linewidth]{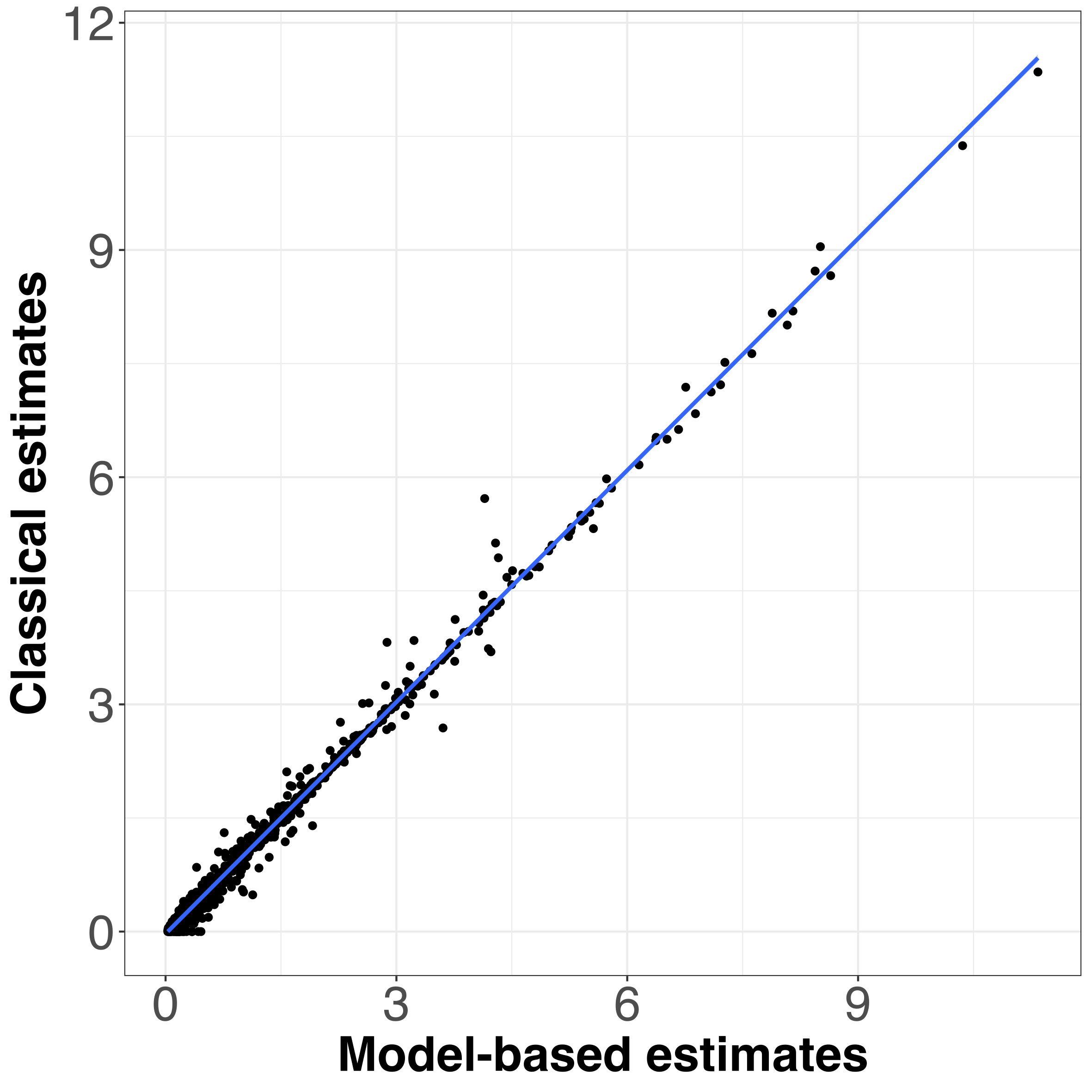}}
    \caption{Comparison of classical SIR estimates and model-based SIR estimates from the health model with relative humidity as climate covariate: (a) plug-in method (b) resampling method}
    \label{fig:RHScatterplotSIR}
\end{figure}

\begin{figure}[!h]
    \centering
    \includegraphics[width=0.8\linewidth]{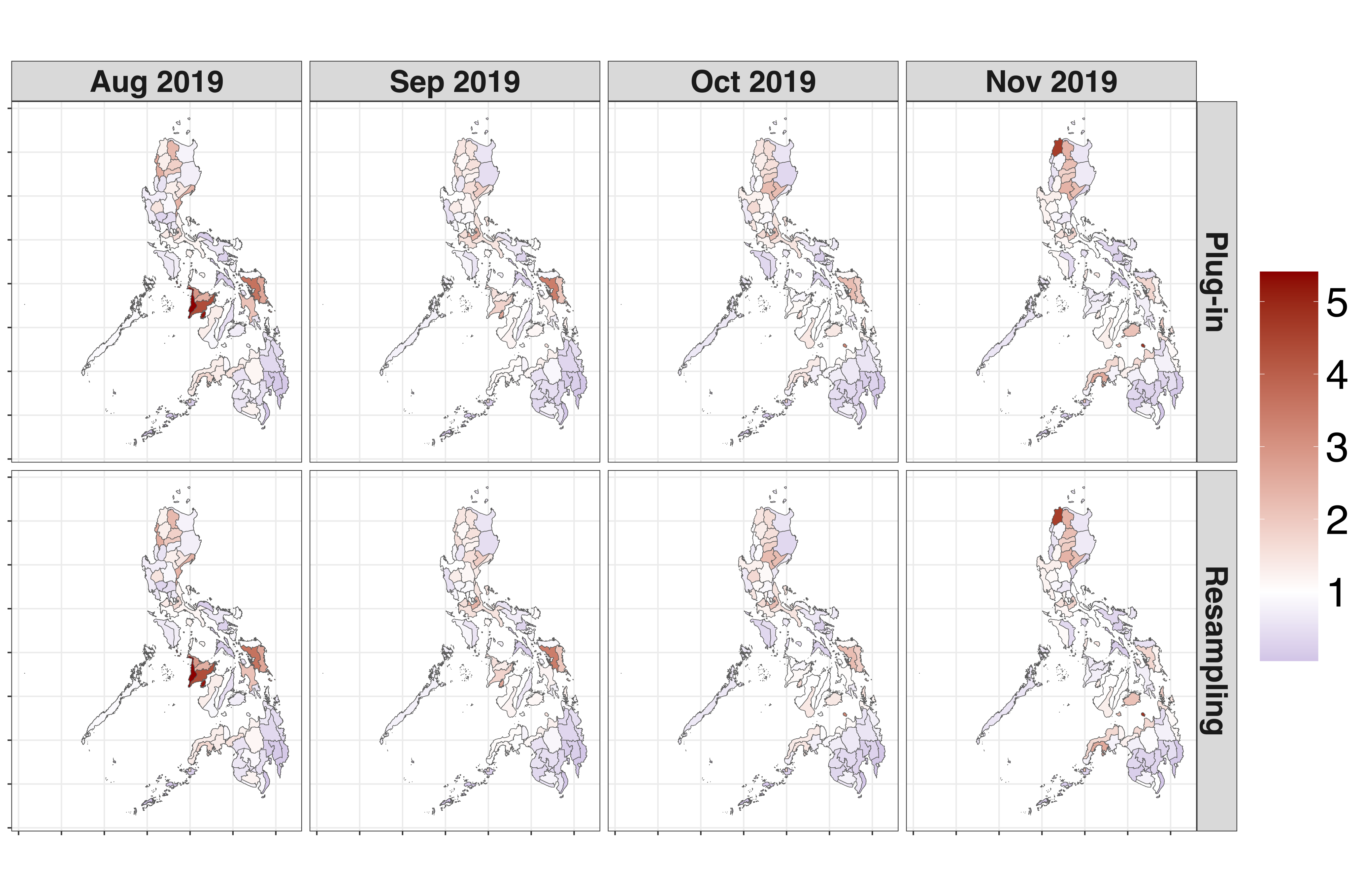}
    \caption{Model-based estimates of dengue risks, $\hat{\lambda}(B_i,t)$, from August 2019 to November 2019, for both plug-in method and resampling method on the dengue model with temperature and log rainfall as climate covariates}
    \label{fig:SIRsDataFusionInput}
\end{figure}

\clearpage

\renewcommand{\kbldelim}{(}
\renewcommand{\kbrdelim}{)}
\begin{equation}\label{eq:varcov_fixed_temprain_datafusion}
  \kbordermatrix{
    & \gamma_1x_1 & \gamma_2x_2 & \gamma_3x_3 & \gamma_4x_4 & \gamma_5x_5 & \gamma_6x_6\\
    \gamma_1x_1 & 1.20517 & -0.00111 & -0.00725 & -0.05684 & -0.21671 & 0.00397 \\
    \gamma_2x_2 & \cdot & 0.00049 & -0.00001 & -0.00005 & 0.00076 & 0.00018 \\
    \gamma_3x_3 & \cdot & \cdot  & 0.00018 & 0.00101 & -0.00352 & -0.00001 \\
    \gamma_4x_4 & \cdot & \cdot & \cdot & 0.10300 & 0.02234 & 0.00311 \\
    \gamma_5x_5 & \cdot & \cdot & \cdot & \cdot & 0.80793 & 0.00035 \\
    \gamma_6x_6 & \cdot & \cdot & \cdot & \cdot & \cdot &  0.00839
  }
\end{equation}
\captionof{mat}{Variance-covariance matrix across several resamples for the fixed effects components of the linear predictor of the health model with temperature and log rainfall as climate covariates}

\renewcommand{\kbldelim}{(}
\renewcommand{\kbrdelim}{)}
\begin{equation}\label{eq:varcov_random_temprain_datafusion}
  \kbordermatrix{
    & \nu(t) & \zeta(t) & \psi(B) & \upsilon(B,t) \\
    \nu(t) & 0.05218 & 0.00006 & 0.00000 & 0.00619   \\
    \zeta(t) & \cdot & 0.00001 & -0.00000 & -0.00000\\
    \psi(B) & \cdot & \cdot & 0.00500 & 0.00107 \\
    \upsilon(B,t) & \cdot & \cdot & \cdot & 0.93642
  }
\end{equation}
\captionof{mat}{Variance-covariance matrix across several resamples for the random effects components of the linear predictor of the health model with temperature and log rainfall as climate covariates}

\renewcommand{\kbldelim}{(}
\renewcommand{\kbrdelim}{)}
\begin{equation}\label{eq:crosscov_temprain_datafusion}
  \kbordermatrix{
    & \gamma_1x_1 & \gamma_2x_2 & \gamma_3x_3 & \gamma_4x_4 & \gamma_5x_5 & \gamma_6x_6\\
    \nu(t) & 0.00471 & -0.00000 & -0.00000 & -0.00019 & 0.00055 & 0.00064   \\
    \zeta(t) & -0.00003 & 0.00000 & 0.00000 & 0.00000 & -0.00001 & -0.00001 \\
   \psi(B) & -0.01027 & 0.00012 & 0.00009 & 0.00732 & 0.01425 & -0.00000 \\
    \upsilon(B,t) & 0.02381 & 0.00023 & -0.00020 & -0.00854 & -0.01481 & 0.00149 
  }
\end{equation}
\captionof{mat}{Cross-covariance matrix across several resamples between the fixed effects and random effects components of the linear predictor of the health model with temperature and log rainfall as climate covariates}

\begin{figure}[H]
    \centering
    \includegraphics[width=0.8\linewidth]{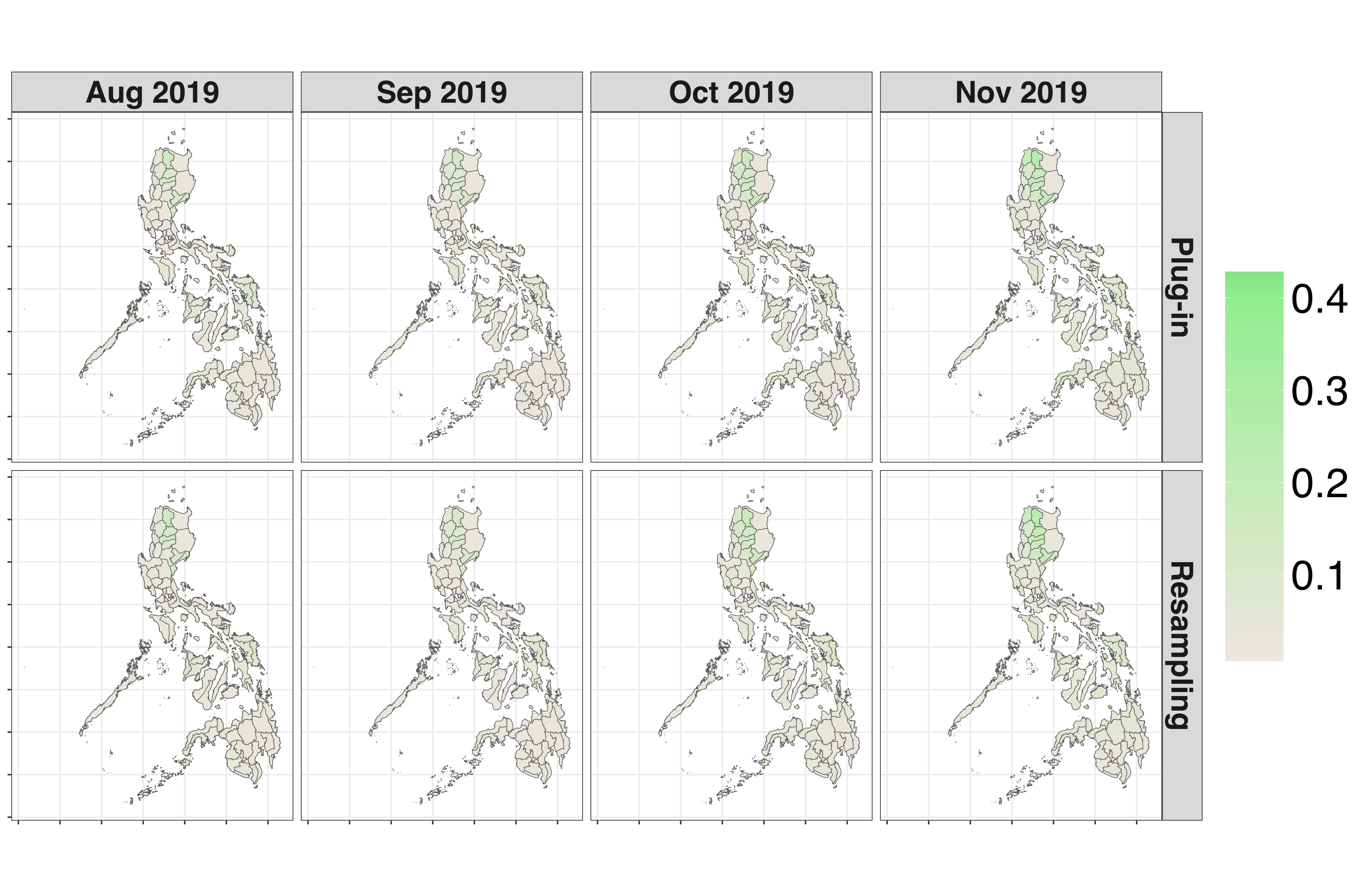}
    \caption{Posterior uncertainty of model-based estimates of dengue risks from August 2019 to November 2019, for both plug-in method and resampling on the dengue model with temperature and log rainfall as climate covariates}
    \label{fig:SIRsDataFusionInputSD}
\end{figure}

\begin{figure}[!h]
    \centering
    \includegraphics[width=0.8\linewidth]{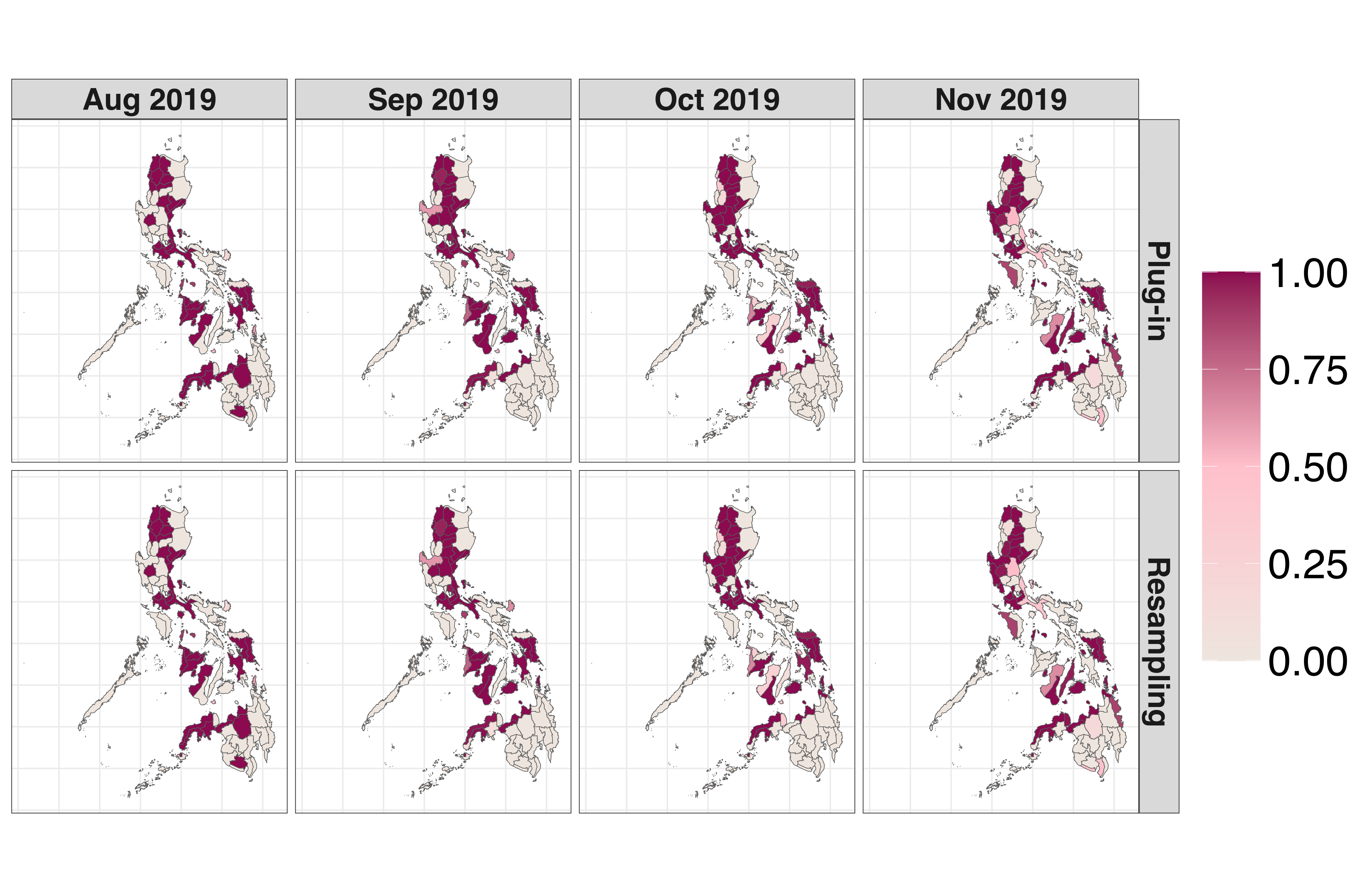}
    \caption{Probability of exceedence, i.e., $\mathbb{P}\big(\lambda(B_i,t)>1 \big)$ from August 2019 to November 2019, for both plug-in method and resampling method on the dengue model with temperature and log rainfall as climate covariates}
    \label{fig:ProbExceedanceDataFusion}
\end{figure}

\end{document}